\documentclass[twocolumn]{aastex631}

\newcommand{\kms}{\ensuremath{\,\rm{km}\,\rm{s}^{-1}}\xspace}
\newcommand{\Rsun}{\ensuremath{\,\rm{R_\odot}}\xspace}
\newcommand{\Msun}{\ensuremath{\,\rm{M}_{\odot}}\xspace}

\newcommand{\Mbh}{\ensuremath{M_{\rm BH}}\xspace}
\newcommand{\Mbheen}{\ensuremath{M_{\rm BH, 1}}\xspace}
\newcommand{\Mbhtwee}{\ensuremath{M_{\rm BH, 2}}\xspace}
\newcommand{\Mco}{\ensuremath{M_{\rm CO}}\xspace}
\newcommand{\Mchirp}{\ensuremath{M_{\rm chirp}}\xspace}
\newcommand{\tdelay}{\ensuremath{t_{\rm delay}}\xspace}
\newcommand{\tform}{\ensuremath{t_{\rm form}}\xspace}

\newcommand{\MbhA}{\ensuremath{M_{\rm BH, A}}\xspace}
\newcommand{\MbhB}{\ensuremath{M_{\rm BH, B}}\xspace}
\newcommand{\MZAMSa}{\ensuremath{M_{\rm ZAMS, A}}\xspace}
\newcommand{\MZAMSb}{\ensuremath{M_{\rm ZAMS, B}}\xspace}

\newcommand{\Zsun}{\ensuremath{\,\rm{Z}_{\rm \odot}}\xspace}
\newcommand{\Lsun}{\ensuremath{\,\rm{L}_{\rm \odot}}\xspace}

\newcommand{\Myr}{\ensuremath{\,\rm{Myr}}\xspace}
\newcommand{\Gyr}{\ensuremath{\,\rm{Gyr}}\xspace}

\newcommand{\Gpc}{\ensuremath{\,\rm{Gpc}}\xspace}

\newcommand{\CEa}{\ensuremath{\alpha_{\rm{CE}}}\xspace}
\newcommand{\RBBH}{\ensuremath{R_{\mathrm{BBH}}(z)}\xspace}
\newcommand{\RBBHzeta}{\ensuremath{R_{\mathrm{BBH}}(z, \zeta)}\xspace}
\newcommand{\RBBHav}{\ensuremath{\overline{R}_{\mathrm{BBH}}(z)}\xspace}
\newcommand{\RBBHdet}{\ensuremath{\mathcal{R}^{det}(z,\zeta)}\xspace}
\newcommand{\RBBHdetU}{\ensuremath{\mathcal{R}^{det}_{\mathrm{Univ}}(\zeta)}\xspace}
\newcommand{\RBBHdetz}{\ensuremath{\mathcal{R}^{det}(z)}\xspace}

\newcommand{\SFRDzZ}{\ensuremath{\mathcal{S}(Z,z)}\xspace} 
\newcommand{\SFRDz}{\ensuremath{\mathrm{SFRD}(z)}\xspace} 

\newcommand{\dpdZ}{\ensuremath{d\mathrm{P}(Z,z)/dZ}\xspace}
\newcommand{\qfinal}{\ensuremath{q_{\mathrm{final}} }\xspace}
\newcommand{\qBBH}{\ensuremath{q_{\mathrm{BBH}} }\xspace}

\newcommand{\COMPAS}{{\tt COMPAS}\xspace}

\usepackage{xspace}
\usepackage{cancel}
\usepackage{amsmath}
\usepackage{hyperref}
\usepackage{lipsum}
\usepackage{amsmath}
\usepackage{graphicx} 
\graphicspath{{figures/}} 

\begin{document}

\title{The redshift evolution of the binary black hole merger rate: a weighty matter}

\correspondingauthor{L.~van Son}
\email{lieke.van.son@cfa.harvard.edu}

\author[0000-0001-5484-4987]{L.~A.~C.~van~Son}
\affiliation{Center for Astrophysics $|$ Harvard $\&$ Smithsonian,60 Garden St., Cambridge, MA 02138, USA}
\affiliation{Anton Pannekoek Institute for Astronomy, University of Amsterdam, Science Park 904, 1098XH Amsterdam, The Netherlands}
\affiliation{Max Planck Institute for Astrophysics, Karl-Schwarzschild-Str. 1, 85748 Garching, Germany}
\author[0000-0001-9336-2825]{S. E. de Mink}
\affiliation{Max Planck Institute for Astrophysics, Karl-Schwarzschild-Str. 1, 85748 Garching, Germany}
\affiliation{Anton Pannekoek Institute for Astronomy, University of Amsterdam, Science Park 904, 1098XH Amsterdam, The Netherlands}
\affiliation{Center for Astrophysics $|$ Harvard $\&$ Smithsonian,
60 Garden St., Cambridge, MA 02138, USA}
\author[0000-0001-9892-177X]{T. Callister}
\affiliation{Center for Computational Astrophysics, Flatiron Institute, New York, NY 10010, USA}
\author[0000-0001-7969-1569]{S. Justham}
\affiliation{School of Astronomy and Space Science, University of the Chinese Academy of Sciences, Beijing 100012, China}
\affiliation{Anton Pannekoek Institute for Astronomy, University of Amsterdam, Science Park 904, 1098XH Amsterdam, The Netherlands}
\affiliation{Max Planck Institute for Astrophysics, Karl-Schwarzschild-Str. 1, 85748 Garching, Germany}
\author[0000-0002-6718-9472]{M. Renzo}
\affiliation{Department of Physics, Columbia University, New York, NY 10027, USA}
\affiliation{Center for Computational Astrophysics, Flatiron Institute, New York, NY 10010, USA}
\author[0000-0001-6147-5761]{T. Wagg}
\affiliation{Department of Astronomy, University of Washington, Seattle, WA, 98195}
\affiliation{Center for Astrophysics $|$ Harvard $\&$ Smithsonian,
60 Garden St., Cambridge, MA 02138, USA}
\affiliation{Max Planck Institute for Astrophysics, Karl-Schwarzschild-Str. 1, 85748 Garching, Germany}
\author[0000-0002-4421-4962]{F. S. Broekgaarden}
\affiliation{Center for Astrophysics $|$ Harvard $\&$ Smithsonian,
60 Garden St., Cambridge, MA 02138, USA}
\author[0000-0002-6056-3070]{F. Kummer}
\affiliation{Anton Pannekoek Institute for Astronomy, University of Amsterdam, Science Park 904, 1098XH Amsterdam, The Netherlands}
\author[0000-0003-3308-2420]{R. Pakmor}
\affiliation{Max Planck Institute for Astrophysics, Karl-Schwarzschild-Str. 1, 85748 Garching, Germany}
\author[0000-0002-6134-8946]{I. Mandel}
\affiliation{Monash Centre for Astrophysics, School of Physics and Astronomy, Monash University, Clayton, Victoria 3800, Australia}
\affiliation{The ARC Center of Excellence for Gravitational Wave Discovery -- OzGrav, Hawthorn VIC 3122, Australia}
\affiliation{Birmingham Institute for Gravitational Wave Astronomy and School of Physics and Astronomy, University of Birmingham, Birmingham, B15 2TT, United Kingdom}

\begin{abstract}
Gravitational wave detectors are starting to reveal the redshift evolution of the binary black hole (BBH) merger rate, \RBBH. 
We make predictions for \RBBH as a function of black hole mass for systems originating from isolated binaries. 
To this end, we investigate correlations between the delay time and black hole mass by means of the suite of binary population synthesis simulations, \COMPAS. We distinguish two channels: the common envelope (CE), and the stable Roche-lobe overflow (RLOF) channel, characterised by whether the system has experienced a common envelope or not.
We find that the CE channel preferentially produces BHs with masses below about 30\Msun and short delay times ($\tdelay \lesssim 1 \Gyr$), while the stable RLOF channel primarily forms systems with BH masses above $30\Msun$ and long delay times ($\tdelay \gtrsim 1 \Gyr$). 
We provide a new fit for the metallicity specific star-formation rate density based on the Illustris TNG simulations, and use this to convert the delay time distributions into a prediction of \RBBH.
This leads to a distinct redshift evolution of \RBBH for high and low primary BH masses.
We furthermore find that, at high redshift, \RBBH is dominated by the CE channel, while at low redshift it contains a large contribution ($\sim 40\%$) from the stable RLOF channel. Our results predict that, for increasing redshifts, BBHs with component masses above 30\Msun will become increasingly scarce relative to less massive BBH systems.
Evidence of this distinct evolution of \RBBH for different BH masses can be tested with future detectors.
\end{abstract}

\section{Introduction}
\label{sec:intro}
The Advanced LIGO \citep{AdvancedLIGO2015}, Advanced Virgo \citep{AdvancedVirgo2015} and KAGRA \citep{KAGRA2021} gravitational wave detectors are revealing gravitational wave events that probe a progressively larger fraction of the Universe   \citep{GWTC1, GWTC2,GWTC2_1_2021, GWTC3} . As the number of gravitational wave detections increases, they unveil the evolution of the binary black hole (BBH) merger rate with redshift.  Current gravitational wave detectors already probe black holes (BHs) with component masses of about $30\Msun$ out to redshifts $z \sim 1$ \citep{2018ApJ...863L..41F,Callister2020_shoutsMurm, GWTC3, GWTC3_popPaper2021,GWTC3_HubbleConst2021}. 
Third-generation detectors, scheduled to start observations in the 2030s, promise to observe stellar mass BBH mergers with component masses in the range $\sim5 - 350\Msun$ out to $z > 10$ \citep[e.g.\ ][]{Sathyaprakash2019_CTET,Sathyaprakash2019,Maggiore2020}.
This means that we are rapidly moving towards a complete picture of both the redshift evolution of the stellar-mass BBHs merger rate, and the redshift evolution of source property distributions.

The redshift evolution of the BBH merger rate contains information on the origin of these BBHs, however, a direct interpretation is complicated. To infer the birth-time and environment of the observed merging BBHs we first need to understand the difference between the time at which the progenitor stars formed and the time of merger of the BBH. This is what we define as the delay time \tdelay. It is the sum of two independent timescales: I) the lifetime of the binary stars up to the moment that both have become compact objects, and II) the inspiral time of the two BHs up to the BBH merger event. The former timescale, i.e.\ the lifetime of massive stars, is typically a few \Myr. The latter timescale depends primarily on the separation between the two BHs at BBH formation \citep{Peters:1964}.
To interpret the BBH merger rate, we first need to understand the impact of the delay time distribution on the observed rate at each redshift.

The delay time of BBHs from isolated binaries of interest can range from \Myr to more than a Hubble time \citep[see e.g.\ ][]{Neijssel+2019, GiacobboMapelli2018}. This implies that BBH mergers observed to merge at a given redshift, $z_{\rm merge}$, formed \Myr to \Gyr earlier. Hence, these mergers are comprised of a mixture of systems that originate from different formation redshifts, and likely probe a range of different formation environments. 

The delay time is thus a very important quantity, which, unfortunately, cannot be observed directly for an individual system. It is possible to make statistical inferences about the delay time distribution using the detections available so far \citep[see e.g.\ ][]{FishbachKalogera2021}. However, inference of the time delay distribution is difficult because it is degenerate with the progenitor formation rate. 
Moreover, we are currently still limited by the low number of sources that are detected out to higher redshifts.

Although the delay time is not directly observable, we will observe the redshift evolution of the source properties, i.e.\ the BH-mass, spin and mass ratio distributions at different redshifts. 
Several earlier studies have investigated the evolution of the BBH merger rate with redshift for the total population of merging BBHs,  \citep[e.g.\ ][]{RodriguezLoeb2018,MapelliGiacobbo2018,Choksi2019,Santoliquido2021}. The redshift evolution of source property distributions remains relatively obscured, though it is actively being studied \citep[see e.g.\ ][]{Neijssel+2019,Mapelli+2021}. Recent work hints towards relations between source properties and redshift evolution. \cite{Mapelli2019_CosmicTime} for example, find that massive BBHs tend to have longer delay times in their models.
An important step to move forward, is thus to associate possible trends in delay time distribution to observable characteristics, while understanding their physical origin.

\begin{figure}
\begin{center}
\includegraphics[width=0.49\textwidth]{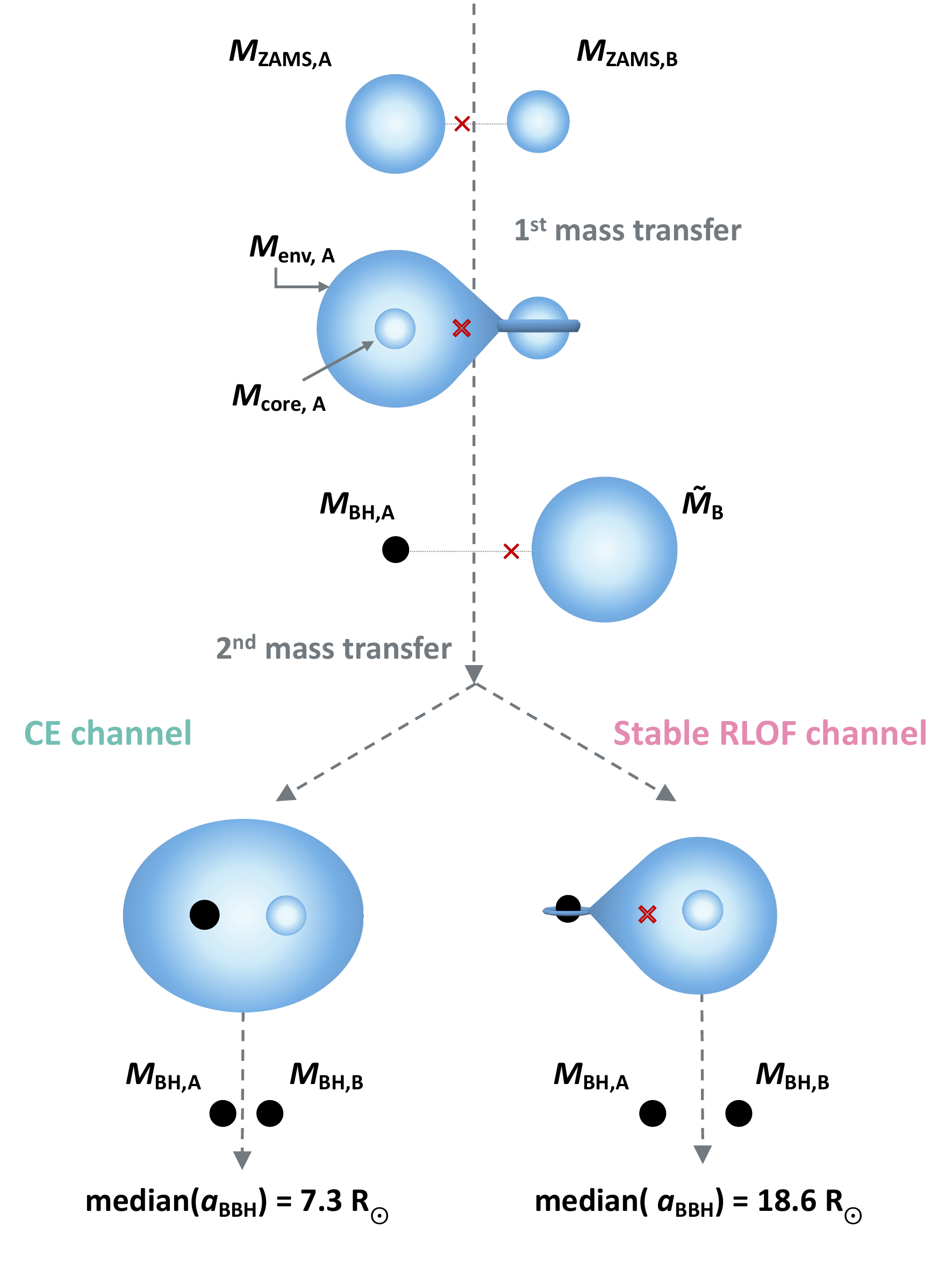}
\caption{Cartoon depiction of the \textit{typical} evolution of a BBH progenitor system through the stable RLOF and CE channel. Annotations refer to masses at zero age main sequence ($M_{\mathrm{ZAMS}}$), the envelope mass ($M_{\mathrm{env}}$), the core mass ($M_{\mathrm{core}}$), mass post mass transfer ($\tilde{M}_{\mathrm{B}}$) and BH mass ($M_{\mathrm{BH}}$). The subscript A (B) denotes the initially more (less) massive star.
The red cross gives an impression of the location of the centre of mass at the onset of the evolutionary phase depicted (not to scale). The median separation at BBH formation is annotated for each channel, considering BBH mergers that can be observed by a `perfect detector' (see text).  \label{fig: channels cartoon}}    
\end{center}
\end{figure}

Here, we inspect the delay time-mass relation for BHs coming from isolated binaries, as predicted by the rapid population synthesis code \COMPAS. We consider two main channels: 1) the common envelope channel \citep[or CE channel, e.g.\  ][]{Belczynski+2007a,Postnov+2014a,Belczynski+2016,Vigna-Gomez+2018}, including BBH systems where the progenitor system has experienced at least one common envelope,
and 2) the stable Roche-lobe overflow channel \citep[or stable RLOF channel, e.g.\ ][]{van-den-Heuvel+2017, Inayoshi2017}. The stable RLOF channel contains all BBH systems that experience \textit{only} stable mass transfer (i.e.\ all systems that \textit{do not} experience CE events, and so it is the complement set of the CE channel).
See also Figure~\ref{fig: channels cartoon} for a cartoon depiction of the most common evolution of these two channels. 
Note that this does not display all possible variations of the CE and stable RLOF channel. However, other sub-channels are rare. For example, the sub-channel where both the first and second mass transfer are unstable (which is one of the mos common sub-channels), contributes only $0.6\%$ to the total rate of BBH mergers as observed by a perfect detector (equation~\ref{eq: perfect detector rate all}). 
The respective contribution of the CE and the stable RLOF channel to the observed population of merging double compact objects is an active area of research  \citep[see e.g.\ ][]{Neijssel+2019, Bavera2021, Marchant2021, GallegosGarcia2021}.  In this work we aim to use characteristic delay time-mass distributions from each channel to make predictions for observables in the gravitational wave distributions.  

This paper is structured as follows: in Section~\ref{sec: method} we describe the population synthesis code \COMPAS used in this work.
We find that massive BHs ($\Mbheen > 30\Msun$, where we define \Mbheen as the more massive BH at BBH merger) predominantly form in BBHs with long delay times ($\tdelay > 1 \Gyr$). We show that this can be explained by differences between the CE channel and the stable RLOF channel in Section~\ref{sec: Mbh-tdelay}.
In Section~\ref{sec: method 2 rate} we describe how we calculate cosmic BBH merger rates. 
We then discuss how the distinct delay times and mass distributions arising from CE and stable RLOF affect the observed merger rate evolution of BBHs in Section~\ref{sec: merger rate redshift}.
In Section~\ref{sec: prospects obs trends} we discuss the prospect of observing trends in the BBH merger rate with current and near-future gravitational wave detectors. Specifically, our models predict that the slope of the intrinsic BBH merger rate density with redshift is more shallow and starts decreasing at lower redshift for higher \Mbheen. 
We discuss the robustness of our main findings and caveats that apply to a population synthesis approach in Section~\ref{sec: discussion}, and summarise our main results in Section~\ref{sec: conclusion}.

\section{Method (I) : Simulating merging BBH populations }
\label{sec: method}
To simulate the evolution of isolated massive binary star progenitors that lead to  merging BBH, we use the rapid population synthesis code that is part of the \COMPAS suite\footnote{see also \url{https://compas.science/}} \citep[version v02.19.04, ][]{COMPAS_method, Stevenson+2017,Vigna-Gomez+2018}. We simulate a total of $10^7$ binaries. To check that our results are converged, we have repeated all analyses for an independent set of $10^6$ binaries, and we found no significant differences. 
In this section we discuss the treatment of stellar evolution and binary interaction processes (Sec.~\ref{ss: method binaries}) and sampling of the initial parameters (Sec~\ref{ss: method sampling}).

\subsection{Binary evolution}
\label{ss: method binaries}
We model the evolution of massive stars in binary systems using fast  algorithms following \citet{Hurley+2000, Hurley+2002}, based on detailed evolutionary models by \citet{Pols+1998}. Here we summarise the treatment of the physical processes that are most relevant for this study. For a full description of the code we refer to the references mentioned above. 

\paragraph{Winds}
For hot O and B type stars (with effective temperatures $T_{\mathrm{eff}} \geq 12500$K), we follow \citet{Vink+2000,Vink+2001} to account for metallicity-dependent stellar wind mass loss. 
For cooler, more evolved stars ($T_{\mathrm{eff}} \leq 12500$K) the mass-loss prescription from \cite{KudritzkiReimers1978} and the prescription from \cite{Nieuwenhuijzen+1990}, modified by a metallicity dependent factor from \cite{Kudritzki+1989}, are compared and the maximum is adopted. The latter mass-loss prescription is only assumed to be non-zero for stars with luminosity $L > 4000 \,\Lsun$. 
For low mass stars that evolve towards the asymptotic giant branch, the prescription from  \cite{Vassiliadis+1993} is added to this comparison.
For hot Wolf-Rayet-like stars, we use the empirical mass loss prescription from \cite{Belczynski+2010}, that is adapted from \citet{Hamann+1998} but scaled by metallicity following \cite{Vink+2005}.
For very luminous stars, that lie above the Humphreys-Davidson limit, i.e.\ if the luminosities $L$ and stellar radii $R$ fulfil the condition $L> 6\times 10^{5} \Lsun$ and $(R/\Rsun) (L/\,\Lsun)^{1/2}>10^5$ \citep{Humphreys+1979}, we assume enhanced mass loss rates following \cite{Hurley+2000}, motivated by the scarcity of observed stars in this regime and the observed Luminous Blue Variables (LBV) phenomenon.
This additional mass loss is metallicity independent \citep[in line with recent results from, e.g.\ ][]{DaviesBeasor2020}, and is meant to mimic eruptive mass loss.  

\paragraph{Stable mass transfer and common envelope phases}
We account for mass transfer when a star overflows its Roche lobe \citep{Eggleton1983}. To determine whether Roche-lobe overflow is stable we use an estimate for the response of the radius of the donor star, $R$ and its Roche lobe, $R_{\mathrm{RL}}$ as a result of mass transfer.
\COMPAS determines stability by comparing estimates of the adiabatic response of the donors radius and the response of the donors Roche-lobe radius \citep[see e.g.\ ][and references therein]{Vigna-Gomez+2018,VignaGomez2020}. This procedure depends crucially on the assumed value of $\zeta_{*} \equiv (\partial \log R / \partial \log M)_{\rm ad}$, with $R$ and $M$ the radius and mass of the donor star,  for different types of donor stars \citep[e.g.\ ][]{Soberman+1997}. We assume $\zeta_{\rm ad} = 2$ for main sequence donors, $\zeta_{\rm ad} = 6.5$ for Hertzsprung gap donor stars \citep{Ge+2015} and follow \cite{Soberman+1997} for donor stars post helium ignition. 

During stable mass transfer onto a stellar companion we assume that the accretion rate is limited to ten times the thermal rate of the accreting star \citep{Neo+1977,Hurley+2002}. If the accreting component is a BH, the accretion is assumed to be Eddington limited.
Material lost from the system during non conservative mass transfer,  is assumed to carry away the specific orbital angular momentum of the accreting component \citep[e.g.\ ][]{Soberman+1997,van-den-Heuvel+2017}. 
This reduces the orbital angular momentum and can lead to either shrinking or widening of the orbit, depending on the fraction of mass that is accreted and the binary's mass ratio \citep[e.g.\ ][Appendix~A]{vanson2020}.

Unstable mass transfer is assumed to result in CE evolution \citep{Paczynski1976,Ivanova+2013,IvanovaJusthamRicker2020}.
We assume that ejecting the envelope shrinks the binary orbit following the energy considerations proposed by \cite{Webbink1984} and \cite{de-Kool1990}. Here, the pre-CE binding energy of the donor's envelope is equated to the orbital energy that becomes available by shrinking the orbit. How efficiently this orbital energy can be used to eject the envelope is parameterized by the $\CEa$ parameter, which is set to one in this work. For the binding (and internal) energy of the envelope, we use the ``Nanjing" prescription \citep{Dominik+2012}, based on fits provided by \citet{Xu+2010a,Xu+2010}.
We adopt the \textit{pessimistic} CE scenario from \cite{Dominik+2012}, that is, we assume that Hertzsprung Gap donor stars do not survive a CE event.

\paragraph{Supernovae, kicks and compact remnants}
To model natal supernova kicks, we draw kick velocities with random isotropic orientations and draw the kick magnitudes from a Maxwellian distribution \citep{Hobbs+2005}. BH kicks are reduced by the amount of mass falling back onto the newly-formed BH during the explosion mechanism, following the `delayed' prescription from \citep{Fryer+2012}. This prescription assumes full fallback for BHs resulting from progenitors with a carbon oxygen core mass $\Mco > 11\Msun$, and hence these BHs receive no supernova kick.

The remnant mass is modelled as a function of the estimated $\Mco$ at the moment of core collapse following \cite{Fryer+2012}. 
Stars with helium cores above 35\Msun at the moment of core collapse are assumed to experience pulsational-pair instability following \cite{Farmer+2019}.
Stars with helium core masses between $60-135$\Msun at the moment of core collapse are expected to be completely disrupted by pair instability, and therefore leave no remnant BH.
With this implementation the lower edge of the pair-instability mass gap is located at about $45$\Msun (\citealt{Stevenson+2017,Marchant+2019,Farmer+2019,Farmer2020, Woosley2021}, but see e.g.\  \citealt{Mehta2021}). 
Due to the metallicity dependence of stellar winds and the adopted pulsational-pair instability prescription, the maximum BH mass is also metallicity dependent. The upper limit of about $45\Msun$ is only reached for the lowest metallicity systems (with $Z \lesssim 0.001$). For reference, systems with metallicities of about $Z\sim 0.01$ and $Z\sim 0.0032$ can maximally achieve a BH mass of about $18\Msun$ and $32\Msun$ respectively in our simulations (see Figure~\ref{fig: mass dist by metals} for a decomposition of the BH mass distribution by metallicity).

\subsection{Sampling}
\label{ss: method sampling}

The evolution of a binary system is mainly a function of its initial metallicity $Z$, initial primary and secondary mass $M_1$ and $M_2$, and the initial separation $a$.  

We sample birth metallicities with a probability distribution that is flat-in-log in the range $10^{-4} \leq Z \leq 0.03$. 
Sampling metallicities from a smooth probability distribution is an improvement over discrete sets of metallicity, which is the most common technique in binary population synthesis studies (but see, for example, \citealt{Riley2021} for an exception). Smoothly sampling birth metallicity avoids artificial peaks in the BH mass distribution \citep[e.g.][]{Dominik2015,Kummer_thesis}. The flat-in-log distribution ensures that we sample ample binaries at the low metallicities that are favoured for BBH formation. Later in this paper, when we calculate cosmic merger rates, we re-weight systems to account for the metallicity-dependent star formation (see Section~\ref{sec: method 2 rate}). We adjust the normalisation of this re-weighting over the metallicity range of our simulations to preserve the correct total star-formation rate, i.e., star formation at more extreme metallicities is not discarded.

We assume the masses of the initially more massive stellar components (the primary, $M_{1}$) are universally distributed following a \cite{Kroupa2001} initial mass function and draw masses in the range 10 - 150\Msun, in order to focus on stars that evolve into BHs. 
The binary systems are assumed to follow a uniform distribution of mass ratios ($0.01 \lesssim q \equiv M_{2}/M_{1} < 1.0$, with $M_2$, the mass of the secondary star). We require $M_{2} \geq 0.1\Msun$. The initial binary separations are assumed to follow a distribution of orbital separations that is flat in the logarithm \citep{Opik1924} in the range $0.01-1000\,$AU. Binary systems that fill their Roche lobe at zero age main sequence are discarded. 
All binary orbits are assumed to be circular at birth. 

If a zero age main sequence (ZAMS) star is rotating faster than the metallicity-dependent rotational frequency threshold described in \cite{Riley2021}, the binary is assumed to evolve chemically homogeneously. In this work, we focus on the `classical' pathway of isolated binaries towards merging BBHs and thus we exclude chemically homogeneously evolving stars from our sample.

Because BBH mergers are intrinsically very rare events, direct sampling of the birth distributions is very inefficient and time consuming. We therefore make use of the adaptive importance sampling code STROOPWAFEL. This algorithm consists of an initial exploration phase to find regions of interest in the binary parameter space. In a subsequent  adaptive refinement phase we optimise the simulations by sampling near the regions of interest \citep[see][for details]{Broekgaarden2019}.

\begin{figure*}
\centering
\includegraphics[width=0.325\textwidth]{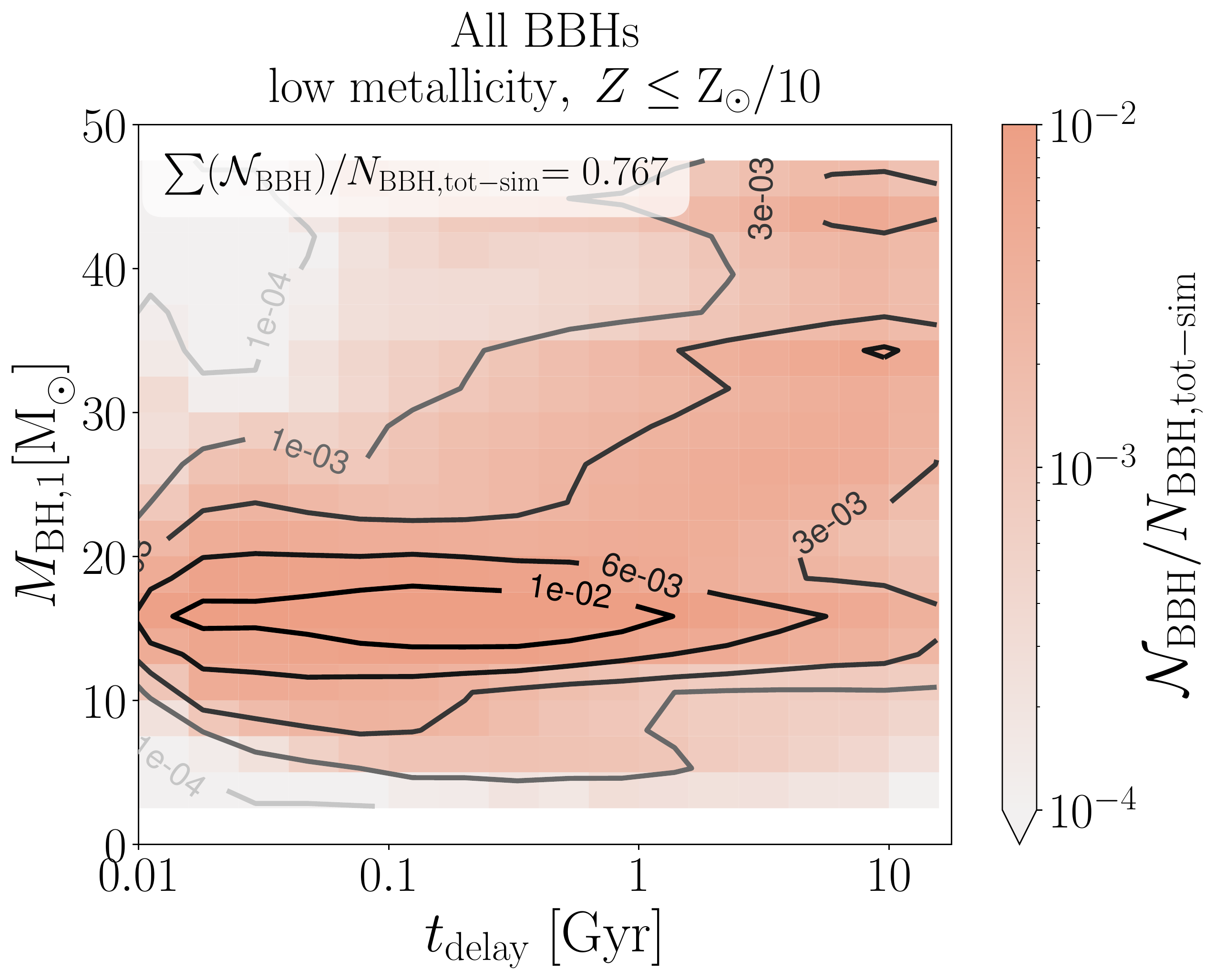}
\includegraphics[width=0.325\textwidth]{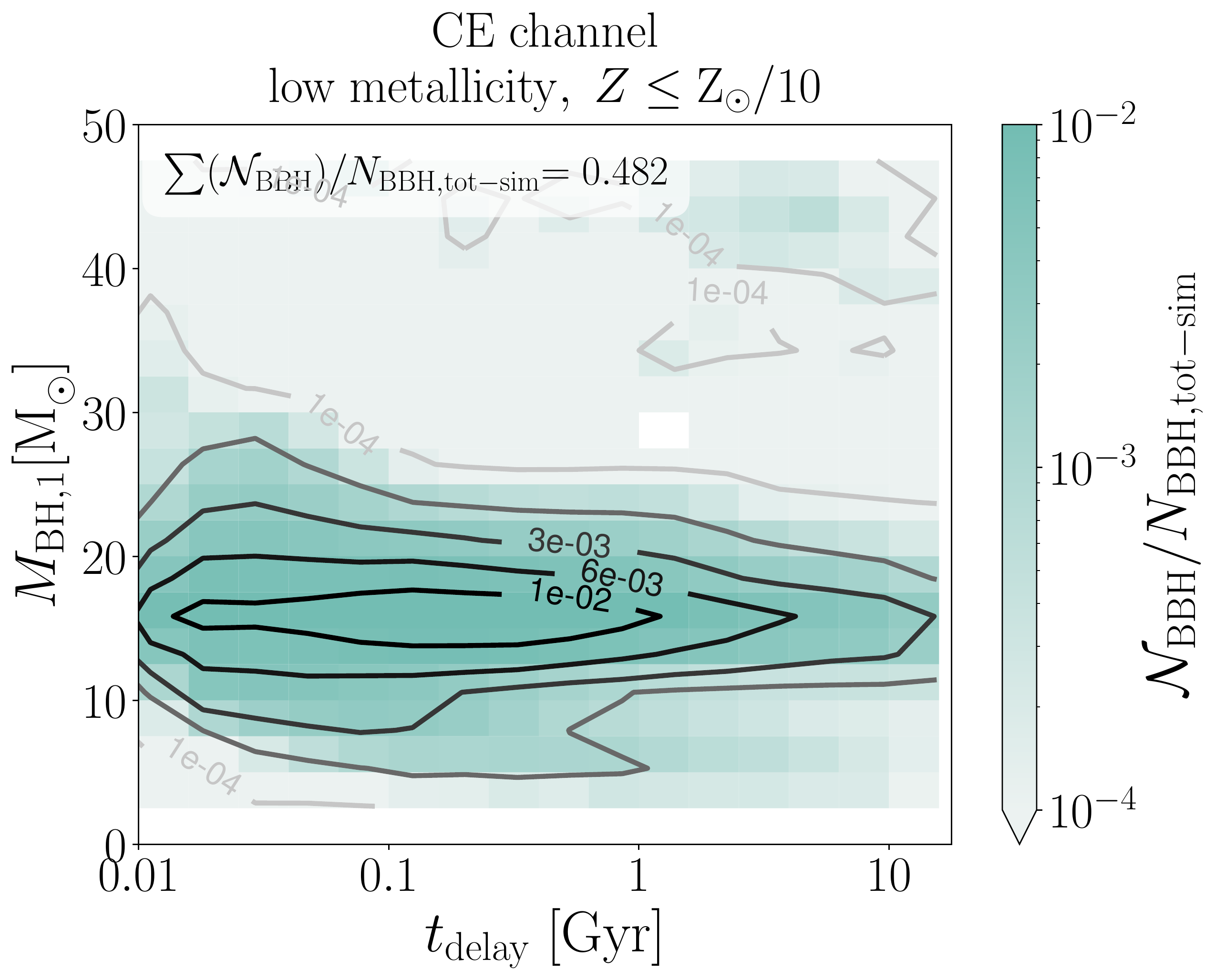}%
\includegraphics[width=0.325\textwidth]{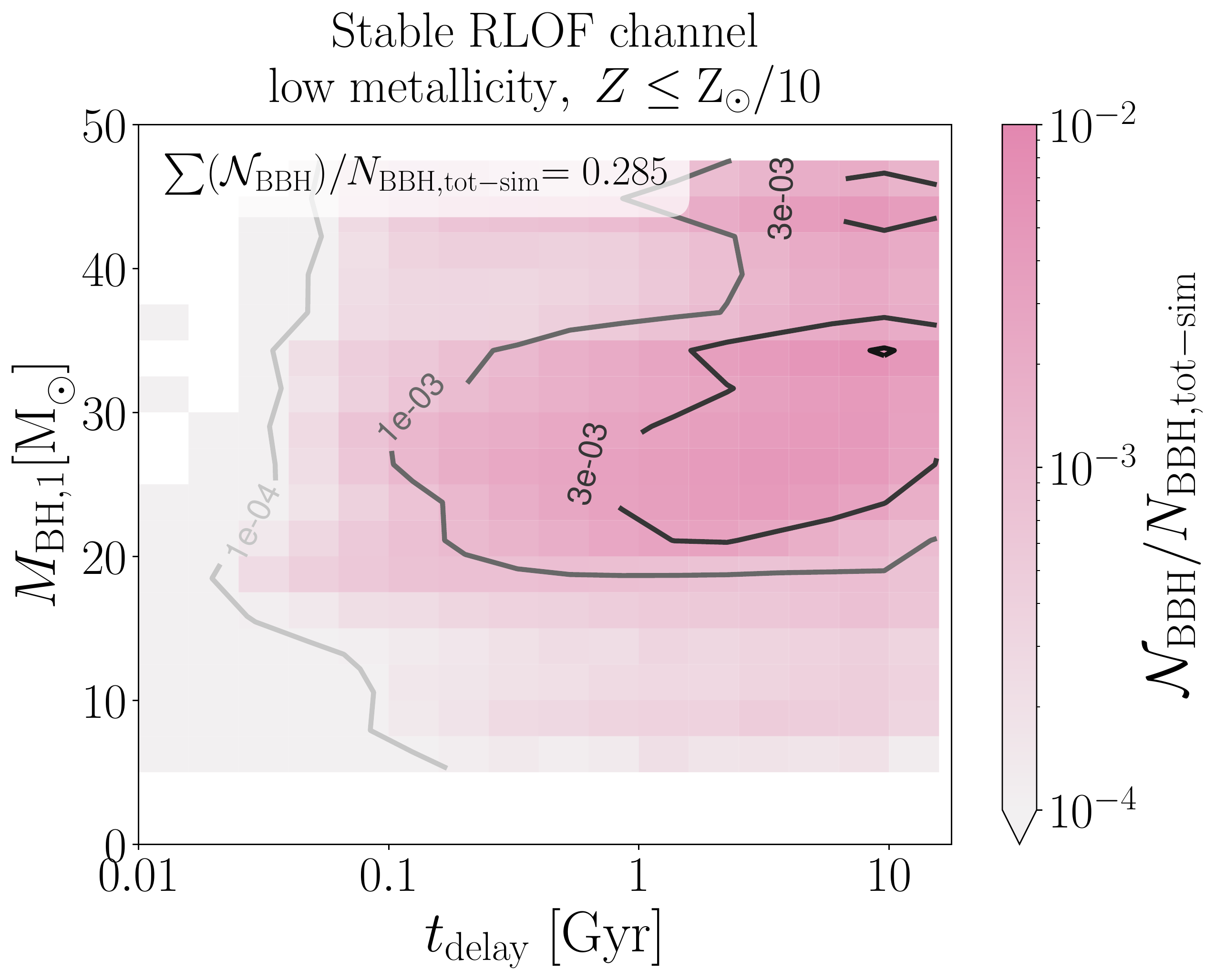}

\includegraphics[width=0.325\textwidth]{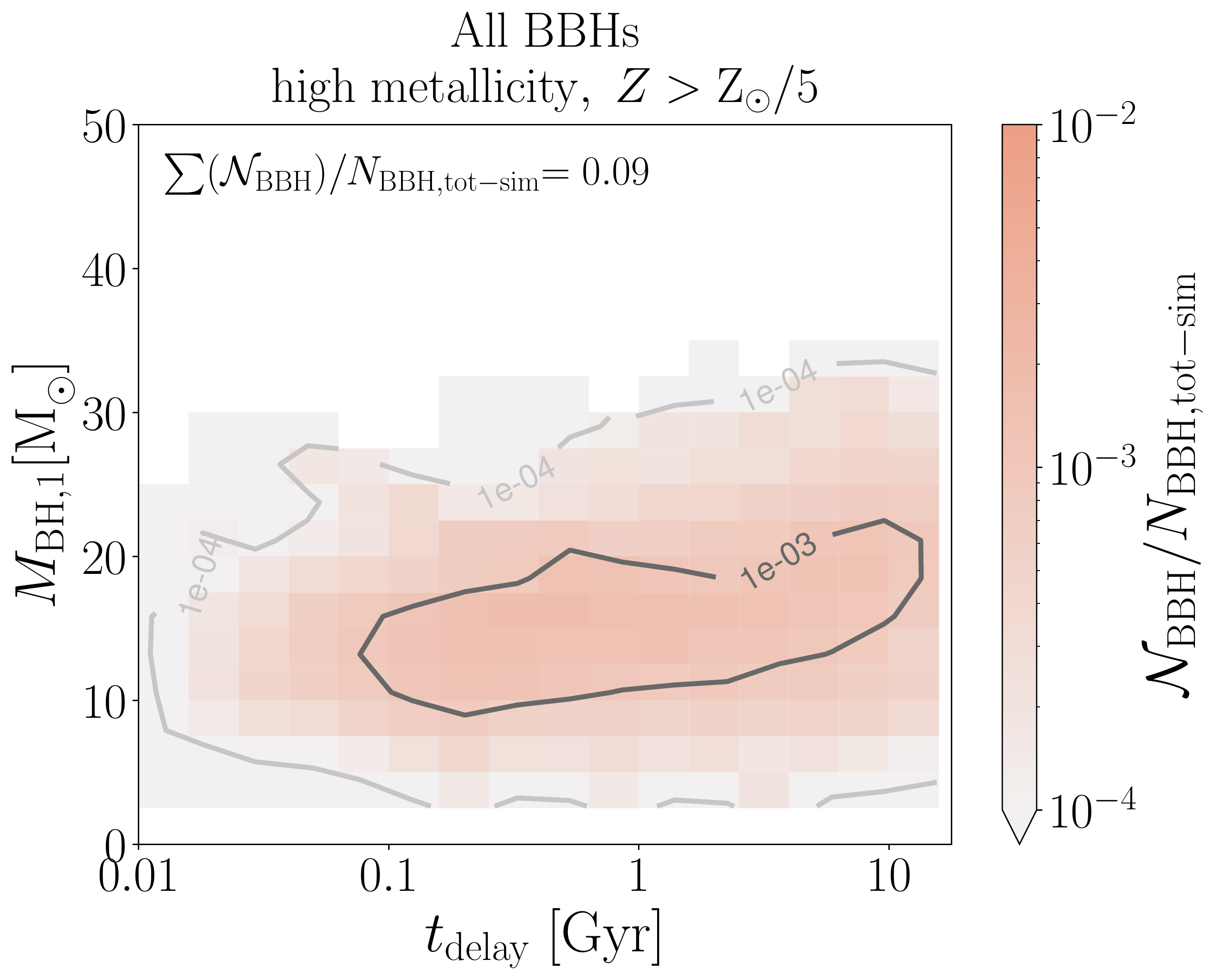}
\includegraphics[width=0.325\textwidth]{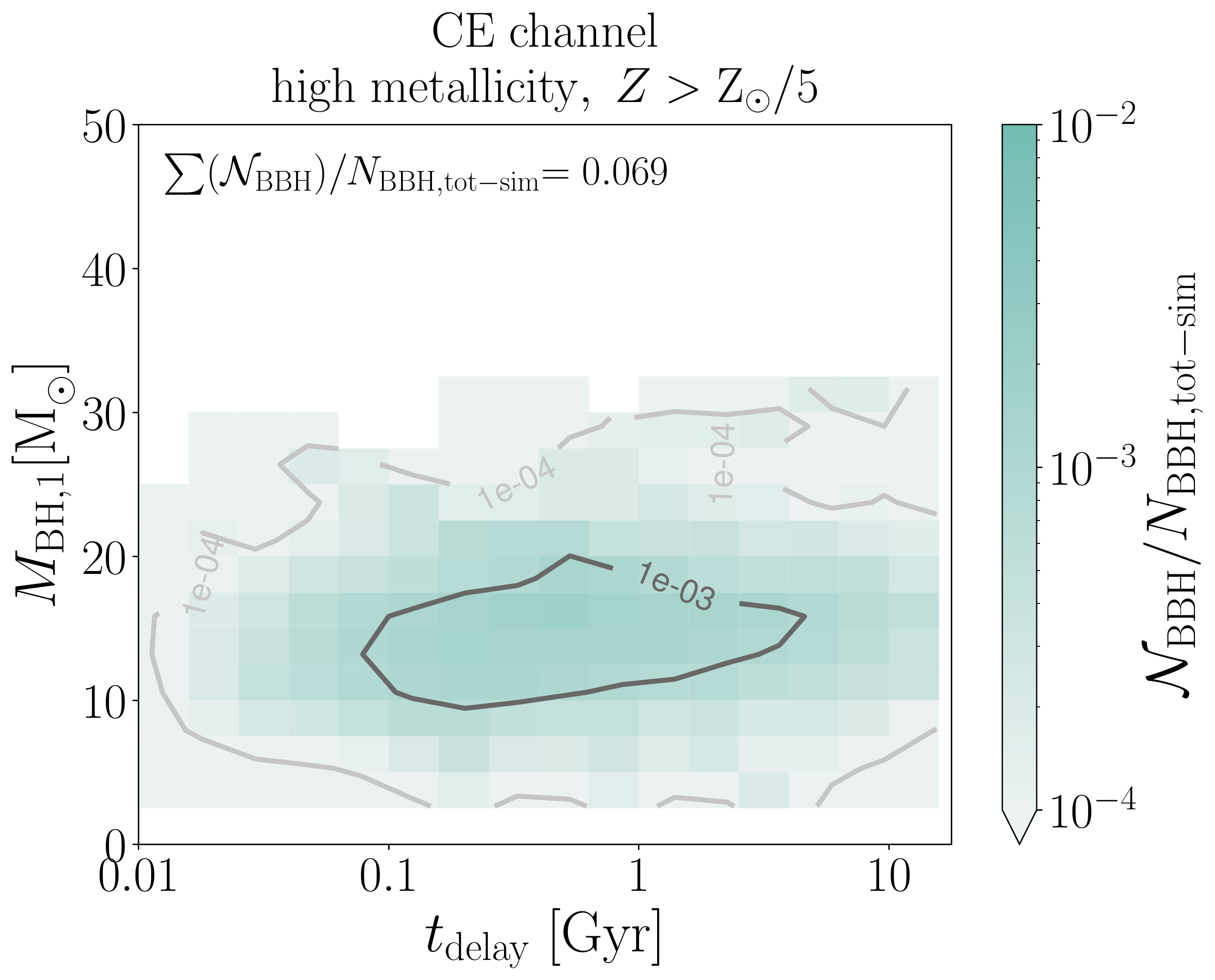}%
\includegraphics[width=0.325\textwidth]{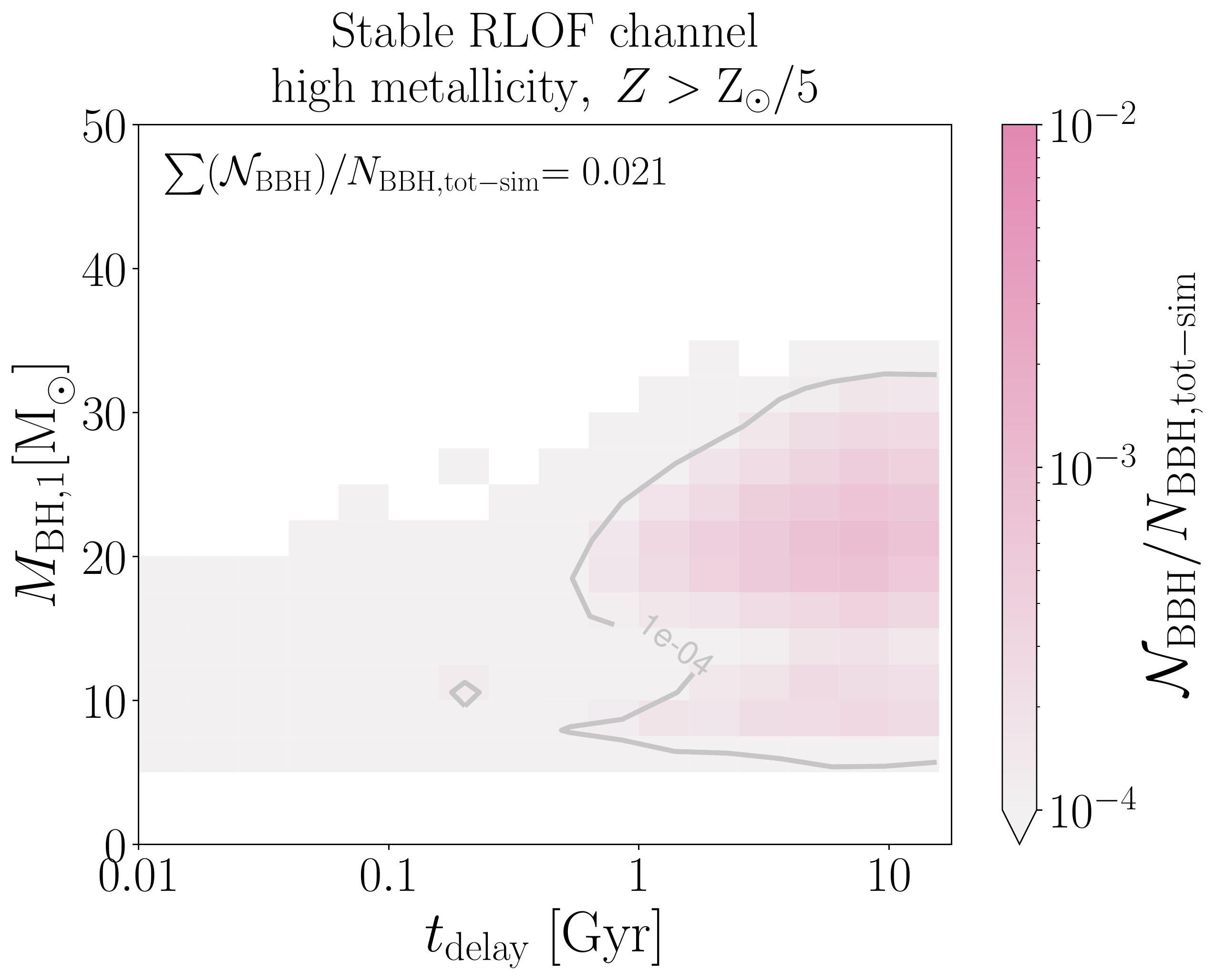}

\caption{Two-dimensional histograms of the distribution of delay times and primary masses for BBHs in our simulation. The top and bottom row show results for low ($\leq\Zsun/10$) and high ($>\Zsun/5$) metallicity, respectively.
The left most panels show all BBHs, while the middle and right most panels are split by formation channel. All histograms are normalised relative to the total simulation including all simulated metallicities. The colour-bar and contours thus indicate the relative frequency of occurrence in our simulations. We use bin sizes of $\Delta \log_{10}(\tdelay) = 0.2$ and $\Delta\Mbheen=2.5\Msun$. 
All panels reveal a lack of BBH systems with high mass ($\Mbheen \gtrsim 30\Msun$) and short delay time ($\tdelay \lesssim 0.1 \Gyr$).
\label{fig: Mbh_tdelay}}
\end{figure*}

\section{BH mass-delay time relations}
\label{sec: Mbh-tdelay}
In this section, we first explore the type of BBHs that can be produced by the isolated channel according to our simulations.  
We aim to find links between the delay time \tdelay and observable properties, such as BH masses and spins. Of these, the BH mass is observationally the best constrained source property. Hence our main focus is on the BH mass. 
While we do not discuss BH spins here, previous studies have argued that tidal spin-up is most likely in close binaries with short delay times \citep[e.g.][]{Kushnir+2016,Zaldarriaga+2018,Bavera2020}. In appendix \ref{sec: q per delay time}, we additionally investigate the correlations between BBH mass ratios and \tdelay.

In Figure~\ref{fig: Mbh_tdelay} we show two-dimensional histograms of \tdelay, and the mass of the heavier BH, \Mbheen, for BBHs in our simulations. 
In the top row we show results for low metallicity, which is representative for the majority of BBH formation \citep[defined as $Z \leq \Zsun/10$, with solar metallicity $\Zsun=0.014$,][]{Asplund+2009}. To elucidate the impact of metallicity, we show results for the highest metallicities ($Z > \Zsun/5$) in the bottom row.
In the left column we show the result for all BBHs in the selected metallicity range. In the middle and right column we show the separate contributions of the CE and stable RLOF channel respectively.
All histograms shown are normalised relative to the number of merging BBHs in our full simulation, combining all metallicities. The colour shading and contours thus indicate the relative frequency with which these combinations of primary mass and delay time occur in our full set of simulations. We refer to Sect~\ref{sec: method} for how the progenitors are sampled and weighed in our simulation.  
We note that the underlying distribution in metallicity that is implicitly assumed here, is not representative for star formation in the Universe. Nevertheless, these diagrams are useful to understand trends in the delay times and primary BH masses at low and high metallicity.

When inspecting the left-most-panel in the top row of Figure~\ref{fig: Mbh_tdelay}, which shows the results for all BBH in our simulations for low-metallicity, we observe two main components. Firstly, we see that the histogram peaks at delay times of $\sim 0.1$--$1$\Gyr and primary BH masses of $\sim18\Msun$. This peak comes predominantly from systems formed through the CE channel (as can be seen from the top middle panel). Secondly, we see a noticeable tail of more massive systems $\Mbh \gtrsim 20 \Msun$ with longer delay times around $\sim 10$\Gyr, which predominantly come from the stable RLOF channel (as can be seen in the top right-most panel).  Finally, we see a dearth of BBH systems with high masses ($\Mbheen \geq 30\Msun$) and short delay times ($\tdelay~\leq~0.1~\Gyr$), which are not formed by either of the channels considered here. 

Comparing low and high metallicity (top and bottom row respectively), we see that the same two components are present, but the systems with highest mass are absent at high metallicity. This result is understood as the effect of the metallicity dependent stellar winds, which are stronger for higher metallicity \citep[e.g.\ ][]{Vink+2005}. The high metallicity systems thus also display a lack of BH systems with high masses ($\Mbheen \geq 30\Msun$) and short delay times ($\tdelay~\leq~0.1~\Gyr$).

In the following subsections we discuss the origin for these features. 

\subsection{Why the CE channel does not produce high-mass black holes}
\label{ss: CE channel reasoning}
We find that the massive progenitor stars that lead to BHs with masses $\Mbheen~>~30\Msun$ are disfavoured from engaging in, and surviving, CE events in our simulations because of a variety of effects.
To form such BHs, we need stars that form helium cores of at least $M_{\mathrm{He}}\gtrsim 30\Msun$. Such cores can only be formed in the most massive stars in our simulations, typically with zero-age main sequence masses of 60\Msun and higher, although we note that the exact value is considerably uncertain. Such massive stars are unlikely to engage in, and survive a CE for several reasons.

First of all, the massive progenitors of heavy black holes are thought to experience heavy mass loss, which can remove a large part of the hydrogen envelope before the stars initiates interaction with its companion.  Although mass loss by radiatively driven winds is thought to be reduced at low metallicity, mass loss by LBV eruptions is likely to still be very significant also at low metallicity \citep[e.g.\ ][]{Smith2014, Sanyal+2017, Kalari+2018, Davies+2018, Higgins+2020, Sabhahit+2021, Gilkis+2021}. In fact, such heavy mass loss can prevent massive stars in wider binaries from ever filling their Roche lobe \citep[][]{Mennekens2014, Belczynski2016_COmergerRate}. In our simulations this is the dominant reason for the suppression of the CE channel at higher masses.  

Secondly, even if a massive progenitor would fill its Roche lobe, it is unlikely to do so while it has a convective envelope. It is generally thought that donor stars with extended convective envelopes are favoured for successful ejection of a common envelope. This is mainly because convective stars have large dimensions, and a relatively large fraction of the mass is located at large radii. The binding energy of the envelopes of such stars is thus low with respect to radiative counterparts, and it is thought that the envelope can therefore more easily be removed by an inspiraling companion, as recently emphasised by  \citet{Klencki2021} and \citet{Marchant2021}.
Very massive stars typically do not grow to the dimensions needed to cool their envelope sufficiently to become unstable against convection.
Even though some massive stars may manage develop a deep convective envelope, they do not significantly expand further in radius (in contrast to less massive stars that will ascend the giant branch). Hence very massive stars generally fill their Roche lobe at an earlier point in their evolution, when the envelope was still radiative. 
Overall, the occurrence of successful CE is therefore very rare for such massive stars.

Thirdly, closely related to the second effect, mass transfer from high-mass donor stars is preferentially stable and hence it does not initiate a CE phase. This is especially true for radiative donors, as the early adiabatic response of radiative envelopes to mass loss is contraction \citep[see, e.g.\ ][]{Hjellming+1987}. Recent studies, based on simulations with a more sophisticated treatment of the physics, tend to emphasize this finding, also for convective donors \citep[e.g.\ ][]{PavlovskiiIvanova2015,Pavlovskii2017,Marchant2021}. In addition, albeit more speculatively, this effect may be enhanced by the role of envelope inflation. This occurs in massive stars that are close to the Eddington limit. They can develop extended halos \citep[e.g.\ ][]{Sanyal+2015,Jiang+2015, Jiang+2018}. This can likely cause stable mass exchange before the star has really filled its Roche lobe.  
Although our simulations treat the stability criteria in a very simplified way, the recent studies mentioned above tend to strengthen our findings that mergers involving more massive BHs are unlikely from the CE channel. 

We remind the reader that, in the CE channel, it is normally the second phase of mass transfer where the common envelope phase occurs, see Fig.~\ref{fig: channels cartoon}. The considerations above thus primarily concern the initially less massive star in the binary system. 
In principle, it is possible to form BBH mergers with at least one heavy BH from binary systems with a very massive primary ($\gtrsim 60\Msun$) and significantly less massive secondary ($\lesssim 40 \Msun$). The heavy BH then originates from the primary star, while the secondary star is of low enough mass to initiate a CE phase in which the envelope is ejected successfully. However, we find that such systems are extremely rare.  The secondary typically accretes during the first mass transfer phase and becomes massive enough to be subject to the first two effects mentioned above. This scenario thus only works for systems with extreme initial mass ratios. Such systems tend to merge upon the first mass transfer phase and will thus not be able to form BBHs that merge within a Hubble time.   

Overall we find that the formation of BBHs with at least one heavy BH is not impossible through the CE channel, but very unlikely in our simulations.  More detailed recent studies on partial aspects of the problem strengthen this finding. 

\subsection{Why the stable RLOF channel does not produce short delay times}
\label{ss: stable RLOF channel reasoning}
%
We find that the stable RLOF channel leads to longer delay times than the CE channel, due to longer inspiral times. These longer inspiral times are caused by wider separations (larger semi-major axis) at BBH formation.
We find that the median separation at BBH formation is about $7\Rsun$ for systems that came from the CE channel, and about $20\Rsun$ for systems that come from the stable RLOF channel, when considering all systems that can be observed by a `perfect detector' (see Eq.~\ref{eq: perfect detector rate all}). Wider separations lead to longer inspiral times because the orbital decay time from gravitational-wave emission scales with the fourth power of the separation \citep[][]{Peters:1964}. We find that the effect of the component masses and eccentricity of BBH systems are typically subdominant to the effect of the separation. 

To understand why the CE channel produces shorter separations we consider the difference in orbital evolution for both channels. For stable mass transfer, whether the orbit widens or shrinks depends on the mass ratio, the amount of mass lost from the system, and the assumed angular momentum that is carried away by the mass that is lost \citep[e.g.\ ][]{Soberman+1997}. To produce merging BBH systems through stable RLOF we typically need to considerably shrink the orbit during reverse mass transfer \citep{van-den-Heuvel+2017}. The accretor is already a BH at this time and its accretion is assumed to be limited to the Eddington accretion rate. This means that most of the mass that is transferred is lost from the system.  For highly non-conservative mass transfer, the orbit shrinks  (when $M_{\mathrm{acc} }/M_{\mathrm{donor} } \le 0.79$, for which see e.g.\ Appendix~A from \citealt{vanson2020}) under the assumption that mass is lost from the vicinity of the accreting companion and has the specific angular momentum of the accretor's orbit.
This criterion may be fulfilled when the secondary star fills its Roche lobe at first and lead to shrinking of the orbit, but as more mass is lost, the orbital evolution can reverse from shrinking to widening. 
In contrast, CE evolution exclusively shrinks the orbit in our simulations, in agreement with general expectation (e.g.\ \citealt{Paczynski1976}, \citealt{Ivanova+2013}).

Even though many of the details regarding orbital shrinking are uncertain in both scenarios, these mechanisms are so different that we can robustly expect substantial differences in the resulting final separations. Since the separation is the dominant term in the expression for the inspiral time, we are confident that our finding that the two channels lead to a difference in their delay times is robust, at least qualitatively.
For completeness, we show the delay times distributions, similar to Figure~\ref{fig: Mbh_tdelay}, but for all metallicities and integrated over \Mbheen in Appendix \ref{s: delay times}.

\section{Method (II) : Calculating Intrinsic  merger rates}
\label{sec: method 2 rate}
To place our results into cosmological context we need to integrate over the metallicity-dependent star formation rate density, \SFRDzZ \citep[see also ][]{Dominik2013, Dominik2015, Belczynski2016_COmergerRate, MandelMink2016, Chruslinska+2018}. 
This results in an intrinsic BBH merger rate density, \RBBH, that we will discuss in Sections~\ref{sec: Mbh redshift plane} and \ref{sec: prospects obs trends}. 
Throughout this work we adopt cosmological parameters consistent with the WMAP9--cosmology \citep{Hinshaw2013}  including $h = \rm H_0$ /(100  km s$^{-1}$  Mpc$^{-1}$) $= 0.693$, where $\rm H_{0}$ is the Hubble constant. 

\subsection{Estimating the intrinsic BBH merger rate}
\label{ss: intrinsic rate}

We follow the method described in  \cite{Neijssel+2019} and \cite{Broekgaarden+2021a} to calculate the BBH merger rate%
\footnote{The scripts to compute the rates are available as part of the \COMPAS suite  \url{https://github.com/TeamCOMPAS/COMPAS}.}.

The number of detections that occur during the active observing time ($T_{\rm obs}$, measured in the detector frame at $z=0$) of an infinitely sensitive gravitational-wave detector is given by
\begin{equation}
    \frac{d^2N_{\mathrm{det}}}{d\zeta dz} = \frac{\RBBHzeta}{d\zeta} \left[ \frac{dV_c}{dz}(z) \right] \frac{T_{\rm obs}}{1+z} ,
\end{equation}

\noindent where $N_{\mathrm{det}}$ is the number of detectable BBH mergers, $\zeta$ is the set of parameters that describe a BBH, and $\frac{dV_c}{dz}(z)$ is the differential co-moving volume per redshift \citep[see e.g.\ ][]{LIGOScientific:2018jsj}. 

Our goal is to estimate the intrinsic merger rate density of all BBHs in the source frame, \RBBH:
\begin{equation}
    \RBBH = \int \ d\zeta \ \RBBHzeta = \frac{d^2N_{\mathrm{BBH}}}{dV_c dt}(z) \ \mathrm{[cGpc^{-3} yr^{-1}]}, 
    \label{eq: rate dens BBH}
\end{equation}
which is the number of mergers $N_{\mathrm{BBH}}$ per co-moving volume $V_c$ in co-moving gigaparsec, cGpc$^{-3}$ per year, with $t$ the time in the source frame.

Often, we would like to evaluate the intrinsic rate density over larger redshift bins. For that purpose, we define the volume averaged intrinsic merger rate density:
\begin{equation}
    \RBBHav = \frac{ \int_{z_{\mathrm{min} }}^{z_{\mathrm{max} }} \ \RBBH \frac{dV_c}{dz} \ dz }{\int_{z_{\mathrm{min} }}^{z_{\mathrm{max} }} \ \frac{dV_c}{dz} \ dz} \ \mathrm{[cGpc^{-3} yr^{-1}]}, 
    \label{eq: av rate dens BBH}
\end{equation}

To approximate the intrinsic merger rate density at redshift $z$, we convolve the number of BBH mergers per unit star-forming mass with the star-formation rate density over the merger time $t_m(z)$, and integrate this over all metallicities:

\begin{equation}
\centering 
\begin{split}
     & R_{\mathrm{BBH}}(z, \zeta) = \
     \int dZ' \int\displaylimits_{0}^{t_{\mathrm{m}(z)}} d\tdelay' \\
    &  
    \underbrace {\frac{d^2N_{\mathrm{form}}}{dM_{\mathrm{SF}} \ d\tdelay} (Z', \tdelay', \zeta) }_{\mathrm{BBH \ formation \ rate}} \, \ast 
        \underbrace{\frac{}{} \mathcal{S}\mathrm{(} Z',z(\tform) \mathrm{)} }_{Z\mathrm{-dependent \ SFRD}} ,
    \label{eq: rate dens broken down}
\end{split}
\end{equation}

\noindent where the time of merger, $t_m(z)$, delay time, \tdelay, and formation time, \tform, are related by $t_{\mathrm{form}} = t_m - \tdelay$. 
We adopt the redshift of first star-formation $z_{\mathrm{first \ SF}} = 10$ in our work.
Equation~\ref{eq: rate dens broken down} is evaluated  at redshift steps of $dz = 0.001$. 

Our choice for the metallicity-dependent star formation rate at the formation redshift, $\mathcal{S}\mathrm{(} Z, z_{\mathrm{form}}$(\tform)) is detailed and discussed in Appendix~\ref{s: fitting SFRD}. $d^2N_{\mathrm{form} }/(dM_{\mathrm{SF} } \ d\tdelay)$ is the number of BBH systems that form with delay times in the interval $d\tdelay$ per unit of star forming mass $dM_{\mathrm{SF}}$. 
Because we model only a small fraction of the total star forming mass, we need to re-normalise our results, given the initial distributions of primary masses and mass ratios (see \S\ref{ss: method sampling}). In our simulations we neglect single stars, only draw primary masses in the range $10-150\Msun$ and apply adaptive importance sampling. When re-normalising, we assume that the Universe has a constant binary fraction of $f_{\rm bin} =0.7$ \citep{Sana+2012}, and stars are formed with initial masses in the range $0.1 - 200\Msun$.

\section{The merger rates and mass function at different redshifts}

\label{sec: merger rate redshift}

\subsection{The role of the two formation channels}
\label{sec: Mbh redshift plane}

\begin{figure*}
\centering
\includegraphics[width=0.32\textwidth]{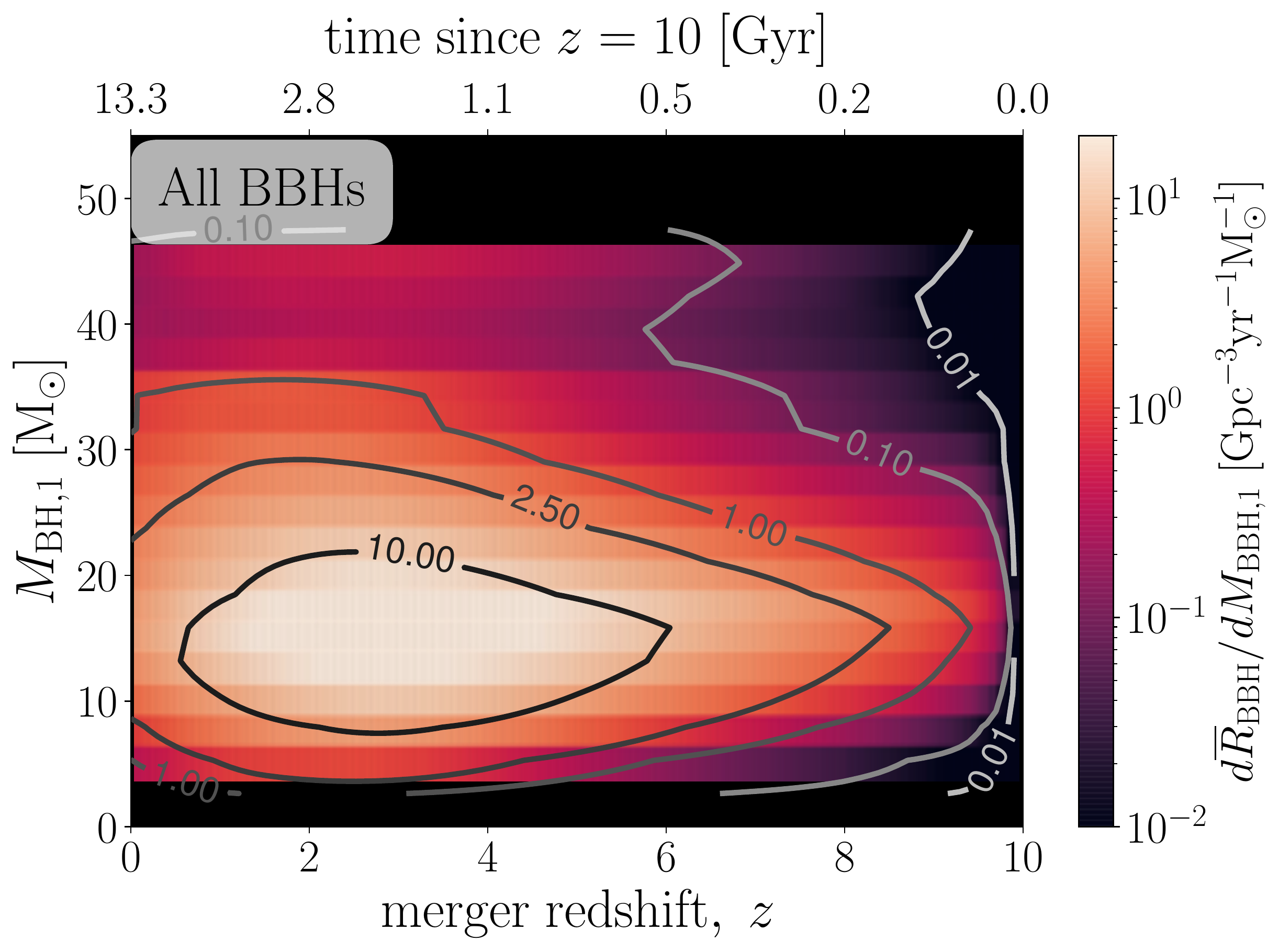}
\includegraphics[width=0.32\textwidth]{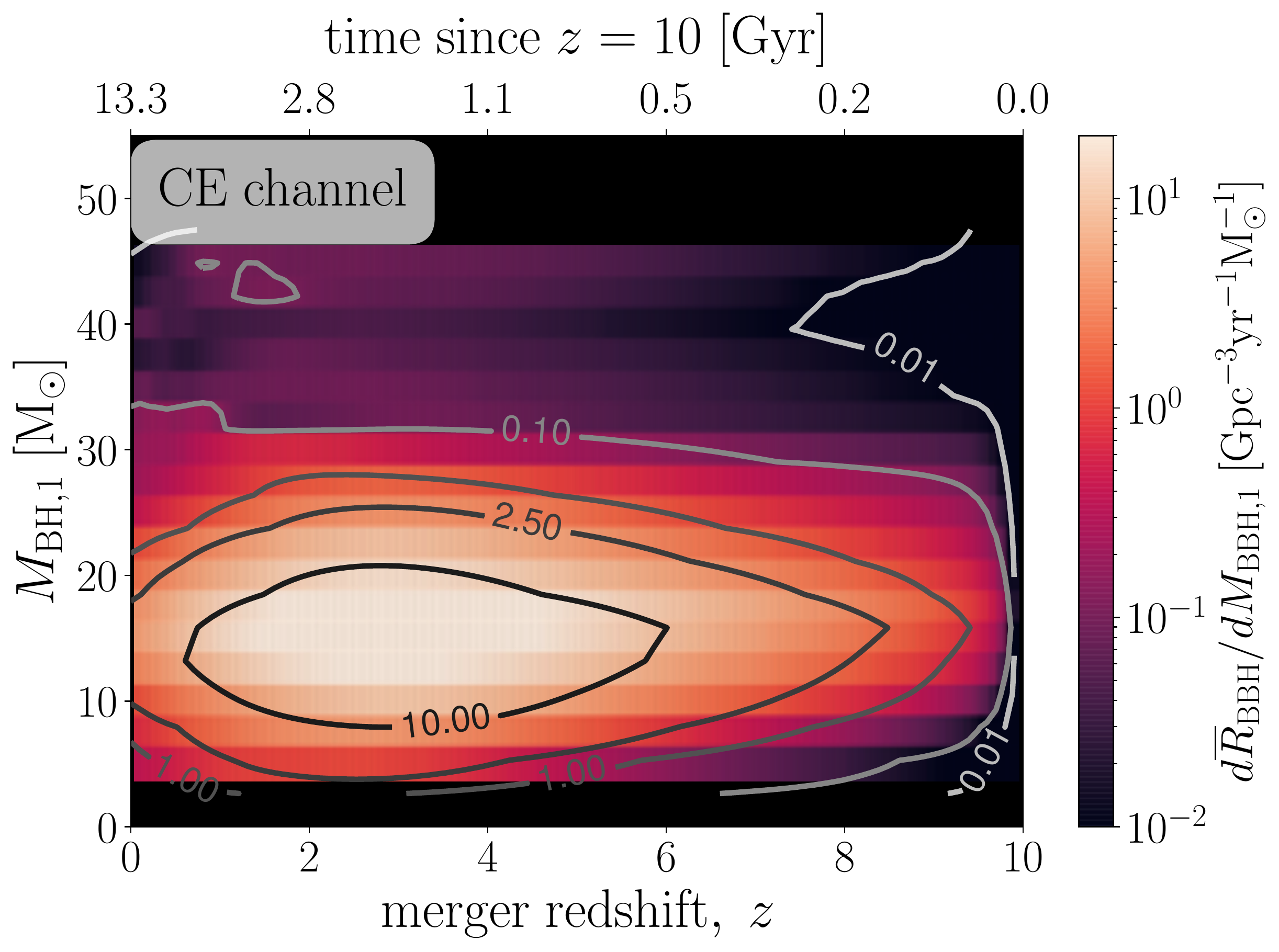}
\includegraphics[width=0.32\textwidth]{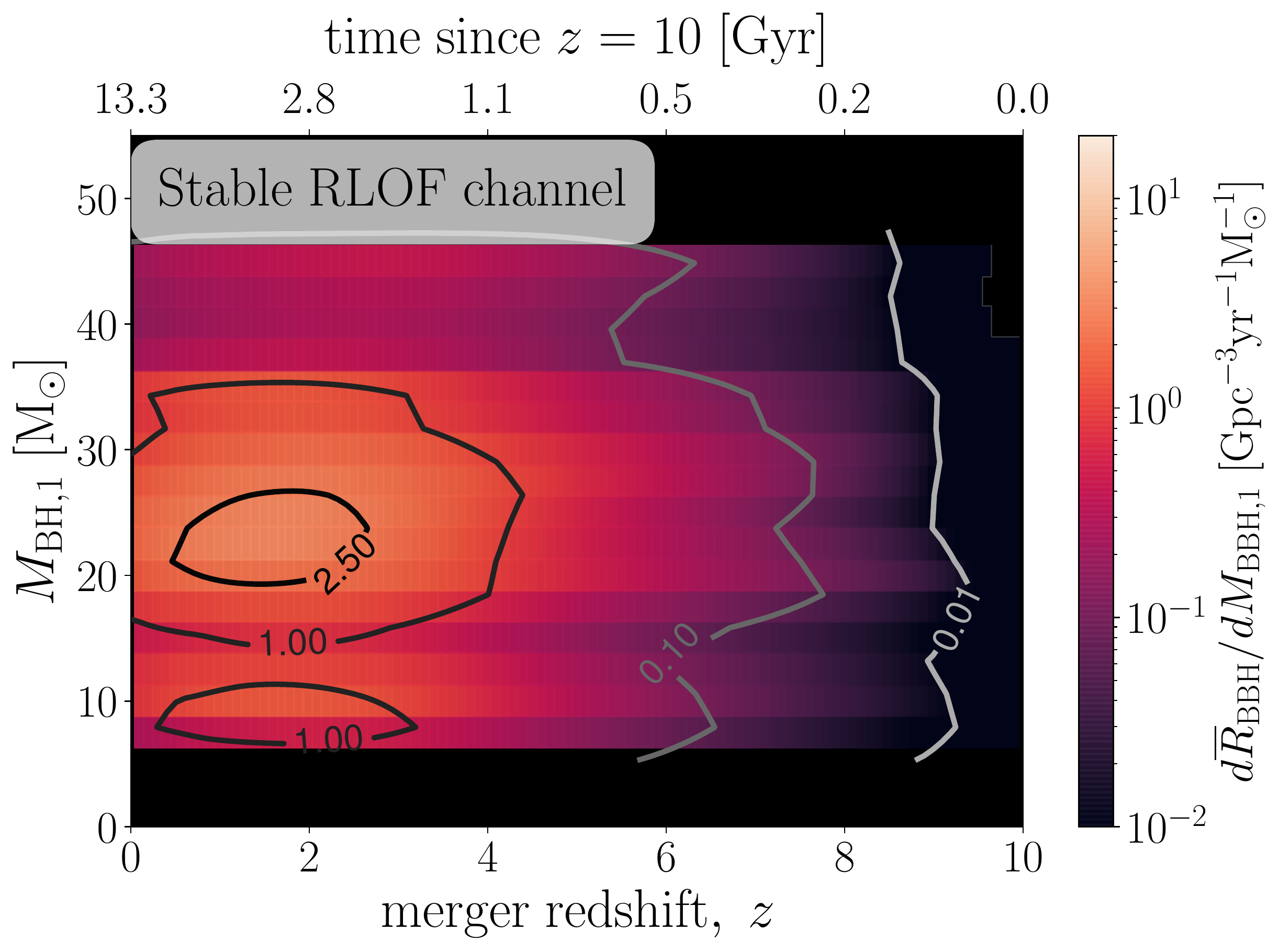}
\caption{The averaged intrinsic merger rate density \RBBHav, for redshift bins of $dz=0.2$, and primary BH mass bins of $d\Mbheen = 2.5\Msun$. The top axis shows the time passed since $z = 10$, which we have chosen as the redshift of first star formation. The left panel shows the full distribution. The middle panel shows mergers of systems that have experienced at least one CE during their evolution, while the right panel shows mergers of systems that formed through the stable RLOF channel. All panels show a dearth of high mass BHs ($\Mbheen\gtrsim30\Msun$) merging at higher redshifts ($z>6$). \label{fig: MBH-z}}
\end{figure*}

\begin{figure*}
\includegraphics[width=0.48\textwidth]{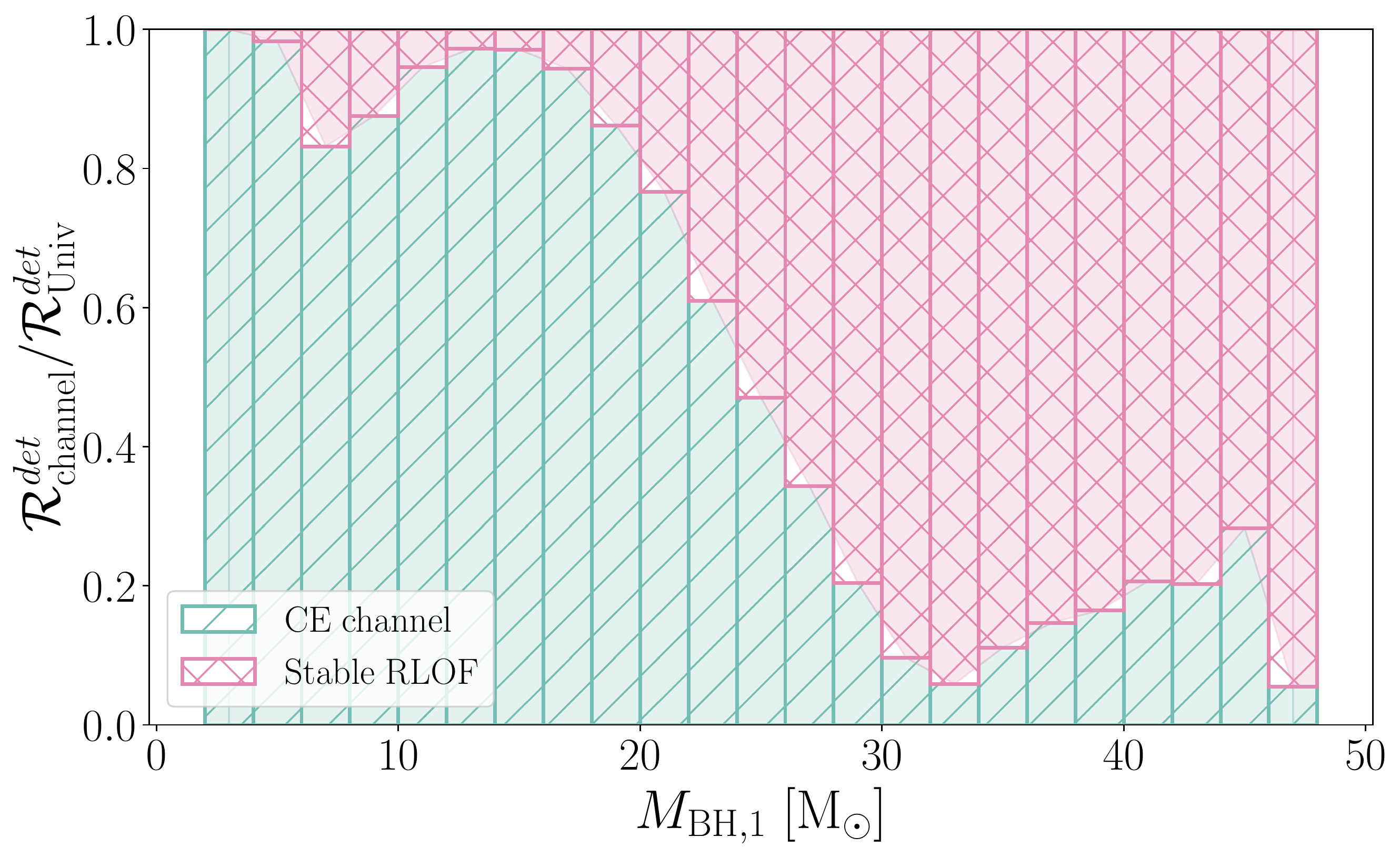}
\includegraphics[width=0.48\textwidth]{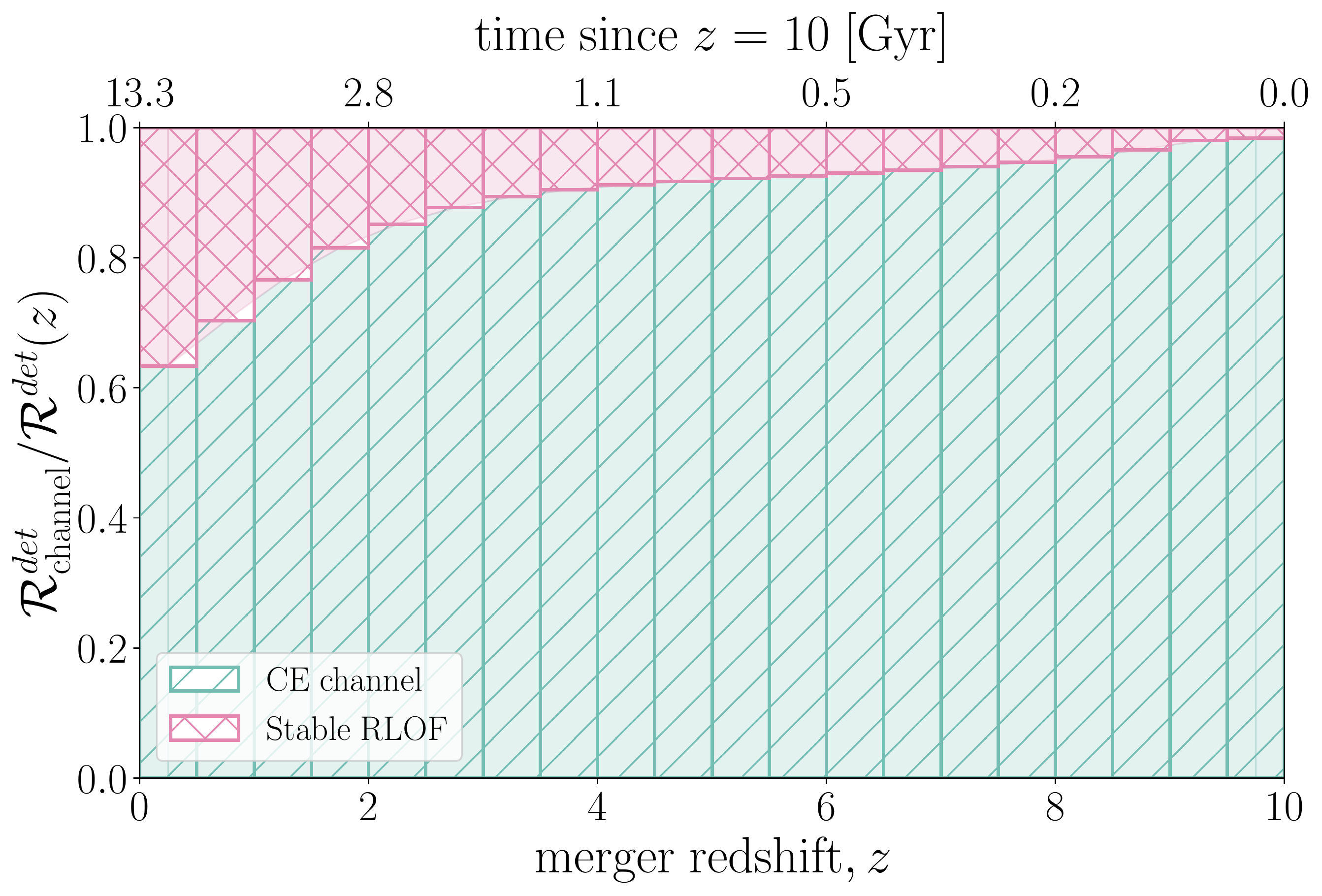}
\caption{ Fractional contribution of the CE channel (green hatched) and the stable RLOF channel (pink cross hatched) to \RBBHdet. Left panel shows the relative contributions to \RBBHdet per mass bin after integrating over all redshifts.  Right panel shows the fractional contribution to \RBBHdet integrated over all \Mbheen, as a function of redshift. 
\label{fig: fractional contribution}}
\end{figure*}

In Figure~\ref{fig: MBH-z} we show the averaged intrinsic merger rate density \RBBHav, as a function of redshift, $z$, and per primary BH mass, $\Mbheen$. We split the rate by channel, showing the CE and stable RLOF channel in the bottom left and right panel respectively. 

In the left hand panel, we see that the overall BBH merger rate density peaks around redshift $2-3$, and at a mass of about 15\Msun for the most massive BH. The merger rate decreases towards higher mass and higher redshift. 
Comparing the middle and right panels, we see that the CE channel and RLOF channel contribute to the rate in distinct ways.

We would like to quantify the relative contribution of each channel to the production of \Mbheen.
For this purpose we define the total rate of BBH mergers in the detector frame as: 
\begin{equation}
    \RBBHdet =  \frac{\RBBHzeta}{1+z} \frac{dV_c}{dz} .
    \label{eq: perfect detector rate}
\end{equation} 
Integrating this from redshift zero to the redshift of first star formation, we obtain the total rate of BBH mergers throughout the Universe:
\begin{equation}
    \RBBHdetU = \int_{0}^{z_{\mathrm{first \ SF}}} dz \  \RBBHdet 
    \label{eq: perfect detector rate all}
\end{equation}
This is the same as the BBH merger rate as observed by an infinitely sensitive detector at redshift zero. 
In the left-hand panel of Figure~\ref{fig: fractional contribution} we show what fraction of \RBBHdetU derives from which channel for different values of \Mbheen.
This emphasises how the stable RLOF channel dominates \RBBHdetU at higher masses, while the CE channel dominates for primary BH masses below 25\Msun.

The formation channels differ in how they contribute to the intrinsic merger rate density as a function of redshift. Specifically, the contribution of the stable RLOF channel decreases faster towards higher redshifts than the CE channel.  As a result, the CE channel becomes increasingly dominant towards higher redshifts. 
To show this more clearly, we again integrate \RBBHdet, but now over all \Mbheen to obtain \RBBHdetz. We show what fraction of \RBBHdetz derives from which channel for different redshift bins in the right-hand panel of Figure~\ref{fig: fractional contribution}.  Overall the CE channel is dominant, but the stable RLOF channel becomes more important at low redshift, and is responsible for about 40\% of BBHs merging in the local Universe. 

The reduced contribution of the stable RLOF channel at higher redshifts is a result of the scarcity of short delay times in this channel, as shown in Fig~\ref{fig: Mbh_tdelay}. Systems coming from the stable RLOF channel generally have delay times $\gtrsim 1 \Gyr$. At redshift 6, only 0.5\Gyr has passed since our adopted redshift of first star formation ($z=10$). This means that systems coming from the stable RLOF channel have typically not had enough time to merge at these high redshifts. 
For completeness, we show the distributions similar to Figure~\ref{fig: MBH-z}, but for chirp mass \Mchirp in Appendix~\ref{s: rate chirp mass}.

In Figure \ref{fig: mass dist channels} we display the distribution of \Mbheen split by formation channel, for merger redshifts between $0$ and $0.5$ (see equation \ref{eq: av rate dens BBH}). 

The results in Figure~\ref{fig: mass dist channels} imply that the high-mass merger events that have been detected so far at relatively low redshift, primarily come from the stable RLOF channel (assuming that the observed BBH merger rate is dominated by these two channels). This is in contrast to the results in, e.g., \citet{Belczynski+2016} and \citet{Stevenson+2017}, but agrees with findings in more recent work from e.g.\  \cite{Neijssel+2019} and \cite{GallegosGarcia2021}.

\begin{figure}
\includegraphics[width=0.49\textwidth]{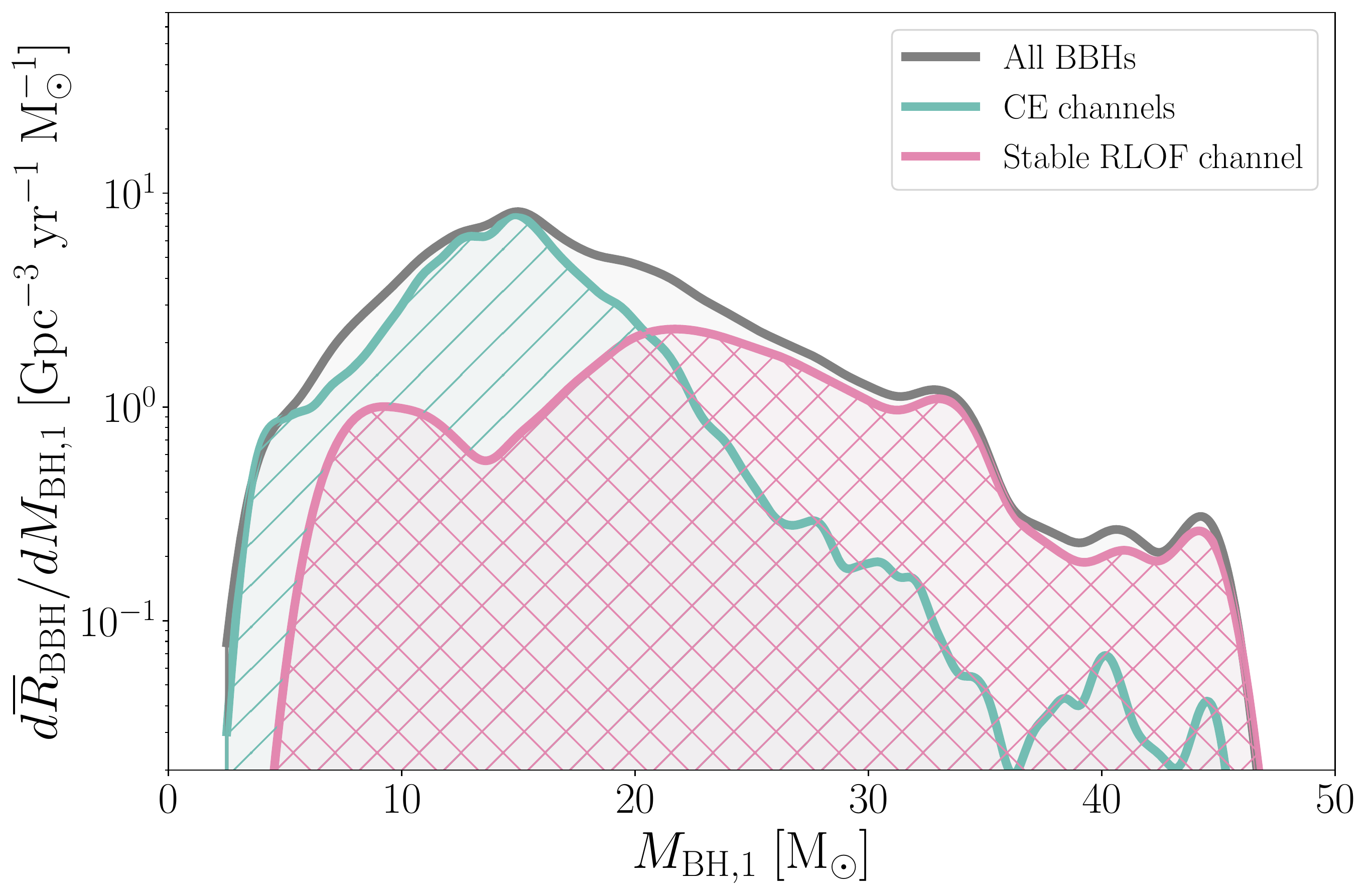} 
\caption{Distribution of primary BH masses \Mbheen split by formation channel, for merger redshifts between $0 \leq z <0.5$.  \label{fig: mass dist channels}  }
\end{figure}

\subsection{The shape of the mass function at different redshifts.}
\label{ss: rate binned by redshift}
\begin{figure*}
\includegraphics[width=0.49\textwidth]{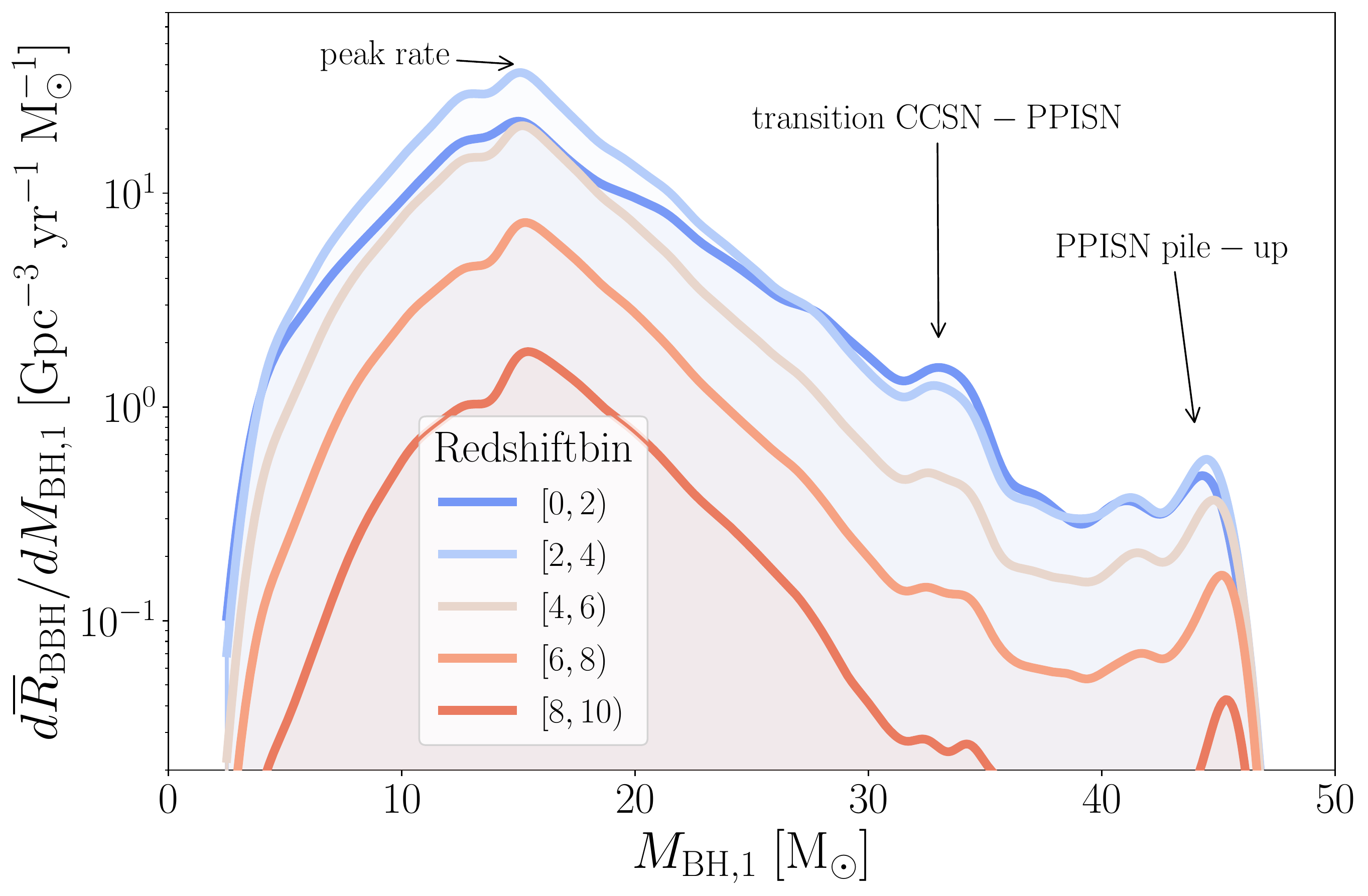} 
\includegraphics[width=0.49\textwidth]{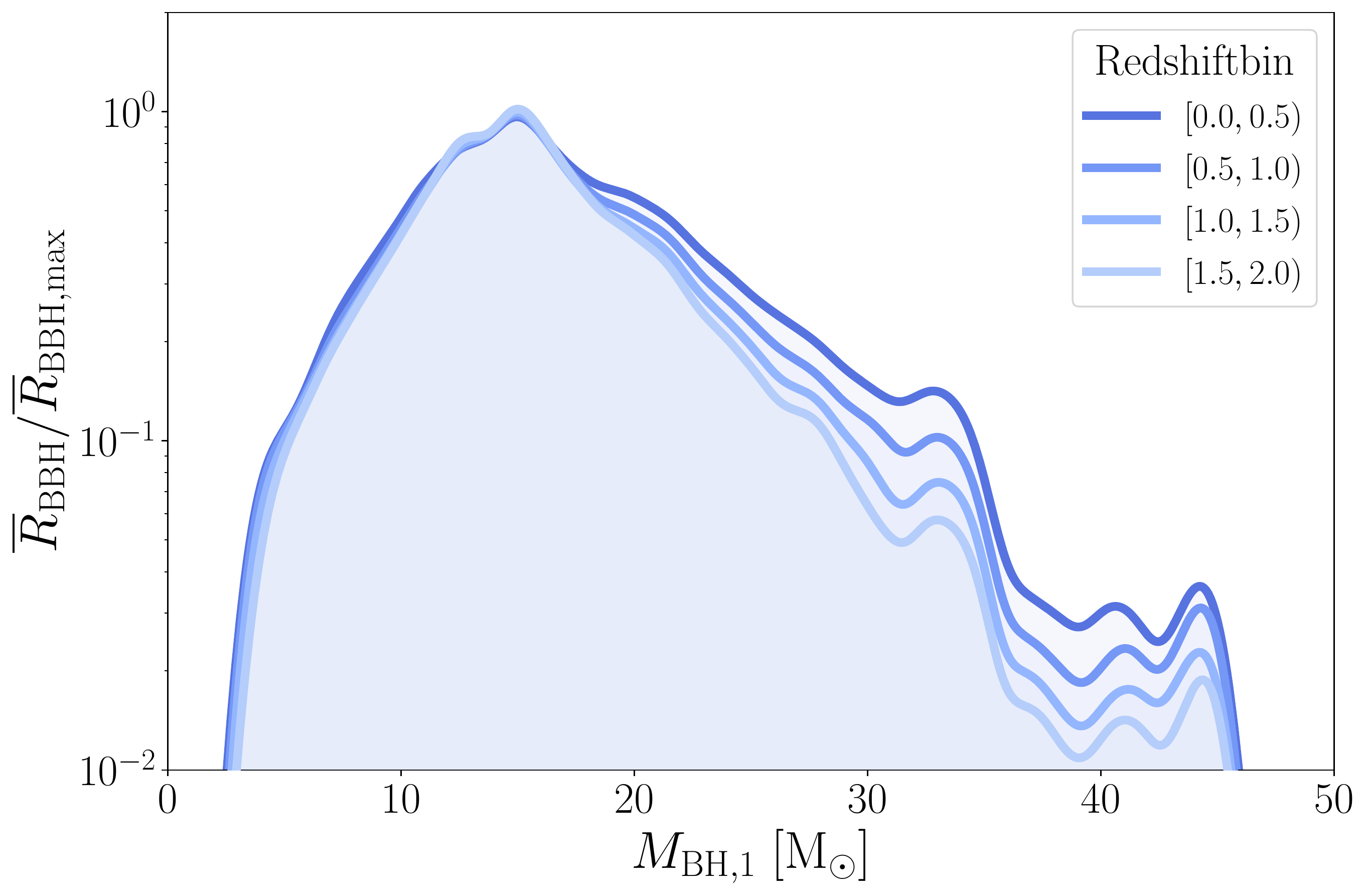}
\caption{Distribution of primary BH masses \Mbheen for several redshift bins. The left panel shows the general trend for different redshift bins. The right panel shows the same distribution normalised by the peak rate value for the given redshift bin, with a focus on redshifts up to $z = 2$. Both distributions are shown down to $\Mbheen=2.5\Msun$, which is our minimum allowed BH mass. This shows that the distribution of primary BH masses evolves with redshift.  \label{fig: mass dist redshifts}}
\end{figure*}

In the left panel of Figure~\ref{fig: mass dist redshifts} we show the \Mbheen distribution for different redshift bins (again adopting the averaged intrinsic merger rate density \RBBHav for every redshiftbin). We see that there are features of the mass distribution that persist in all redshift bins. Firstly, the peak of the distribution occurs at $\sim 18\Msun$. From Figures~\ref{fig: Mbh_tdelay} and \ref{fig: MBH-z} we find that this peak originates from the CE channel. 

\begin{figure}
\centering
\includegraphics[width=0.49\textwidth]{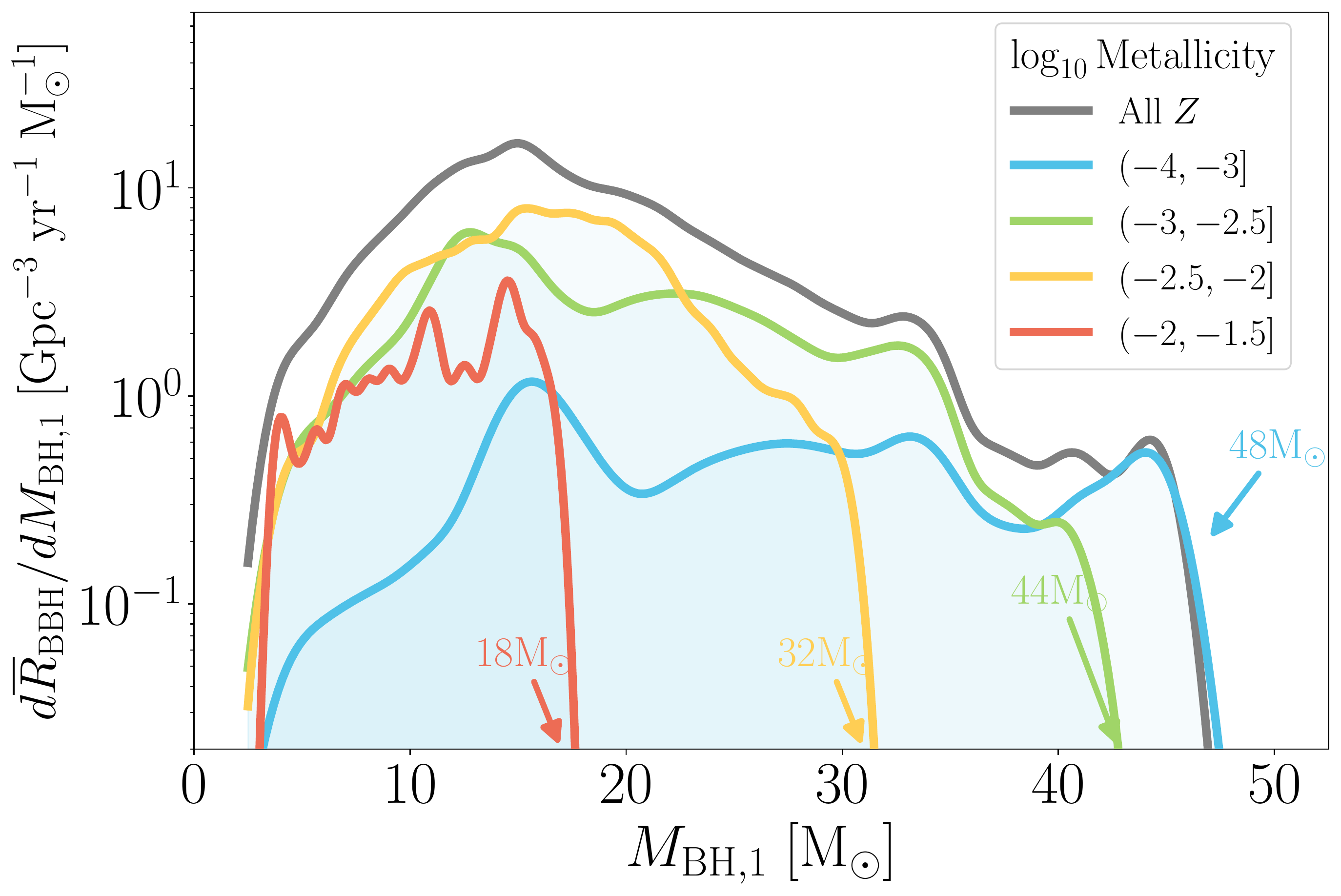}
\caption{Breakdown of the \Mbheen mass distribution by birth metallicity for all BBH mergers between redshifts $0\leq~z~<~0.5$. The maximum BH mass that contributes to each metallicity bin is annotated. \label{fig: mass dist by metals}}    
\end{figure}

In every redshift bin, \RBBHav decays for BH masses above $\sim 18\Msun$. 
In part, the slope on the right side of $\sim 18\Msun$ is steepened due to the decay of the initial mass function towards higher mass stars.
However, the primary driver behind the decay towards higher masses is the effect of metallicity: higher metallicities lead to more mass loss through stellar winds, and therefore shift the maximum possible \Mbheen to lower values. In Figure~\ref{fig: mass dist by metals} we show this shift in the maximum BH mass by dissecting the \Mbheen distribution for $0<z<0.5$ into bins of different formation metallicities. This shows that the maximum BH mass is about 18\Msun  in our simulations for the high metallicities ($Z \gtrsim 0.01$) that dominate the metallicity dependent star formation rate density, \SFRDzZ. 
For completeness, we show the \Mbheen distribution split by both formation channel and formation metallicity in appendix Figure \ref{fig: mass dist channels and Z}. This shows that the stable RLOF channel dominates the higher mass end of the distribution at every metallicity.

The decay of the distribution for BH masses below $\sim 18\Msun$ in Figure~\ref{fig: mass dist redshifts}, can be understood as a combination of our adopted SN kick and CE physics.
Firstly, above carbon oxygen core masses of $\Mco=11\Msun$, BHs are assumed to experience full fallback, and hence receive no kick. BHs from lower-mass progenitors are expected to receive higher SN kicks \citep[given the adopted BH-kick prescription from][]{Fryer+2012}. These higher SN kicks can unbind the binary system and thus prevent the formation of a merging BBH system (see also panels M, N and O in  Figure~\ref{fig: physics variations2}).. 
Secondly, for the same change in orbital separation, lower-mass BHs can provide less orbital energy to help unbind the common envelope. This means that progressively lower-mass BHs will fail to eject their companion`s envelope at a given CE efficiency $\alpha_{\mathrm{CE} }$. Increasing $\alpha_{\mathrm{CE} }$ will allow successful CE ejection for lower-mass BHs, thus pushing the peak of the mass distribution to lower-mass BHs (see also panels F-I Figure~\ref{fig: physics variations2}).

Apart from the peak in Figure~\ref{fig: mass dist redshifts}, two other distinct features persist in all redshift bins. The first is the rise in \RBBH just before the edge of the distribution at $\Mbheen \approx 45\Msun$. This feature is caused by the prescription for pair pulsations.  Specifically, we adopted the prescriptions from \cite{Farmer+2019} (see Section~\ref{sec: method}). This is also called the `pulsational pair-instability supernova' (or PPISN) pile-up \citep[e.g.\ ][]{Talbot:2018cva,Marchant+2019}.
Secondly there is a bump at $\Mbheen \sim 35\Msun$. This bump is an artefact of the transition between prescriptions for remnant masses from core collapse supernovae \citep[CCSN, following ][]{Fryer+2012}, to remnant masses from pair pulsational instability supernovae \citep[from ][]{Farmer+2019}. Though the bump in our results is an artificial feature, it is not clear that the transition between core-collapse supernovae and pair pulsational supernovae should be smooth. For example, \cite{Renzo2020_convection} argue that such a discontinuity can occur if convection is not efficient at carrying away energy for the lowest mass systems that experience pair pulsations. Furthermore, \cite{GWTC3_popPaper2021} find evidence for an overdensity in the merger rate ($>99\%$ credibility) at $\Mbheen = 35^{+1.5}_{-3.1}\Msun$ . It is difficult to attribute this observed peak to the PPISN pile-up at the lower-edge of the PISN mass gap, since stellar models predict this pile-up to occur at masses of about $40-60\Msun$ \citep[see e.g.\ ][ and references therein]{Marchant+2019,Farmer+2019, Renzo2020_ejecta, Renzo2020_convection, Marchant2020, Woosley2021, Costa2021}. 

\begin{figure*}
\includegraphics[width=\textwidth]{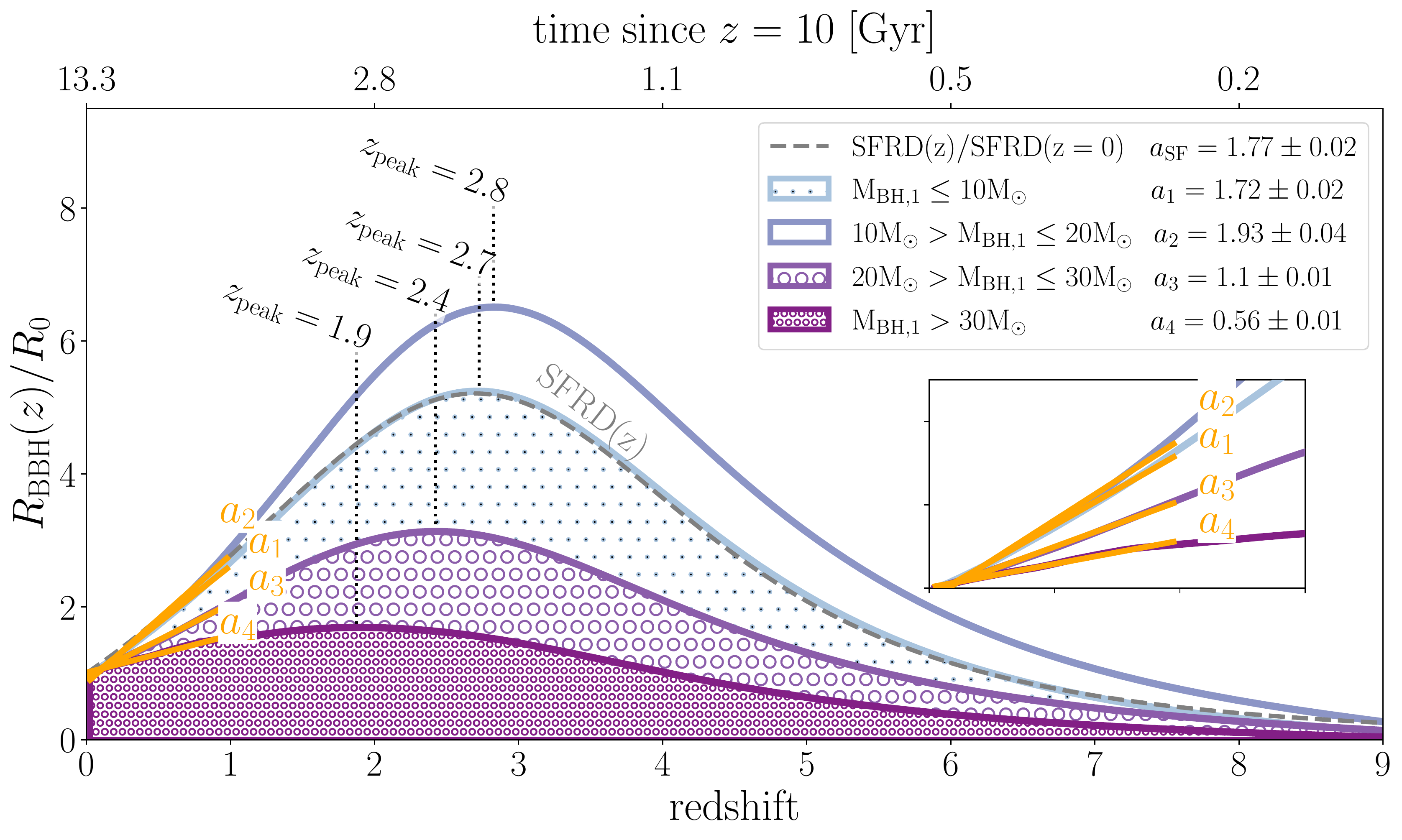}
\caption{Intrinsic BBH merger rate density as a function of redshift, $z$ (\RBBH, Eq.~\ref{eq: rate dens BBH}), normalised by the rate at redshift zero ($R_0$), for several bins in primary BH mass. The top axis shows the time since $z=10$, which we have chosen as the redshift of first star formation. The dashed grey line shows the star formation rate density as a function of redshift, SFRD(z), normalised by  the star formation rate density at redshift 0, SFRD($z=0$). The redshift at which the merger rate peaks is annotated with a dotted line for each mass bin. A linear fit to the merger rate density between $0 \leq z \leq 1$ is shown with an orange line for each mass bin (these are also highlighted in the inset). The respective slopes of these fits are annotated in the legend. This shows that, at low redshift, the slope of \RBBH is more shallow for higher \Mbheen. \label{fig: rate dens - z}}

\end{figure*}

To investigate redshift evolution of the primary BH mass distribution,  in the right panel of Figure~\ref{fig: mass dist redshifts} we show the intrinsic distribution normalised by the peak rate for each redshift bin. We focus on redshifts in the range $0 < z \leq 2$, because a large absolute change in \RBBH is contained in this redshift range (see Figure~\ref{fig: rate dens - z}), while the contribution from different metallicities to \SFRDzZ does not vary greatly up to $z\sim 1.5$.
The right panel of Figure~\ref{fig: mass dist redshifts} shows that the high mass end ($\Mbheen > 18\Msun$) decays faster at higher redshifts than the low mass end ($\Mbheen \leq 18\Msun$) of the distribution. 
We find that the ratio of $\Mbheen > 18\Msun / \Mbheen \leq 18\Msun$ is about 0.7 in the redshift bin $0-0.5$, while it is about 0.45 in the redshift bin $1-1.5$.
The steeper decay of the high mass end of the mass distribution for higher redshifts can be explained by the scarcer contribution of the stable RLOF channel (which is responsible for the high mass end of the mass distribution) towards higher redshifts, as discussed above in Section \ref{sec: Mbh redshift plane}.

\section{Prospects for observing trends with redshift in the intrinsic merger rate density}
\label{sec: prospects obs trends}

Third-generation detectors promise to probe BBH mergers across all redshifts of interest, but these instruments are still at least a decade away \citep[e.g.\ ][]{Sathyaprakash2019_CTET}. Present-day detectors are, however, already beginning to probe the evolution at low redshift. 

In the previous section we found evolution of the high-mass slope of the predicted \Mbheen distribution for redshifts in the range $0-2$.  Since current ground based detectors already detect many systems with $\Mbheen > 20\Msun$, it is possible to start probing this mass-specific redshift evolution of the merger rate \RBBH  \citep{GWTC2_popPaper2021,GWTC3_popPaper2021}.  

In this section we explore the possibility of probing trends of the rates separated by mass bin as a function of redshift.  In Section~\ref{ss: rate binned by mass} we show our predictions and in Section~\ref{ss: observing slopes in LIGO} we discuss whether these effects are observable in the second gravitational-wave transient catalogue (GWTC-2).

\subsection{The slope of the intrinsic rates per mass bin at low redshift}
\label{ss: rate binned by mass}
In Fig.~\ref{fig: rate dens - z} we show how the intrinsic BBH merger rate density, \RBBH, evolves as a function of redshift for four different \Mbheen mass bins.
In each mass bin we have normalised the merger rate to the rate at redshift zero, to emphasize different trends at low redshifts. We see clear differences in the evolution of the rate at low redshift and the overall redshift evolution. These differences are highlighted by the orange lines, that show linear fits in the range $0\leq~z\leq~1$, with the slopes $a_i$ provided in the legend.

For the lowest-mass BHs ($\Mbheen \leq 10\Msun$ and $10\Msun \leq \Mbheen \leq 20\Msun$), our models predict a steep increase of the BBH merger rate density with increasing redshift, with a slope that is very similar to the slope of  \SFRDz/SFRD($z=0$). 
The peak of the merger rate of the lowest \Mbheen bin coincides with the peak of \SFRDz/SFRD($z=0$), as adopted in our models (at $z=2.7$).  
The merger rate for slightly higher masses ($10\Msun \leq \Mbheen \leq 20\Msun$), peaks at slightly higher redshifts, around $z=2.8$.
The redshift evolution of $\RBBH/R_0$ follows the shape of \SFRDz/SFRD($z=0$) for these mass bins, because the lowest-mass events are formed predominantly through the CE channel, which produces short delay time systems.
On top of this, these lower-mass events can form from almost all metallicities, as opposed to the high-mass systems that only form from the lowest metallicities (see Figure~\ref{fig: mass dist by metals}). 

In contrast, for BHs with masses in the range $20\Msun~<~\Mbheen~\leq~30\Msun$ we find that the evolution of the merger rate with redshift is much less steep in the low-redshift regime than the merger rate for lower-mass BHs. Moreover, the merger rate of these events starts to decline at redshift $z = 2.4$, lower than the redshift of peak \SFRDz.
The rate density for the most massive BHs ($\Mbheen > 30\Msun$) exhibits the flattest slope and peaks at the lowest redshift ( at $z=1.9$).
In other words, in order to capture the peak of the BBH merger rate density for BHs with $\Mbheen \gtrsim 30\Msun$ we need gravitational wave detectors that can observe out to redshift $z\sim2$ (depending on the exact location of the peak of star formation). 
This peak at lower redshift can be understood from the characteristics of the stable RLOF channel, which is the primary producer of such massive events. As discussed in Sections~\ref{sec: Mbh-tdelay} and \ref{sec: merger rate redshift}, these events primarily form with long delay times.  
Hence, at progressively higher redshifts, the fraction of systems formed through the stable RLOF channel BBHs that can contribute to the merger rate decreases. The systems that don't contribute at higher redshift have not had sufficient time since the adopted moment of first star formation to merge as a BBH.

This implies that mergers of massive BHs are relatively less common at higher redshifts. This may at first sight seem counter intuitive, considering that at higher redshifts, the low metallicities that allow for the formation of massive BHs are more common \citep[see Figure~\ref{fig: mass dist by metals} and, e.g.\ ][]{Vink+2005,Belczynski_2010,Spera+2019}.

\subsection{Observing the different slopes in GWTC-2}
\label{ss: observing slopes in LIGO}
To test our prediction of a distinct redshift evolution for different \Mbheen as discussed in Section~\ref{ss: rate binned by mass}, we look for observational evidence of a different slope in \RBBH in the open data from the first, second, and half of the third observing runs of Advanced LIGO and Advanced Virgo \citep{GWTC2_OpenDataSoftX_2021}, also presented in the gravitational-wave transient catalogues GWTC-2 \citep{GWTC2} and GWTC-2.1 \citep{GWTC2_1_2021}. 
To this end, we use the observed BBH mergers to hierarchically infer their underlying mass and spin distributions \citep[e.g.\ ][]{MandelFarrGair2019}.

Contrary to our predictions here, analyses of the BBH population typically assume that BBHs have independently distributed masses and redshifts, with $p(\Mbheen,z) = p(\Mbheen)p(z)$.
Here, we will explore several alternative models for the joint distribution $p(\Mbheen,z)$ of BBH masses and redshifts.
Our method closely follows that of \cite{Callister2021}. We assume that the distribution of mass ratios $p(q|\Mbheen, \gamma )$ follows a power-law with index $\gamma$ and that the distribution of effective spins, $p(\chi_{\mathrm{eff} }|\mu_{\chi}, \sigma_{\chi})$, follows a Gaussian with mean $\mu_{\chi}$ and variance $\sigma_{\chi}$ \citep{Roulet2019,Miller2020}. 

For primary masses and redshifts, we take as a baseline the \textsc{Powerlaw + Peak} model from \cite{GWTC2_popPaper2021}, with an overall merger rate that is allowed to evolve as a function of $z$:
\begin{equation}
\label{eq:mass-z-model}
\begin{aligned}
&\frac{dN_\mathrm{BBH}}{dt\, d\Mbheen\, dz} = R_0 
    \frac{dV_c}{dz} (1+z)^\kappa
    \Big[f_p P(\Mbheen|\lambda, m_{\mathrm{max} }) 
        \\&\hspace{2cm} + (1-f_p)N(\Mbheen|\mu_m, \sigma_m, m_{\mathrm{max}})
        \Big].
\end{aligned}
\end{equation}
Here, the assumed primary mass distribution is a mixture between a power law $P(\Mbheen|\lambda, m_{\mathrm{max}}) \propto \Mbheen^{\lambda}$ (for $\Mbheen$ between $5\,M_\odot$ and $m_\mathrm{max}$) and a Gaussian peak $N(\Mbheen|\mu_m, \sigma_m, m_{\mathrm{max}})$, with mean $\mu_m$ and variance $\sigma_m$, which is needed to fit an observed excess of BBHs with primary masses near $\Mbheen\approx 35\,M_\odot$.
$R_0$ is the local rate of BBH mergers per co-moving volume at $z=0$.

We inspect several variations of this model in an attempt to identify any relationship between BBH masses and their redshift distribution.

First, we expanded Eq.~\eqref{eq:mass-z-model} such that the parameter $\kappa$, governing the BBH rate evolution, is a function of $\Mbheen$.
We considered several possibilities, including a piecewise function cut at $30\,M_\odot$,
    \begin{equation}
    \kappa(\Mbheen) = \begin{cases}
        \kappa_\mathrm{low} & (\Mbheen<30\,M_\odot) \\
        \kappa_\mathrm{high} & (\Mbheen\geq 30\,M_\odot),
        \end{cases}
    \end{equation}
a piecewise function in which the cut location $m_\mathrm{cut}$ itself varies as a free parameter,
    \begin{equation}
    \kappa(\Mbheen|m_\mathrm{cut}) = \begin{cases}
        \kappa_\mathrm{low} & (\Mbheen<m_\mathrm{cut}) \\
        \kappa_\mathrm{high} & (\Mbheen\geq m_\mathrm{cut}),
        \end{cases}
    \end{equation}
and a case in which $\kappa$ is a linear function of $\Mbheen$:
    \begin{equation}
    \kappa\big(\Mbheen|\kappa_0,\kappa'\big) = \kappa_0 + \kappa' \left(\frac{\Mbheen}{30\Msun} - 1 \right).
    \end{equation}
In Fig.~\ref{fig: rate dens - z}, we also saw that $dR_\mathrm{BBH}/dz$ is not a strictly monotonic function of mass.
Instead, this slope reaches a maximum in the range $10\,M_\odot < \Mbheen \leq 20\,M_\odot$, below which it again decreases.
To capture this possibility, we additionally considered a \textit{three}-bin piecewise model,
\begin{equation}
    \kappa(\Mbheen) = \begin{cases}
        \kappa_\mathrm{low} & (\Mbheen<10\,M_\odot) \\
        \kappa_\mathrm{mid} & (10\,M_\odot\geq\Mbheen<30\,M_\odot) \\
        \kappa_\mathrm{high} & (\Mbheen\geq 30\,M_\odot),
        \end{cases}
    \end{equation}
We do not consider more complex models, given the relative scarcity of the data available at the time of writing.
In all four cases above, we find no evidence for a varying redshift distribution as a function of mass.

As mentioned above, the BBH primary mass distribution in GWTC-2 is well-modelled as a mixture between a broad power law and an additional peak between $30$ to $35\,M_\odot$.
As an alternative test, we allow the rates of BBHs comprising the broad power law and those situated in the peak to each evolve independently as a function of redshift: 
\begin{equation}
\label{eq:mass-z-model-2}
\begin{aligned}
&\frac{dN_\mathrm{BBH}}{dt\, d\Mbheen\, dz} = \frac{dV_c}{dz}
    \Big[
    R_0^\mathrm{pl} (1+z)^{\kappa_\mathrm{pl}} P(\Mbheen|\lambda, m_{\mathrm{max} }) 
    \\&\hspace{0.5cm}
    + R_0^\mathrm{peak} (1+z)^{\kappa_\mathrm{peak}} N(\Mbheen|\mu_m, \sigma_m, m_{\mathrm{max}}).
        \Big],
\end{aligned}
\end{equation}
in which $R_0^\mathrm{pl}$ and $R_0^\mathrm{peak}$ are the local merger rate densities of BBHs in the power law and peak, respectively, with $\kappa_\mathrm{pl}$ and $\kappa_\mathrm{peak}$ governing the redshift evolution of each rate.
We find very marginal evidence that the BBH mergers comprising these two components obey different redshift distributions; we measure  $\kappa_\mathrm{pl} = 2.7^{+3.2}_{-3.5}$ and $\kappa_\mathrm{peak} = 0.7^{+4.0}_{-5.8}$, with $\kappa_\mathrm{peak} < \kappa_\mathrm{pl}$ for about 70\% of the posterior samples.
However, our large uncertainties mean we cannot draw any conclusions about differing rate evolution (or lack thereof).

We conclude that we find insufficient evidence in GWTC-2 \citep{GWTC2} for a distinct redshift evolution of \RBBH for different \Mbheen. 
This is consistent with \cite{Fishbach2021_when_bigBH}, who find no strong evidence in GWTC-2 that the BBH mass distribution evolves with redshift.
Specifically, they find that the detections in GWTC-2 are consistent with a mass distribution that consists of a power law with a break that does \textit{not} evolve with redshift, as well as with a mass distribution that includes a sharp maximum mass cutoff, if this cutoff \textit{does} evolve with redshift. 
Furthermore, \cite{FishbachKalogera2021} found no strong evidence for the time delay distribution to evolve with mass. They did find a mild preference for high mass ($\Mbheen \sim 50\Msun$) BBH to prefer shorter delay times than the low mass ($\Mbheen \sim 15\Msun$) BBH systems. However, they also argue that this preference could be an effect of higher mass BHs forming more strictly at the lowest metallicities (which is consistent with our findings in Figure \ref{fig: mass dist by metals}). Alternatively, these high mass mergers with masses of about $50\Msun$ could be probing hierarchical mergers. 

At the time of writing, finding evidence for a distinct redshift evolution in GWTC-2 is difficult, considering that observed BBHs with lower mass primary BH masses ($\Mbheen\sim 10\Msun$) only probe the very local Universe ($z\lesssim 0.4$). As can be seen from Figure~\ref{fig: rate dens - z}, this redshift range encompasses only a small fraction of the BBH merger rate evolution. 
Given the prospects of observing BBH mergers out to increasingly high redshifts with Advanced LIGO, Advanced Virgo and KAGRA \citep{LIGO2018_obsprospects}, second- \citep[Voyager][]{Adhikari2020}, and third-generation detectors like the Einstein telescope \citep{Punturo_2010,Hild_2011,Sathyaprakash2019,Maggiore2020} and the Cosmic Explorer \citep{Abbott2017_CE,Reitze2019} we expect our predicted different evolution of the BBH merger rate to be either confirmed or disproven within the coming decades.

\section{Discussion}
\label{sec: discussion}
In the previous sections we showed our prediction that the mass distribution of merging BBH systems varies with redshift. Specifically, we showed that the evolution of the merger rate with redshift, \RBBH, is more shallow and peaks at lower redshifts for systems with higher primary BH masses compared to systems with lower primary BH masses. 
This difference is the result of the contribution of two different formation channels. The CE channel predominantly forms lower mass BBH systems ($\Mbheen \lesssim 30\Msun$) and allows for very short delay times ($\tdelay < 1 \Gyr$).  In contrast, the stable RLOF channel is the main source of massive systems ($\Mbheen \gtrsim 30\Msun$) and primarily forms systems with longer delay times ($\tdelay \gtrsim 1 \Gyr$). 

The quantitative predictions presented in this work are subject to several major uncertainties and we discuss the key ones in the remainder of this section. Throughout this section we also argue why we expect our qualitative findings to be robust. 

\subsection{ The relative contribution of the CE and stable RLOF channel} \label{ss: contrib. stable and CE channel}

The prediction that merging BBHs can be formed through both the CE and stable RLOF channels has been reported by various groups \citep[e.g.\ ][]{van-den-Heuvel+2017, Bavera2021, Marchant2021, Broekgaarden+2021a, GallegosGarcia2021, Shao2021, Olejak2021}.
However, the relative contribution of both channels is uncertain. This is mainly due to uncertainties in the treatment of stability of mass transfer, and whether or not the ejection of a common envelope is successful (\citealt{Ivanova+2013}, \citealt{IvanovaJusthamRicker2020}, and references therein).

Recent work by e.g. \citet{Pavlovskii2017}, \citet{Klencki2021}, \citet{Marchant2021} and \citet{GallegosGarcia2021} have questioned whether the CE channel plays a prominent role, based on results obtained with the 1D detailed binary evolutionary code MESA \citep{Paxton+2015}. They argue that systems that are typically assumed to lead to successful CE ejection in rapid population synthesis simulations (such as ours), will instead fail to initiate and survive a common envelope phase.  
If true, this would potentially drastically reduce the relative contribution of the CE channel. This would have major implications for the field and implies that the contribution of the CE channel is over estimated in our work.

Despite all off the above, it seems unlikely that the CE channel does not operate at all. Various compact binary systems containing double white dwarfs and double neutron stars exist, which are hard to all explain through other formation channels \citep{RebassaMansergas+2007,RebassaMansergas+2012,Nebot+2011,Ivanova+2013}.    
As long as the CE channel plays a non-negligible role, we believe that at least our qualitative conclusions will hold.

\subsection{Are the delay time and mass distributions of the two channels distinguishable? \label{ss: tdelay dist both channels}} 

Although the detailed shape of the delay time and mass distributions are uncertain, we believe that our finding that these two channels lead to distinct delay time distributions is robust for the following reasons. 

The first reason is that the CE channel and stable RLOF channel lose angular momentum through intrinsically different mechanisms as explained in  Section~\ref{ss: stable RLOF channel reasoning}. Because of this, it is reasonable to expect a difference in the distributions of final separations and thus inspiral times. In fact, fine tuning would be required to avoid significant differences. Similar arguments can be made for the mass distribution  \citep[see e.g.][]{Dominik+2012,EldridgeStanway2016, Bavera2021,GallegosGarcia2021}.

To better understand the impact of our (uncertain) model assumptions on the resulting delay time and mass distributions we have analysed the suite of models presented in \citet{Broekgaarden+2021b} (see Appendix~\ref{s: physics variations}). A relative lack of high mass BHs with
short delay times was found in all model variations. Furthermore, we find significant differences in the delay-time and mass distributions for the two channels for almost all variations.

Exceptions concern the models where we assume high values for the CE ejection efficiency $\alpha_{\mathrm{CE}}$ (panels~H and I in Figure~\ref{fig: physics variations1}).
In these simulations the number of short delay-time systems resulting from the CE channel is reduced (for $\alpha_{\mathrm{CE}}=2$) or disappear entirely (for $\alpha_{\mathrm{CE}} = 10$).  
The latter assumption results in delay-time distributions for the CE and RLOF channel that are practically indistinguishable, but we consider such high efficiencies unrealistic.

The distinction in the \Mbheen distribution diminishes in the models where a fixed accretion efficiency during stable Roche-lobe overflow involving two stellar companions is considered, $\beta = 0.25$ and $\beta = 0.5$, where $\beta$ denotes the fraction of the mass lost by the donor that is accreted by the companion (see panels~B and C in Figure~\ref{fig: physics variations1}). 
In these models, we find that the RLOF channel is less efficient in producing BBH mergers, especially in the case of systems with high-mass \Mbheen. We still find significant differences in the delay times between the two channels, but the RLOF and CE channel can no longer be clearly distinguished in the $\Mbheen$ distribution.  While the mass accretion efficiency is an important uncertainty in our simulations, we do not believe that assuming a fixed accretion efficiency is realistic.

%
\subsection{Alternative observables to distinguish the two channels}
\label{sss: other obs}

We are not able to directly observe whether a BBH was formed through the CE channel or the stable RLOF channel. Hence we need characteristic observable source properties to expose the distinct rate evolution. 
In this work we have focused on BH mass as this can be inferred relatively well from observations. Possible other observables that could be used are the distribution of the BH spins, the secondary masses, and the mass ratio.

\begin{figure*}
\centering
\includegraphics[width=0.45\textwidth]{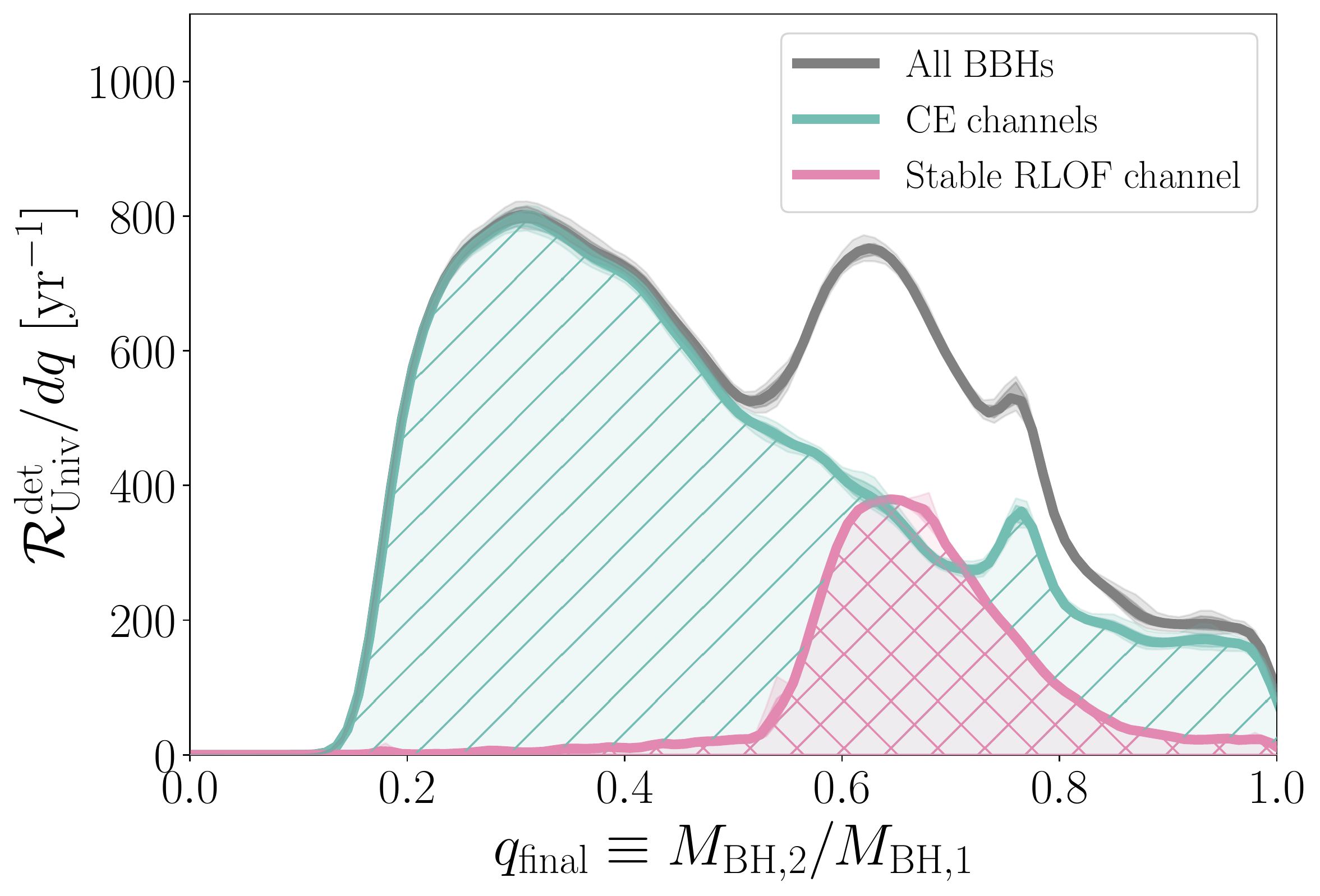}
\includegraphics[width=0.45\textwidth]{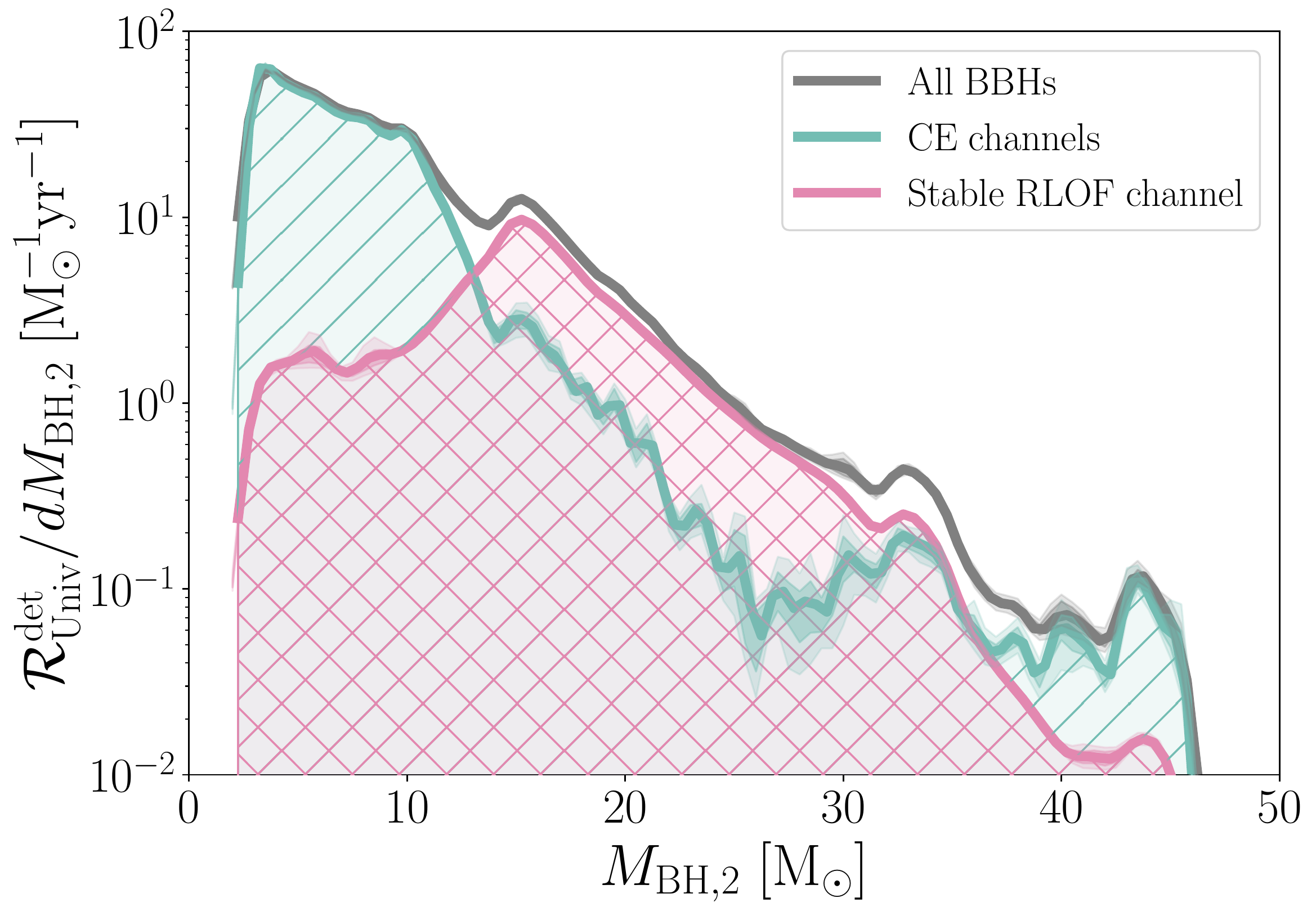}
\caption{Distributions of  mass ratios, $\qfinal$, and secondary masses, \Mbhtwee, for BBHs seen by a hypothetical perfect detector (\RBBHdetU, equation~\ref{eq: perfect detector rate all}). Each panel shows the distribution for all systems in grey, the stable RLOF channel in cross hatched pink, and the CE channel in line hatched green. The dark and light shaded areas shows the 1- and 2-$\sigma$ sampling uncertainties respectively, obtained through bootstrapping. \label{fig: other obs}}
\end{figure*}

\paragraph {Mass ratios}
In the left panel of Figure~\ref{fig: other obs} we show our predictions for the distribution of mass ratios as seen by a hypothetical perfect detector (equation~\ref{eq: perfect detector rate all}), which are very different for both channels. The CE channel preferentially produces systems with unequal masses ($\qfinal \approx 0.3$) but the distribution is broad and spans from $0.2 \lesssim \qfinal \lesssim 1$.  In contrast, we find that the stable RLOF channel predominantly forms merging binaries with $0.6 \lesssim \qfinal \lesssim 0.8$ in our simulation. The distinct shape of this distribution is the result of the requirement of the stability of mass transfer, the total-mass to core-mass relation, the mass transfer efficiency (see Appendix~\ref{sec: q per delay time} for an analytical derivation of the low \qfinal end). The clear difference in the two distributions is promising, but we note that at the time of writing the mass ratios inferred for the detected systems are typically not well constrained \citep[e.g.][]{GWTC3}.

\paragraph{Secondary masses} The distribution of secondary masses, \Mbhtwee, is shown in the right panel of Figure~\ref{fig: other obs}.  
The CE channel dominates the formation of low secondary BH masses $ \Mbhtwee < 15\Msun$, while the stable RLOF channel dominates in the range $15\Msun < \Mbhtwee < 40\Msun$. The reason for this is the same as discussed in  Section~\ref{ss: CE channel reasoning}. The CE channel dominates again for the highest secondary mass BHs ($36\Msun < \Mbhtwee < 46\Msun$). The contribution of the stable RLOF drops quickly here due to a lack of equal mass systems and the PISN mass limit of about $46\Msun$.  We caution not to over interpret the features of the highest mass BHs as the uncertainties in the evolution of the progenitor systems are the largest here.

 
\paragraph{Spins} Gravitational wave observations provide constraints on the mass weighted effective spin, $\chi_{\rm eff}$ and for some events on the individual spin magnitudes and their orientation. The constraints on the spin have been  suggested as a promising diagnostic to  distinguish formation scenarios
\citep[e.g.][]{Kushnir+2016,Hotokezaka+2017,Zaldarriaga+2018}

Our simulations do not provide predictions for the spin, but 
\citet{Bavera2020} showed that, in case of the CE channel, the post-CE separation may well be small enough to allow for tidal spin up of the He core that is the progenitor of the second born BH \citep[e.g.\ ][]{Bavera2020, Mandel2020}.  In the case of the stable RLOF channel, final separations are expected to be too wide for tidal spin-up \citep[e.g.][]{Bavera2021}, but one might expect spin-up of the first born BH through mass transfer \citep[e.g.\ ][]{Bardeen1970}, although this is matter of debate. 
In case of Eddington limited accretion, spin up may not be significant \citep{Bavera2021}.
In the case of super-Eddington accretion it remains unclear whether one can significantly spin up the accreting BH \cite[e.g.\  ][]{Tchekhovskoy2012} and in this case the orbit widens preventing the formation of a GW source \citep{vanson2020}. Furthermore, large uncertainties remain in the angular momentum transport of massive stars, which makes it difficult to accurately translate stellar spins to BH spins (see e.g.\ \citealt{Fuller2015}, \citealt{OlejakBelczynski2021} and \citealt{Steinle2021} for a discussion of possible pathways to spinning BHs from the isolated binary channel).

\subsection{ The uncertain metallicity dependent cosmic star formation history  \label{ss: effect of SFRD}}
In general, variations in the assumed \SFRDzZ have a large impact on \RBBH, and the shape of the BH mass distribution \citep[e.g.\ ][]{Chruslinska+2018, Neijssel+2019, Broekgaarden+2021b, Briel+2021}.
 Because the highest mass BHs can only form from the lowest metallicities (see Figure \ref{fig: mass dist by metals}), the stable RLOF channel will only play a significant role in the BBH merger rate if there is sufficient star formation at low metallicity, and the stable RLOF systems have had enough time to coalesce since this low metallicity star formation. 

 To test the effect of the \SFRDzZ on our main results, we repeated our complete analysis while adopting the phenomenological model from \cite{Neijssel+2019}. 
This \SFRDzZ forms fewer stars at low metallicity ($Z <0.01$) for the majority of our simulated star-forming universe, but forms a significantly larger amount of low-metallicity stars at the highest redshifts.
Because this model is very sharply peaked around the mean metallicity at each redshift there is almost no star formation at low metallicities for all redshifts lower than $z\approx1$.
In contrast, in our fiducial model we adopt a skewed distribution to capture the tail of low metallicity star formation at low redshifts. 

With this \SFRDzZ, we still retrieve the distinct redshift evolution for different BH mass bins, similar to the trends discussed in Sections~\ref{sec: merger rate redshift} and \ref{sec: prospects obs trends} . Specifically we find a steep positive slope for \RBBH between $0<z<1$ for BBHs with $\Mbheen < 20\Msun$, and a more shallow slope for  BBHs with $\Mbheen \geq 20\Msun$. This causes the high mass end ($\Mbheen \gtrsim 20\Msun$) of the \Mbheen mass distribution to decay faster at higher redshifts than the low-mass end ($\Mbheen \lesssim 18\Msun$) of the distribution. This is in line with \cite{Neijssel+2019}, who also found evidence for evolution of the BBH mass distribution with redshift. 

Our estimate of the total intrinsic BBH merger rate is $R_0 = 73 \Gpc^{-3} \rm{yr^{-1}}$  at redshift zero, and $R_{0.2} = 94 \Gpc^{-3} \rm{yr^{-1}}$ at $z = 0.2$. Although this rate prediction is not an outlier in the recent review of local BBH merger rate predictions for isolated binaries from \cite{MandelBroekgaarden2021}, it is a factor 2-5 higher than  the most recent estimates from the LIGO/Virgo/Kagra collaboration \citep[$R_{0.2}=17.3-45\Gpc^{-3}\mathrm{yr^{-1}}$,][]{GWTC3_popPaper2021}.
Our setup and binary physics assumptions are similar to those in \cite{Neijssel+2019}, who predict a local rate of $R_0 \approx 22 \Gpc^{-3} \rm{yr^{-1}}$. The difference in our rate prediction stems from our updated prescription for the metallicity-dependent star-formation rate density  as described above, \SFRDzZ (see also Appendix~\ref{s: fitting SFRD}).

 Although we acknowledge the large uncertainties in \SFRDzZ, we note that if we are sufficiently confident in the delay time distributions of observed BBH mergers, the redshift evolution of the BBH merger rate can be used to measure the star formation rate with gravitational waves \citep[][]{2019Vitale}. Therefore, detecting evolution in the BH mass distribution as described in Section~\ref{sec: prospects obs trends} could help us constrain \SFRDzZ through gravitational waves.

 \subsection{ Further caveats of rapid population synthesis \label{ss: caveats} }

All uncertainties that apply to rapid population synthesis simulations also apply to this work  \citep[see e.g. ][]{AblimitMaeda2018,Belczynski+2021,Broekgaarden+2021b}.   Above, we already discussed the main uncertainties related to mass transfer stability and the treatment of common envelope phases.  Below, we highlight further known shortcomings and uncertainties that are expected to impact our quantitative predictions

A major uncertainty for the evolution of massive stars concerns internal mixing and, specifically, mixing beyond the boundaries of the convectively unstable regions. This directly impacts the core masses. In our simulations we use prescriptions from \cite{Hurley+2000} that are fitted against models by \citet{Pols+1997}. For stars with initial masses higher than $50\Msun$ these fits are extrapolated.  The core masses in our simulations turn out to be substantially smaller than those predicted in more recent grids of detailed evolutionary models that were calibrated against observations \citep[e.g.\ ][]{Brott+2011}. 
Overall, we expect that our core masses for high mass stars to be underestimated (as is true for all simulations that apply the original Hurley formulae). This will affect the quantitative predictions for the BH mass, and mass ratio distributions. This includes our predictions for the maximum BH mass that is efficiently formed through the CE channel ($\sim 30\Msun$ in this work). 

The post-supernova remnant mass, including the amount of fallback, is uncertain.
In particular, stars that retain a significant fraction of their envelope up to the moment of core collapse have been hypothesised to produce massive BHs if the envelope is assumed to entirely fall back onto the newly formed BH \citep[e.g.\ ][]{Fernandez2018,DiCarlo2019a, DiCarlo2019b}.
This way, relatively low mass stars ($M_{\mathrm{ZAMS}} \lesssim 40\Msun$) that are expected to more easily lead to successful CE events (following our arguments as stated in Section~\ref{ss: CE channel reasoning}), can still form high BH masses \citep[$\Mbheen \gtrsim 30\Msun$,][]{DiCarlo2019a, DiCarlo2019b,DiCarlo2020,Kremer2020}.
However, for red supergiant stars, the envelope is expected to be sufficiently loosely bound that the change in gravitational mass due to neutrino losses when a core collapses likely unbinds the envelope \citep{Nadezhin1980,Lovegrove+2013, Adams+2017}. Complete fallback is expected only for blue and yellow supergiants \citep{Fernandez2018, IvanovFernandez2021}.
Moreover, in this work we only study isolated binaries, which are not able to form BBH progenitors that merge within the age of the Universe without the system transferring or losing angular momentum as a consequence of mass transfer. Mass transfer, whether stable or unstable (CE) leads to significant mass loss in our simulations.
Therefore, we find that forming merging BBHs with a massive primary BH through the fallback of a hydrogen envelope only works if there is an external mechanism that brings the BH progenitors closer together. 

Lastly, in this work we have assumed a universal initial mass function (IMF). However, recent studies suggest that the IMF might be more top-heavy at low metallicity \citep[e.g.\ ][]{Geha2013,Navarro2015,Schneider2018, Gennaro2018}. Although uncertainties in the IMF can have a large impact on rate predictions \citep{de-Mink+2015, Chruslinska+2021}, to first order, we expect to still retrieve a distinct redshift evolution, \RBBH for low and high mass BHs because the existence of the CE channel and stable RLOF channel is not affected by IMF changes.  A full study of the effect of a non-universal IMF is outside the scope of this paper.

\subsection{Contribution from other formation channels}
\label{ss: other channels}
In this work, we focus on predictions from the isolated binary channel. However, the observed population of merging BBHs is most likely a mixture of several channels \citep{Zevin2021,Wong2021_fmix}.
The variety of physics involved is vast, and hence the span of predictions for merging BBH properties is equally large. See also \citet{Mapelli2021} and \citet{MandelFarmer2018} for reviews of proposed formation channels, and \cite{MandelBroekgaarden2021} for a review of predictions for the merger rates from said formation channels.
Below we summarise findings for other formation channels, with an emphasis on delay-time predictions, the slope of \RBBH, and the predicted mass distribution (see also, \citealt{FishbachKalogera2021} for an overview of delay time predictions from several different formation channels).

Two formation channels which exhibit a preference for the formation of more massive BBHs are chemically-homogeneous evolution \citep[CHE; e.g.\ ][]{de-Mink+2009, Song+2013,Song+2015,MandelMink2016,Marchant+2016,Riley2021} and Population III binaries \citep[e.g.\ ][]{Marigo:2001,Belczynski:2004popIII,Kinugawa:2014,Inayoshi2017}.
\cite{Riley2021} find that CHE binaries have quite short delay times (between $0.1-1\Gyr$), causing the redshift evolution of \RBBH to be fairly similar between CHE binaries and the full population of isolated binaries.  
\cite{duBuisson2020} furthermore find that the intrinsic BBH merger rate from CHE binaries evolves less steeply at low redshift than their adopted SFRD. 
\cite{Ng+2021} compare the intrinsic BBH merger rate density from formation in isolated binaries and dynamical formation in globular clusters, to predictions for BBH mergers formed from Population III stars. They find that Population III remnants should result in a secondary peak of \RBBH around $z\approx12$ (beyond what we have adopted as the redshift of first star formation). 

Several formation channels have been proposed where the BBH merger is assisted by dynamical encounters. These include BBH formation in nuclear star clusters \citep[e.g.\ ][]{Antonini+2016,PetrovichAntonini2017,Antonini+2019,ArcaSedda+2020,ArcaSedda2020,FragioneSilk2020}, globular clusters \citep[e.g.\ ][]{Downing2010,Bae2014,Askar2017,Fragione2018,Rodriguez:2019huv} and young stellar clusters \citep[e.g.\ ][]{PortegiesZwart2000,Mapelli2013,Ziosi2014,Mapelli2017, Bouffanais2019,Fragione2021}.
For globular clusters, \cite{Choksi2019} find a merger rate that is weakly increasing out to $z=1.5$ and drops at higher redshift. This behaviour is driven by dynamical processes within the cluster, which introduce a significant delay between cluster formation and BBH mergers.

Recent studies aim to compare the redshift evolution of the intrinsic BBH merger rate between different formation channels. 
\cite{Zevin2021} investigate the local source properties for the CE channel, stable RLOF channel, globular clusters and nuclear clusters. Their Figure 1 shows evidence that the stable RLOF channel preferentially forms higher chirp masses than the CE channel. 
\citet{Mapelli+2021} compare the rate evolution of the intrinsic BBH merger rate from isolated binaries to the rate from nuclear star clusters, globular star clusters and young stellar clusters. They find that the primary BH mass function is more top heavy at high redshift for both globular and nuclear star clusters.
%
In contrast to our work, they find that the mass distribution from isolated binaries does not vary greatly with redshift, because the majority of systems in their isolated binary channel is formed through CE, which results in short delay times. 
However, the mass distribution of isolated binaries in their Figure 5 appears to contain fewer primary BH masses of $\gtrsim 20\Msun$ at redshift 4 relative to redshift 0 (although this effect is smaller than the variation with redshift that they retrieve for nuclear and globular clusters). 

Lastly, AGN disks \citep[e.g.\ ][]{Baruteau2011,Bellovary2016,Leigh2018,Yang2019,Secunda2019,McKernan2020}, and mergers in hierarchical systems assisted by dynamical interactions  \citep[e.g.\ ][]{Kimpson2016,Antonini+2017,RodriguezAntonini2018,Hoang2018} have also been proposed as promising formation channels for BBH mergers.

At the time of writing, the estimates for the relative contribution of formation channels are highly uncertain.
However, linking source properties to predictions for the rate evolution with redshift, such as in this work, could help distinguish between the many possible origins of merging BBH systems.

\section{Conclusions and summary }
\label{sec: conclusion}
We discuss the implications of relations between the delay time and BH mass for BBH systems that originate from isolated binaries. 
We explore the origin of these relations by dividing our simulations into two main formation channels: BBH systems that have experienced at least one common envelope (the `CE channel') and systems that did not experience a CE, i.e.\ that only experienced stable Roche-lobe overflow (the `stable RLOF channel').
We discuss how our findings affect the redshift evolution of the BBH mass distribution.
Specifically, we find a distinct redshift evolution of the BBH merger rate, \RBBH, for different primary BH masses, \Mbheen. 
Below we summarise our main findings.

\paragraph{ \textbf{The CE channel predominantly forms BBH systems with masses $\Mbheen \lesssim 30\Msun$ and typically short delay times ($\tdelay < 1 \Gyr$)}}
    The CE channel typically leads to shorter separations at BBH formation than the stable RLOF channel. This causes on average shorter inspiral times and thus shorter delay times (Figure~\ref{fig: Mbh_tdelay}). 
    The CE channel does not form more massive BHs, because the massive progenitor stars required for these BH masses experience less radial expansion and stronger winds with respect to their lower mass counter parts. This results in conditions that are ill-favoured for successful common-envelope initiation and ejection.

\paragraph{ \textbf{The stable RLOF channel generally forms BBH systems with longer delay times ($\tdelay \gtrsim 1 \Gyr$) and it is the main source of BBH systems with $\Mbheen \gtrsim 30\Msun$. }}    
    The stable RLOF channel primarily produces larger separations at BBH formation than the CE channel, which result in longer delay times. Because high mass stars are ill-favoured for successful common-envelope initiation and ejection, the highest mass BHs are almost exclusively formed through the stable RLOF channel.

\paragraph{\textbf{The redshift evolution of the intrinsic BBH merger rate density is different for low and high \Mbheen}}
    Due to the relations between the delay time and BH mass, we find distinctly different slopes in the BBH merger rate density \RBBH for different mass ranges of \Mbheen (see Figure~\ref{fig: rate dens - z}).
    The merger rate density of the lowest mass BHs ($\Mbheen \leq 20\Msun$) is dominated by the CE channel. For these BH masses, the merger rate density has a slope at low redshift that is similar to the slope of the star formation rate.
    The merger rate density of the highest mass BHs ($\Mbheen \geq 30\Msun$) is dominated by the stable RLOF channel. These higher mass systems have relatively longer delay times ($\tdelay > 1 \Gyr$), causing the rate density to peak at lower redshift than the peak of the star formation rate. We find that in the low-redshift regime that current detectors probe, the evolution of the merger rate density is less steep for higher-mass \Mbheen than for lower-mass BHs.

    Although we cannot find significant evidence for this relation in the observed data at the time of writing, if isolated binaries contribute significantly to the BBH merger rate density, we expect that the distinct redshift evolution of the intrinsic merger rate density for different BH masses will be verifiable with near-future detectors (see Section~\ref{ss: observing slopes in LIGO}).

\paragraph{\textbf{The contribution of different formation channels to \RBBH varies with redshift.}}
    While the CE channel dominates the production of merging BBHs in the Universe, we predict that almost half of the systems we see merging at redshift 0 come from the stable RLOF channel (Figure~\ref{fig: fractional contribution}). Conversely, in the high redshift Universe, the contribution to \RBBH from the stable RLOF channel will be negligible.

\section*{Acknowledgements}
We thank Charlie Conroy and Eva Laplace for useful discussions and support. We thank Will Farr for his suggestion to investigate trends in the rate per mass bin and Maya Fishbach for discussing the impact of the results at an early stage. The authors thank Lokesh Khandelwal for his invaluable work on STROOPWAFEL.
The authors are furthermore grateful for stimulating conversations with and Katie Breivik and members of the BinCosmos, COMPAS, CCA-GW and MPA stellar groups. LvS performed portions of this study as part of the remote pre-doctoral Program at the Center for Computational Astrophysics of the Flatiron Institute, supported by the Simons Foundation.  LvS and SdM also acknowledge KITP for hospitality. 
The authors acknowledge partial financial support from the  National Science Foundation under Grant No. (NSF grant number 2009131  and PHY-1748958).”
, the Netherlands Organisation for Scientific Research (NWO) as part of the Vidi research program BinWaves with project number 639.042.728 and the European Union’s Horizon 2020 research and innovation program from the European Research Council (ERC, Grant agreement No. 715063). IM is partially supported by the Australian Research Council (ARC) Centre of Excellence for Gravitational-wave Discovery (OzGrav), project number CE170100004. IM is the recipient of the ARC Future Fellowship FT190100574.
This research has made use of NASA’s Astrophysics Data System Bibliographic Services.

\section*{Software and Data}
The data used in this work is available on Zenodo under an open-source 
Creative Commons Attribution license at 
\dataset[doi:10.5281/zenodo.5544170]{https://doi.org/10.5281/zenodo.5544170}.
Simulations in this paper made use of the \COMPAS rapid binary population synthesis code (v02.19.04), which is freely available at \url{http://github.com/TeamCOMPAS/COMPAS}  \citep{COMPAS_method}. The data used in Appendix \ref{s: physics variations} is described in \cite{Broekgaarden+2021b} and is publically available at \url{https://zenodo.org/record/5651073}. 
The authors use the adaptive importance sampling tool STROOPWAFEL from \cite{Broekgaarden2019}, publicly available at \url{https://github.com/lokiysh/stroopwafel}. 

This research has made use of GW data provided by the Gravitational Wave Open Science Center (\url{https://www.gw-openscience.org/}), a service of LIGO Laboratory, the LIGO Scientific Collaboration and the Virgo Collaboration. 
LIGO Laboratory and Advanced LIGO are funded by the United States National Science Foundation (NSF) as well as the Science and Technology Facilities Council (STFC) of the United Kingdom, the Max-Planck-Society (MPS), and the State of Niedersachsen/Germany for support of the construction of Advanced LIGO and construction and operation of the GEO600 detector. Additional support for Advanced LIGO was provided by the Australian Research Council. Virgo is funded, through the European Gravitational Observatory (EGO), by the French Centre National de Recherche Scientifique (CNRS), the Italian Istituto Nazionale di Fisica Nucleare (INFN) and the Dutch Nikhef, with contributions by institutions from Belgium, Germany, Greece, Hungary, Ireland, Japan, Monaco, Poland, Portugal, Spain.  

Further software used in this work: Python \citep{PythonReferenceManual},  Astropy \citep{astropy:2013,astropy:2018} Matplotlib \citep{2007CSE.....9...90H},  {NumPy} \citep{2020NumPy-Array}, SciPy \citep{2020SciPy-NMeth}, \texttt{ipython$/$jupyter} \citep{2007CSE.....9c..21P, Kluyver2016jupyter},  Seaborn \citep{waskom2020seaborn}  and  {hdf5}   \citep{collette_python_hdf5_2019}.

\appendix

\section{  Inspecting mass ratios}
\label{sec: q per delay time}

Below we derive the typical minimum mass ratio of a BBH that forms through the stable RLOF channel, as a function of the uncertain assumptions that go into our population synthesis. 
 We will refer to the the star that is more (less) massive at zero-age main sequence (ZAMS) as the primary (secondary) and with the subscript A (B). 
See Figure~\ref{fig: channels cartoon} for a cartoon example of a stable RLOF system,  including a short definition of the symbols as used in this section. 

\subsection{First mass transfer: from the primary to the secondary}
Since the primary star is more massive, it will evolve on a shorter timescale than the secondary and thus it will be the first to overflow its Roche Lobe. The donor (primary star) typically starts RLOF either at the end of its main sequence,  or during H-shell burning, also known as Case A or early Case B mass transfer. 
We will focus on Case B mass transfer (post core H burning) because, due to the large radial expansion, this is most common case of mass transfer \citep[e.g.\ ][]{Sana+2012}. 
During this phase of stable mass transfer, the primary star will donate at most its envelope to the secondary star. 
We neglect all mass loss due to winds in this simple approximation. 
We capture the mass transfer efficiency in the parameter $\beta$, where $\beta=0$ implies no mass is accreted, while $\beta=1$ implies the complete envelope of the primary is accreted by the secondary. The mass of the secondary after completion of the first mass transfer phase becomes:

\begin{equation}
\label{eq: mtilde}
\begin{aligned}
    \tilde{M_{\mathrm{B}}} & = \MZAMSb + \beta M_{env,A} =  \MZAMSa \cdot q_{\mathrm{ZAMS} } + \MZAMSa \cdot \beta (1 - f_{\mathrm{core} }) \\
    & =  \MZAMSa \cdot \left( q_{\mathrm{ZAMS} } + \beta(1 - f_{\mathrm{core} }) \right),
\end{aligned}
\end{equation}

\noindent where $q_{\mathrm{ZAMS} } \equiv \MZAMSb/\MZAMSa$, and we assume a fraction $f_{\mathrm{core} }$ of the stellar mass is used to form the He core. We implicitly assume the core mass fraction of star A and star B are similar, i.e. $f_{\mathrm{core,A}}/f_{\mathrm{core,B}} \approx 1$.


The primary star will continue to evolve and ultimately form a BH. For the purpose of this argument, we assume the complete core mass of the primary goes into forming the BH mass, i.e.

\begin{equation}
\label{eq: Mbheen}
    \MbhA = \MZAMSa \cdot f_{\mathrm{core} }. 
\end{equation}

\subsection{Second mass transfer: from the secondary to the primary}
When the secondary star ends core-H burning, it will swell up in size and, in our case, start stable mass transfer. 
The second phase of mass transfer is highly non-conservative, since accretion onto the BH is assumed to be Eddington limited. Therefore, \MbhA remains approximately the same, and \MbhB will be approximately;

\begin{equation}
\label{eq: Mbhtwee}
    \MbhB = \tilde{M_{\mathrm{B}} } f_{\mathrm{core} }, 
\end{equation}
where we again assume that the complete He core mass is used to form the BH mass.

\subsection{Final mass ratio}

 We find that for the stable channel, \MbhB typically forms the more massive BH, because in most cases star B accretes a significant fraction of its companions envelope, making it more massive than the primary at ZAMS. 
Hence, we define the typical final mass ratio at BBH formation as:

\begin{equation}
     \qfinal \equiv \frac{\MbhA}{\MbhB} \approx \qBBH.
\end{equation}

Using Equations~\ref{eq: Mbheen} and \ref{eq: Mbhtwee} we find

\begin{equation}
\label{eq: q_final}
\begin{split}
    \qBBH & = \frac{\MZAMSa}{\tilde{M_{\mathrm{B}}} }
    =  \frac{1}{(q_{\mathrm{ZAMS} } + \beta(1 - f_{\mathrm{core}}) )} . 
\end{split}
\end{equation}

 We find that in our simulations, core mass fractions range between about 0.33 and 0.43,  To minimise equation \ref{eq: q_final} we further need to maximise $q_{\mathrm{ZAMS}} = 1$ and $\beta = 1$.  Hence we find $\min(\qfinal) \approx 0.60-0.64$. This agrees broadly with the location of the drop in the distribution of mass ratios that we find in our simulations below around $\qfinal \approx 0.6$, shown in Fig.~\ref{fig: other obs}. 
Understanding the right hand side of the mass ratio distribution is more involved. It is set in part by the requirement that the systems shrinks sufficiently during the second mass transfer, but also by mass transfer efficiency itself.

\begin{figure}
\begin{center}
\includegraphics[width=0.75\textwidth]{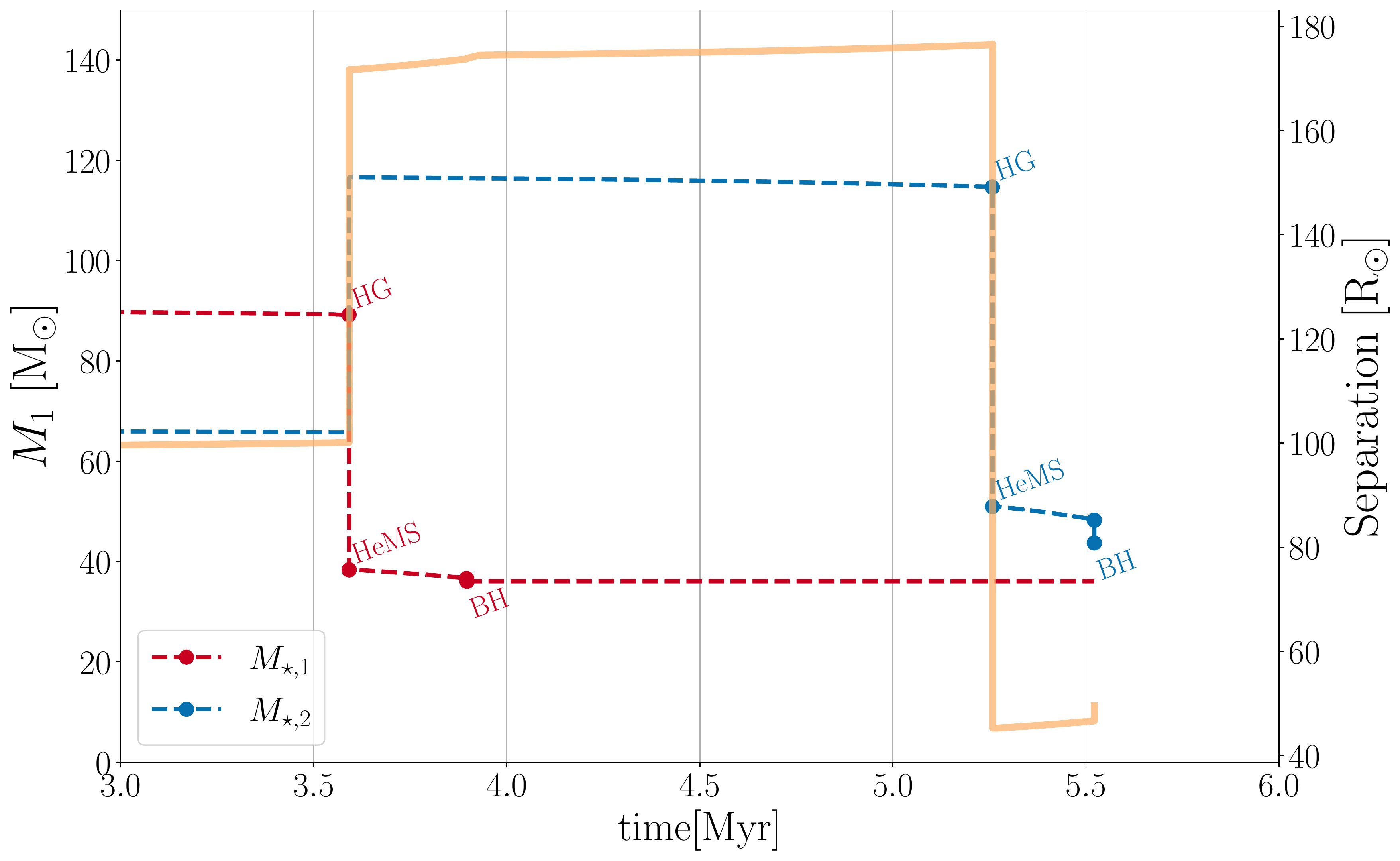}
\caption{Masses (dashed lines, left y-axis) and binary orbital separation (solid yellow line, right y-axis) over time, for a typical BBH progenitor system that evolved through the stable RLOF channel. Transitions to different evolutionary stages are labelled with the following acronyms: HG for Hertzsprung Gap star, HeMS for He Main Sequence star and BH for Black Hole. 
\label{fig: detail RLOF sys}}  
\end{center}
\end{figure}

 For illustration, we also show a typical example system in Figure~\ref{fig: detail RLOF sys}.  This system started with 
$M_{\mathrm{ZAMS,1}} \approx 90\Msun$ and $M_{\mathrm{ZAMS,2}} \approx 70\Msun$ 
and ends with $\MbhA=36\Msun$ and $\MbhB=43\Msun$, hence $\qfinal\approx 0.84$.
 
\section{Delay time distributions}
\label{s: delay times}
\begin{figure}
\includegraphics[width=0.47\textwidth]{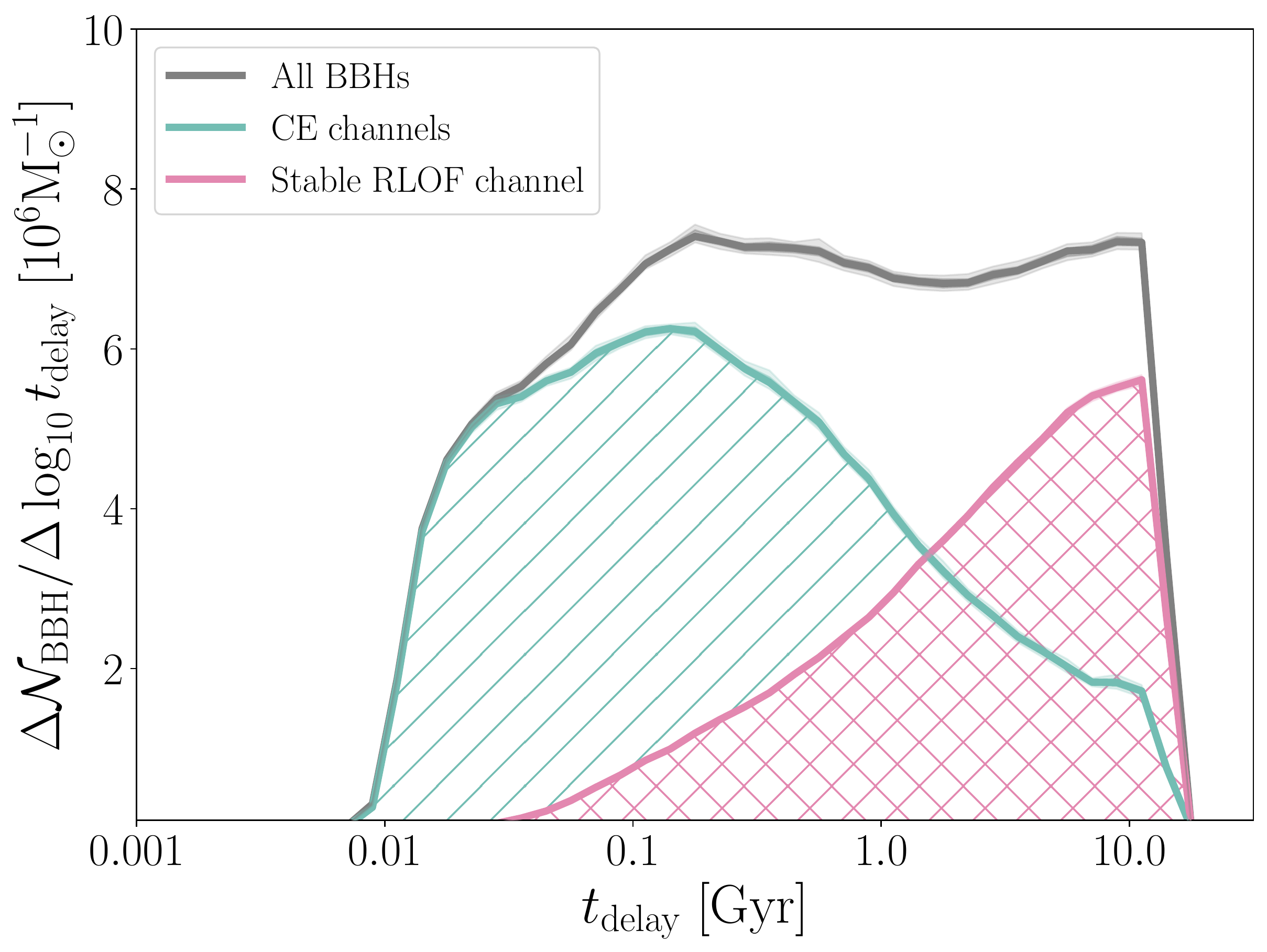}
\includegraphics[width=0.47\textwidth]{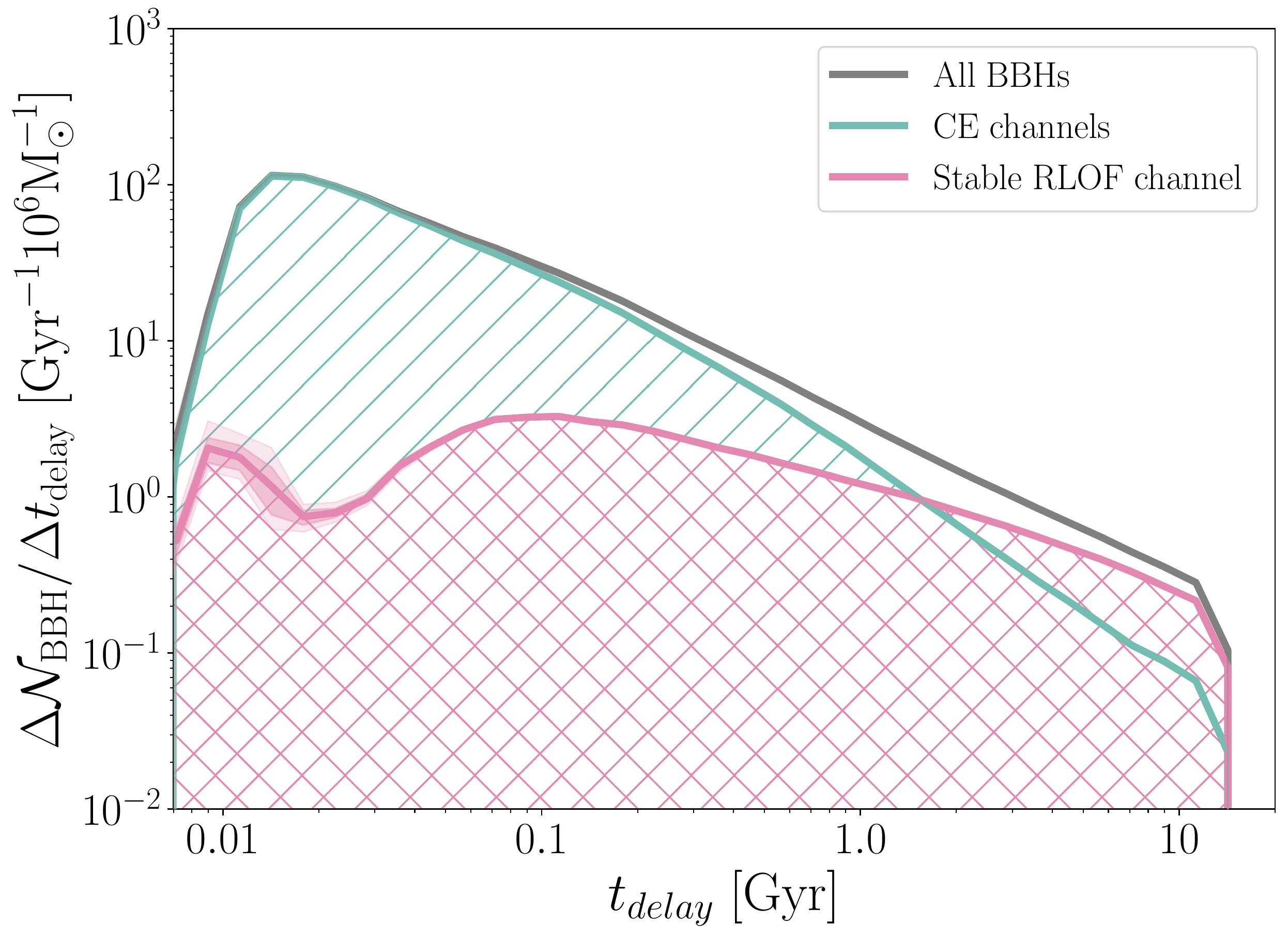}
\caption{Similar to Figure~\ref{fig: Mbh_tdelay}, but integrated over \Mbheen.
The solid line shows the centres of the histogram per $d\log_{10}\tdelay$ (left panel) versus the histogram per $d\tdelay$ (right panel), with bin sizes that are equal size in log-space ($d\log_{10}\tdelay = 0.1$), and hence unequal size in \tdelay.  Both are normalised per $10^6 \Msun$ of star forming mass. This histogram contains a mixture of birth metallicities, that were sampled uniformly in log.  
The dark and light shaded areas shows the 1- and 2-$\sigma$ bootstrapping uncertainties respectively. 
We indicate the stable RLOF channel with pink cross hatched lines, and the CE channel with green line hatches.  \label{fig: tdelay}}
\end{figure}

We emphasise the bimodality in the delay time distribution by plotting the number of merging BBHs per $\log \tdelay$ in the left panel of \ref{fig: tdelay}. This is similar to Figure~\ref{fig: Mbh_tdelay}, but integrated over all BH masses. For completeness, we also show the same distribution, but per \tdelay (i.e.\ not in log space). 

\section{ Metallicity-dependent star formation rate \SFRDzZ}
\label{s: fitting SFRD}
 
Several recent studies have highlighted the importance of the choice of the metallicity dependent cosmic starformation rate density \SFRDzZ and the impact on the final predictions \citep[e.g.][]{Chruslinska2019_effectCO, Chruslinska2019, Neijssel+2019, Broekgaarden+2021b, Briel+2021}.

For the metallicity dependent starformation history assumed in this work we use the IllustrisTNG simulations.  This is a suite of large magneto-hydrodynamical cosmological simulations computed with the moving-mesh code Arepo \citep{Springel2010, Pakmor+2016, Weinberger+2020}. The simulations  follow the formation and evolution of galaxies from high redshift to the current time and solve for the  evolution of dark matter and gas under the influence of feedback from star formation and supermassive blackholes \citep[for details see][]{FirstResTNG_Springel2018,FirstResTNG_Marinacci2018, FirstResTNG_Nelson2018,FirstResTNG_Pillepich2018, FirstResTNG_Naiman2018}.

The simulations were originally calibrated against the observed total cosmic star formation rate density and the stellar mass function of galaxies \citep{Pillepich2018}. They reproduce the evolution of the sizes of galaxies with redshift \citep{Genel+2018} and with observational constraints on the mass-metallicity relation of
galaxies up to z = 2 \citep{Torrey+2019} as well as iron abundances \citep{FirstResTNG_Naiman2018} and
the metallicity gradients within galaxies at low redshift \citep{Hemler+2021}. These simulations have also already been used to make predictions for gravitational wave sources through pairing with  predictions for the outcomes of binary evaluation obtained with the BPASS code \citet{Briel+2021}.

We extract the amount of starformation ongoing at each redshift and metallicity in the IllustrisTNG100 simulations and use this to derive the  metallicity cosmic starformation rate density,  \SFRDzZ.  For this we make use of an analytical fit inspired by \citet{Neijssel+2019}, but adapted to better capture the asymmetry in the metallicity distribution as detailed in Van Son et al. (in prep.). For the simulations presented in this work we use  
\begin{equation}
\label{eq: pure log skew}
 \SFRDzZ = \underbrace{
    a \frac{\left(1 + z\right)^b}{1 + \left[ (1 + z)/c \right]^d}  
 }_{(1) \ \SFRDz} \,\,
    \underbrace {\frac{2}{Z} \phi \left(\frac{\ln Z - \xi(z)}{\omega(z)}\right)
                \Phi\left(\alpha \frac{\ln Z - \xi(z)}{\omega(z)} \right)
                }_{(2) \ \dpdZ } \,\,
            \left[ \Msun \,yr^{-1} \,cMpc^{-3} \right]
\end{equation}
where the first term (1) governs the overall starformation rate density \SFRDz, as a function of redshift $z$ \citep[following the analytical form proposed by ][]{MadauDickinson2014}. The second term (2) governs the metallicity distribution at each redshift, we approximate this with a skewed log-normal distribution written as the product of the standard log-normal distribution, $\phi$, and the cumulative distribution function of the standard log-normal distribution, $\Phi$ \citep{Ohagan+1976}. For the width of the distribution we assume $\omega(z) = \omega_0 \cdot 10^{\omega_z \cdot z}$.
We furthermore ensure that mean of the metallicity distribution has the following simple dependence on redshift  $\langle Z \rangle \equiv \mu(z) = \mu_0 \cdot 10^{\mu_z \cdot z}$ by setting
\begin{equation}
\label{eq mu z}
    \xi(z) = \frac{- \omega(z)^2}{2} \, \ln\left(\frac{  \mu_0 \cdot 10^{\mu_z \cdot z} }{2 \Phi(\beta \omega (z))}  \right)  \quad \text{where} \quad \beta =  \frac{\alpha}{\sqrt{1 + \alpha^2} }.
\end{equation}
This leaves us in total with nine free parameters which are fitted simultaneously. In this work we have used   
$a=0.02$, $b=1.48$, $c=4.45$,  $d=5.9$,  
$\alpha = -1.77$, 
$\mu_0 = 0.025, \mu_z = -0.048$, 
$\omega_0 = 1.125$, 
and $\omega_z = 0.048$  (c.f. Van Son et al. in prep).

We note that our approach differs from the approach taken in some earlier studies that use observed scaling relations to construct a prescription for the metallicity dependent cosmic star formation history, for example as proposed by \citet{Langer2006}. 
Unfortunately, the observational constraints are scarce at high redshift, where simple extrapolations may not be valid. This is problematic for gravitational wave sources, which preferentially form from low metallicity star formation which is most poorly constrained, especially at high redshift \citep[cf. ][]{Chruslinska+2021}.  We have therefore opted instead to make use of current state-of-the-art cosmological simulations \citep[see also][ for a discussion]{Briel+2021}. These provide physically motivated predictions at high redshift and have by now been extensively compared with observational constraints at lower redshift. Despite the large remaining uncertainties in these simulations, we believe this to be our best option at current times.

\section{The redshift dependence of the merger rate as a function of chirp mass}
\label{s: rate chirp mass}
 
\begin{figure}
\centering
\includegraphics[width=0.32\textwidth]{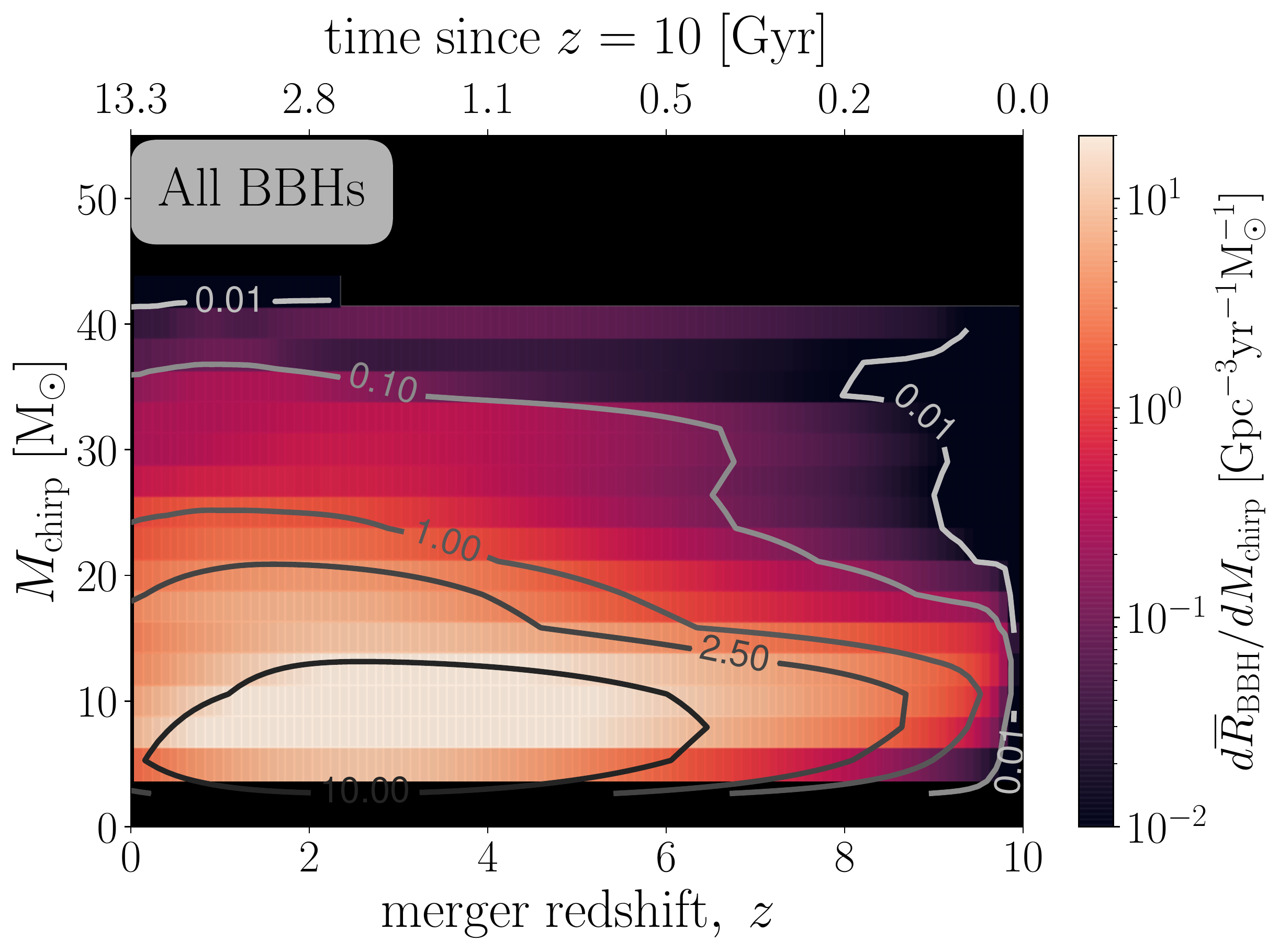}
\includegraphics[width=0.32\textwidth]{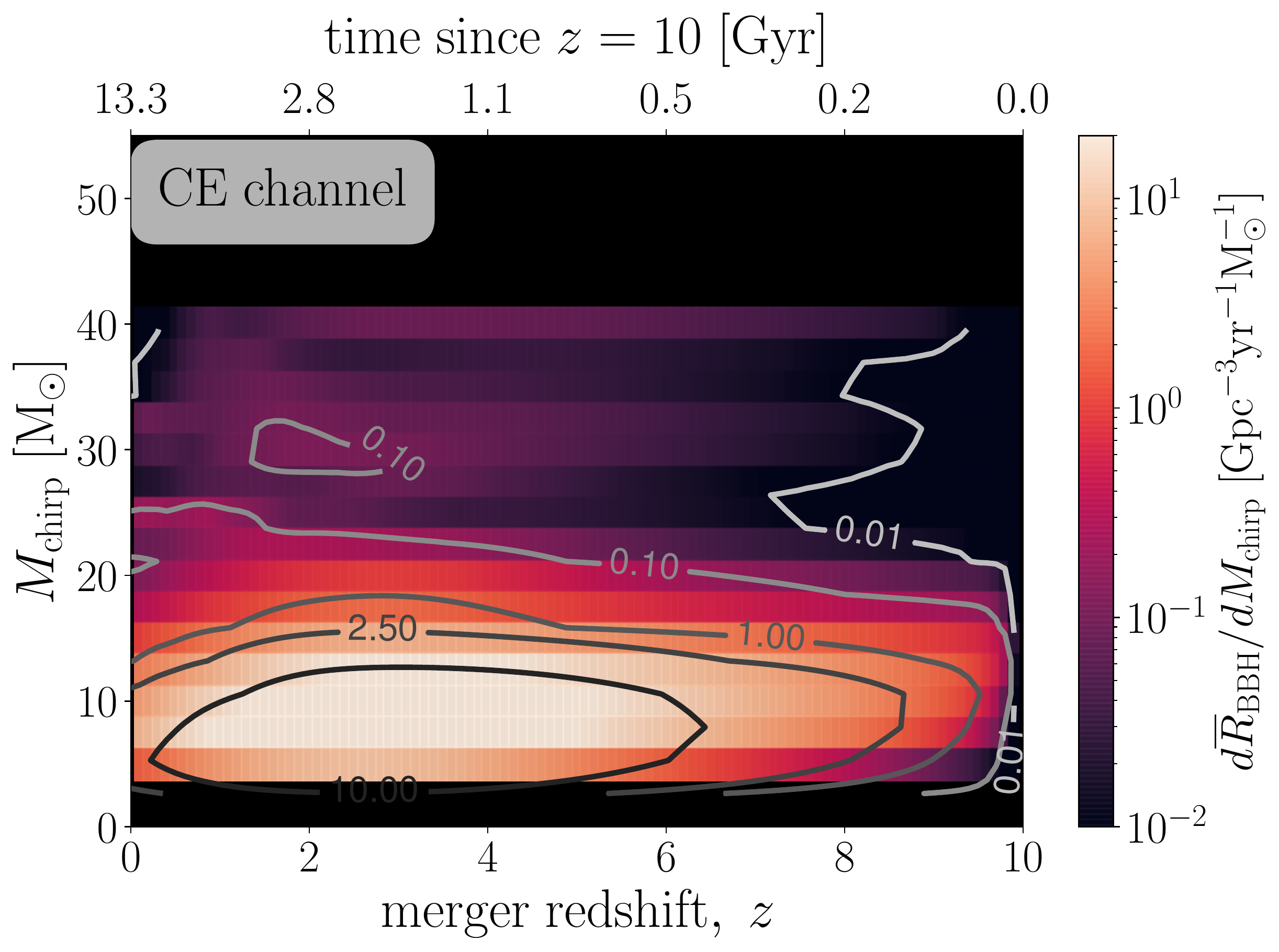}
\includegraphics[width=0.32\textwidth]{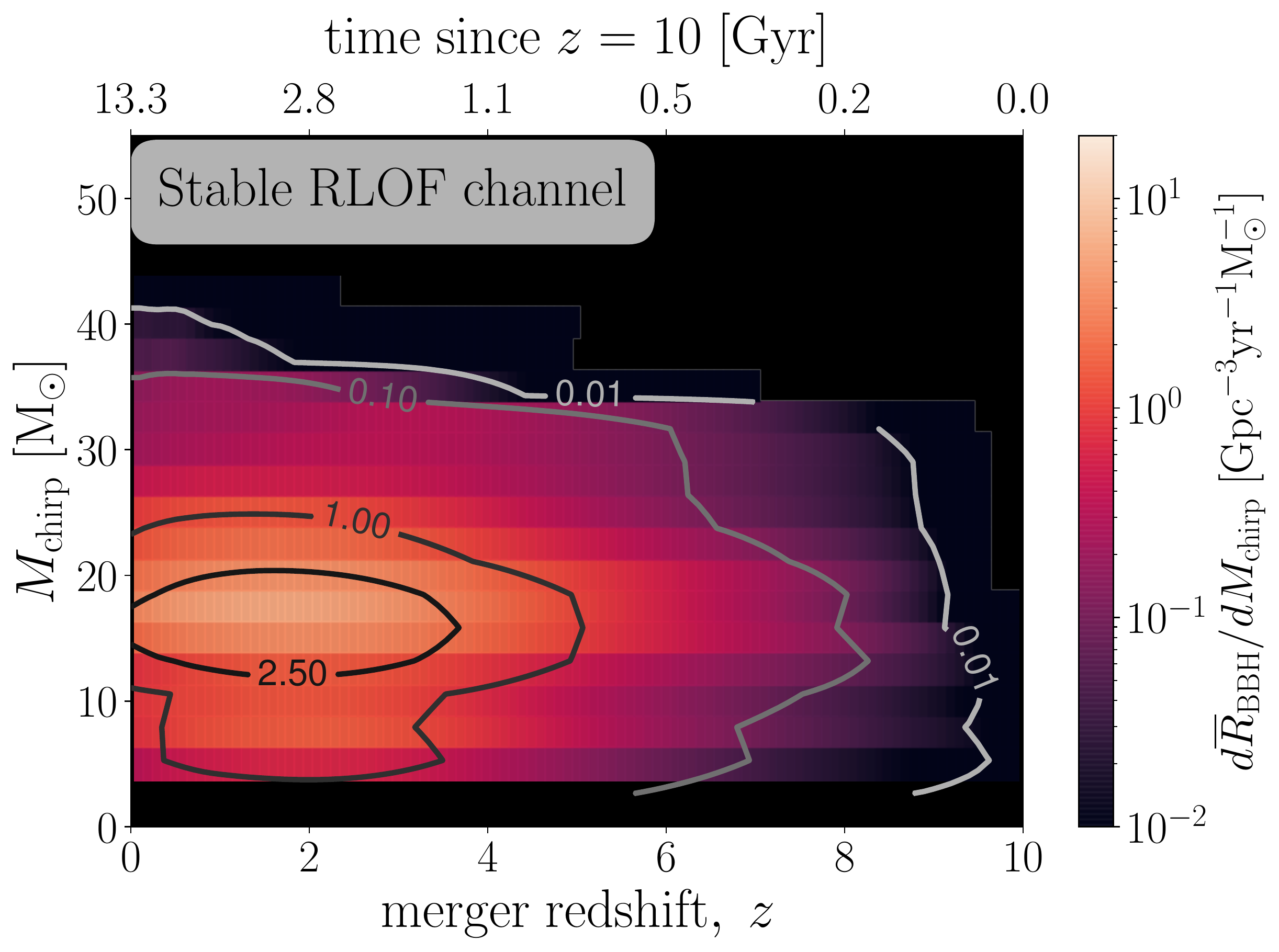}
\caption{The averaged intrinsic merger rate density \RBBHav, for redshift bins of $dz=0.2$, and chrip mass bins of $d\Mchirp=~2.5\Msun$. The colours and symbols are the same as in Figure~\ref{fig: MBH-z}. \label{fig: Mchirp-redshift}}
\end{figure}

In Figure~\ref{fig: Mchirp-redshift} we show the same evolution of \RBBH per primary BH mass, in the merger redshift -- \Mbheen plane as displayed in Figure~\ref{fig: MBH-z}, but as a function of chirp mass, \Mchirp.
 We observe similar trends in the BBH merger distribution when we investigate \Mchirp instead of \Mbheen. Specifically, BBH mergers with high chirp mass ($\Mchirp > 20 \Msun$) originate predominantly from the stable RLOF channel, while the CE channel dominates the BBH merger rate for low chirp mass ($\Mchirp \leq 20 \Msun$).

\section{Mass distribution split by formation channel and metallicity}
\label{s: mass dist channel and Z}

\begin{figure}
\centering
\includegraphics[width=0.49\textwidth]{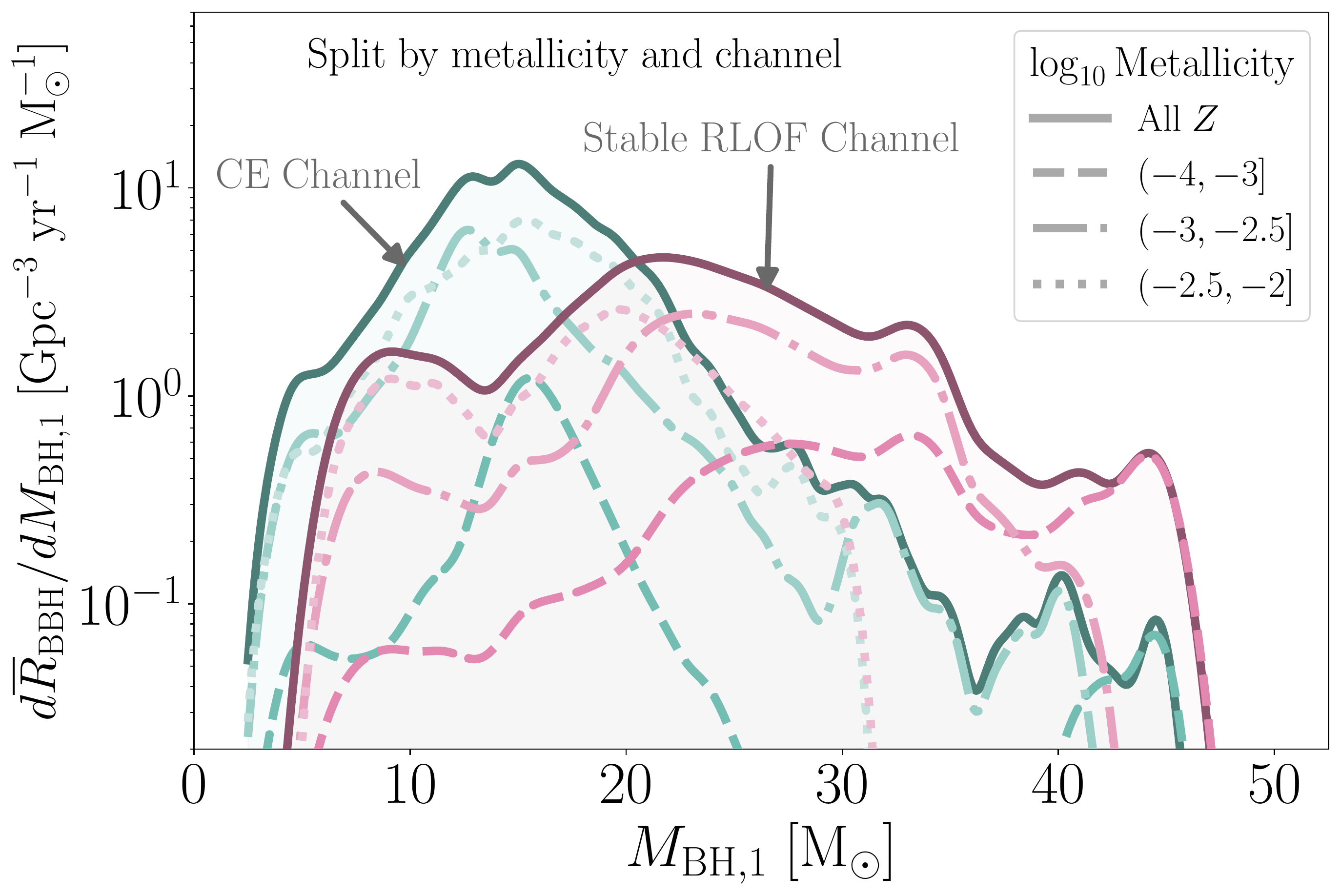} 
\caption{Distribution of primary BH masses \Mbheen split by formation channel and birth metallicity, for merger redshifts between $0 \leq z <0.5$. \label{fig: mass dist channels and Z}}
\end{figure}

 In Figure \ref{fig: mass dist channels and Z} we show the \Mbheen distribution split by both formation channel and formation metallicity. We apply the same metallicity bins as those in Figure \ref{fig: mass dist by metals}, but exclude the highest metallicity bin to focus on metallicities low enough to form BHs with masses above $20\Msun$. This shows that the stable RLOF channel dominates the high mass end of the distribution at every metallicity.

\section{Physics variations}
\label{s: physics variations}
To test the robustness of our finding that the CE channel and stable RLOF channel lead to distinct distributions in delay time and primary BH mass, we use the grid of models presented in \cite{Broekgaarden+2021a} and \cite{Broekgaarden+2021b}. These simulations were performed with a version of \COMPAS that predates the publicly available code (most similar to version 02.13.01 of the publicly available code).  

\begin{figure*}
\centering
\includegraphics[width=0.97\textwidth]{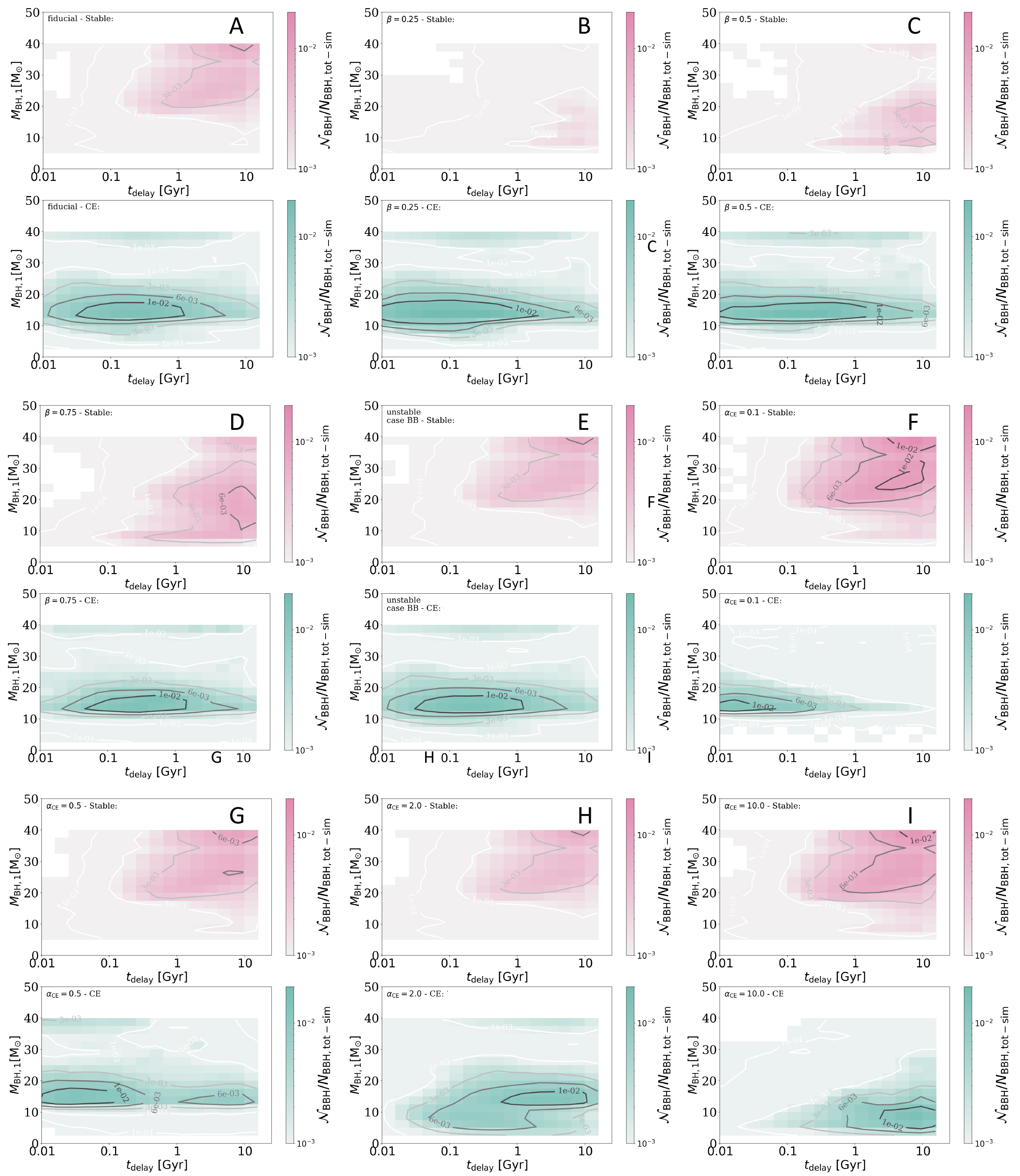}
\caption{Same as Figure~\ref{fig: Mbh_tdelay} but for several variations in the assumed model physics, based on models presented in \cite{Broekgaarden+2021a} and \cite{Broekgaarden+2021b}. The models in each panel are as follows. 
Panel A: the fiducial model (see text). Panels B, C, and D: fixed mass-transfer efficiency of $\beta = 0.25$, $0.5$, and $0.75$ respectively. Panel E: case BB mass transfer is assumed to be always unstable. Panels F, G, H, I: the CE efficiency parameter, $\alpha_{\mathrm{CE} }$, is set to 0.1, 0.5, 2.0, and 10.0 respectively. 
\label{fig: physics variations1}}
\end{figure*}

\begin{figure*}
\includegraphics[width=0.97\textwidth]{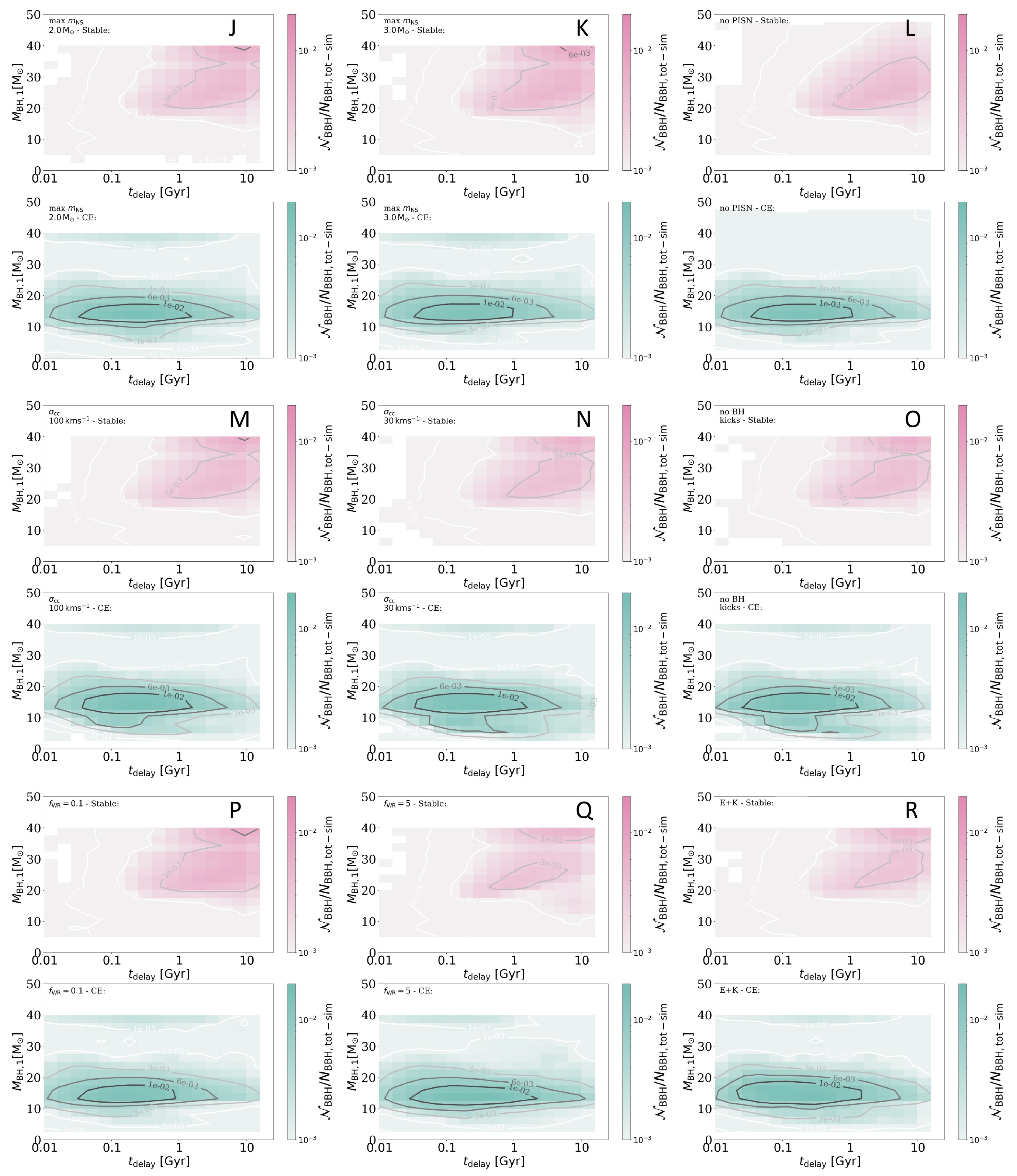}
\caption{Same as Figure~\ref{fig: physics variations1} but for the following model variations:
Panels J and K: maximum neutron star mass is fixed to  $2.0\Msun$ and $3.0\Msun$ respectively. Panel L: no PPISN or PISN implemented. Panels M and N: natal kicks are drawn from a Maxwellian velocity distribution with a one-dimensional root-mean-square velocity dispersion of $\sigma_{CC}$ = 100\kms and 30\kms respectively. Panel O: BHs are assumed to receive no natal kick. Panels P and Q vary the strength of the Wolf-Rayet-like wind mass loss by a constant factor of $f_{\mathrm{WR}} = 0.1$ and 5 respectively. Panel R combines the assumption that case BB mass transfer is always unstable with allowing Hertzsprung-gap donor stars which initiate a CE to survive the CE event (models E and S).\label{fig: physics variations2}}
\end{figure*}

\begin{figure*}
\centering
\includegraphics[width=0.6\textwidth]{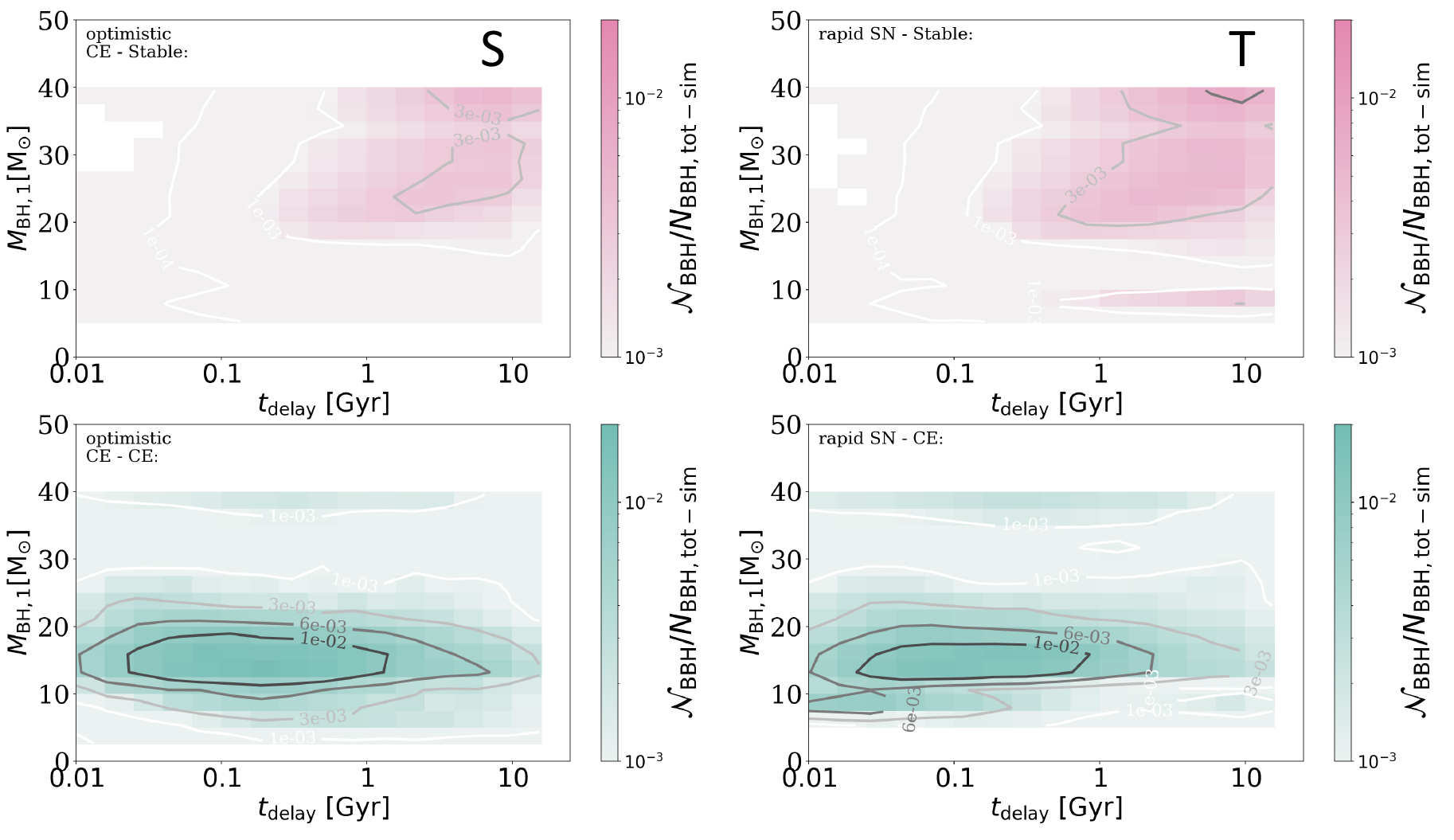}
\caption{Same as Figure~\ref{fig: physics variations1} but for the following model variations:
Panel S: Hertzsprung-gap donor stars initiating a CE are allowed to survive this CE event. Panel T: adopts the \cite{Fryer+2012} ``rapid'' supernova remnant-mass prescription. 
\label{fig: physics variations3}}
\end{figure*}

In Figures~\ref{fig: physics variations1}, \ref{fig: physics variations2}, and \ref{fig: physics variations3}, we show the distribution of primary BH mass ($\Mbheen$) and delay time ($\tdelay$) similar to Figure~\ref{fig: Mbh_tdelay}. Each panel in these Figures displays a separate simulation of $53 \times 10^6$ binaries.
The fiducial model in this grid (panel A in  Figure~\ref{fig: physics variations1}) adopts physics assumptions that are very similar to our model assumptions as described in Section~\ref{sec: method}.
The exceptions are the PPISN prescription \citep[which follows][]{Marchant+2019}, the metallicity sampling (which uses a discrete grid of 53 metallicities between $10^{-4} - 0.03$), and the LBV wind prescription \citep[LBV-type stars, that is, stars above the Humphreys-Davidson limit, are assumed to receive an \textit{additional} wind mass loss of $10^{-4} \Msun yr^{-1}$, inspired by][]{Belczynski_2010}.

Each panel in Figures~\ref{fig: physics variations1}, \ref{fig: physics variations2}  and \ref{fig: physics variations3} considers a physics variation with respect to the fiducial model in panel A. The variations are summarised in the caption of each Figure, and for a full description of the physics assumptions we direct the reader to \cite{Broekgaarden+2021a} and \cite{Broekgaarden+2021b}.

Figures~\ref{fig: physics variations1}, \ref{fig: physics variations2} and \ref{fig: physics variations3} show that the dearth of BBH systems with high mass ($\Mbheen > 30\Msun$) and short delay time ($\tdelay \lesssim 1\Gyr$) is quite robust over numerous physics variations. 
Moreover, as discussed in Section~\ref{sec: discussion}, we retrieve distinct BH-mass and delay-time  distributions for the two channels in almost all variations. 
The exceptions are the models which assume a fixed value for the accretion efficiency $\beta$ of $0.25$ and $0.5$ for episodes of mass transfer with a non-compact accretor (panels B and C in Figure~\ref{fig: physics variations1} ), and the model which assumes a high value for the CE ``efficiency parameter'' ($\alpha_{\mathrm{CE}} = 2$ and $\alpha_{\mathrm{CE}} = 10$; panels H and I in Figure~\ref{fig: physics variations1}).
Those variations in the accretion efficiency $\beta$ diminish the contribution of the stable RLOF channel, and specifically reduce the production of high-mass \Mbheen. This removes the distinction between the channels in the $\Mbheen$ distribution.
Assuming $\alpha_{\mathrm{CE}} = 10$ causes all the short delay-time systems from the CE channel to disappear. This is because at higher $\alpha_{\mathrm{CE}}$, a BH needs to inspiral less deeply into its companion's envelope to achieve envelope ejection. This results in wider post-CE separations and hence more similar delay-time distributions for the two channels.

\bibliography{main}
\bibliographystyle{aasjournal}

\end{document}